\documentclass[prl,nofootinbib,twocolumn,superscriptaddress,preprintnumbers,longbibliography]{revtex4-1} 

\usepackage{amsfonts, amssymb,amsmath,latexsym,graphics, graphicx,epsfig,multirow,comment,
,feyn,slashed,xcolor,afterpage, makecell} 
\usepackage{booktabs}
\usepackage{tabularx,afterpage}
\usepackage[colorlinks=true
,urlcolor=blue
,anchorcolor=blue
,citecolor=blue
,filecolor=blue
,linkcolor=blue
,menucolor=blue
,linktocpage=true
,pdfproducer=medialab
,pdfa=true
]{hyperref}
\usepackage[utf8]{inputenc}
\usepackage{url}

\newcolumntype{C}[1]{>{\centering\let\newline\\\arraybackslash\hspace{0pt}}m{#1}}

\newcommand {\be} {\begin {equation}}
\newcommand {\ee} {\end {equation}} 

\newcommand {\bes} {\begin {equation*}}
\newcommand {\ees} {\end {equation*}}

\newcommand{\es}[2] {\begin{equation} \label{#1} \begin{split} #2 \end{split} \end{equation}}

\usepackage{csvsimple}

\newcommand{\beq}{\begin{equation}}
\newcommand{\eeq}{\end{equation}}

\begin{document}

\title{
A Search for Dark Matter Annihilation in Galaxy Groups}

\preprint{MIT/CTP-4931, PUPT 2534, MCTP 17-15}

\author{Mariangela Lisanti}
\affiliation{Department of Physics, Princeton University, Princeton, NJ 08544}

\author{Siddharth Mishra-Sharma}
\affiliation{Department of Physics, Princeton University, Princeton, NJ 08544}

\author{Nicholas L. Rodd}
\affiliation{Center for Theoretical Physics, Massachusetts Institute of Technology, Cambridge, MA 02139}

\author{Benjamin R. Safdi}
\affiliation{Center for Theoretical Physics, Massachusetts Institute of Technology, Cambridge, MA 02139}
\affiliation{Michigan Center for Theoretical Physics, Department of Physics, University of Michigan, Ann Arbor, MI 48109}

\date{\today}

\begin{abstract}
We use 413 weeks of publicly-available {\it Fermi} Pass 8 gamma-ray data, combined with recently-developed galaxy group catalogs, to search for evidence of dark matter annihilation in extragalactic halos.    In our study, we use luminosity-based mass estimates and mass-to-concentration relations to infer the $J$-factors and associated uncertainties for hundreds of galaxy groups within a redshift range $z \lesssim 0.03$.  We employ a conservative substructure boost factor model, which only enhances the sensitivity by an $\mathcal{O}(1)$ factor.  No significant evidence for dark matter annihilation is found 
and we exclude thermal relic cross sections for dark matter masses below $\sim$30 GeV to 95\% confidence in the $b\bar{b}$ annihilation channel. 
These bounds are comparable to those from Milky Way dwarf spheroidal satellite galaxies.  The results of our analysis increase the tension, but do not rule out, the dark matter interpretation of the Galactic Center excess.  We provide a catalog of the galaxy groups used in this study and their inferred properties, which can be broadly applied to searches for extragalactic dark matter.   \end{abstract}
\maketitle

\noindent {\bf Introduction.}  Weakly-interacting massive particles, which acquire their cosmological abundance through thermal freeze-out in the early Universe, are leading candidates for dark matter (DM).  Such particles can annihilate into Standard Model states in the late Universe, leading to striking gamma-ray signatures that can be detected with observatories such as the {\it Fermi} Large Area Telescope.  
Some of the strongest limits on the annihilation cross section have been set by searching for excess gamma-rays in the Milky Way's dwarf spheroidal satellite galaxies (dSphs)~\cite{Ackermann:2015zua,Fermi-LAT:2016uux}.  In this Letter, we present competitive constraints that are obtained using hundreds of galaxy groups within $z\lesssim0.03$. 

This work is complemented by a companion publication in which we describe the procedure for utilizing  galaxy group catalogs in searches for extragalactic DM~\cite{companion}.  Previous attempts to search for DM outside the Local Group were broad in scope, but yielded weaker constraints than the dSph studies.  For example, limits on the annihilation rate were set by requiring that the DM-induced flux not overproduce the isotropic gamma-ray background~\cite{Ackermann:2015tah}.  These bounds could be improved by further resolving the contribution of sub-threshold point sources to the isotropic background~\cite{Zechlin:2016pme,Lisanti:2016jub}, or by  
looking at the auto-correlation spectrum~\cite{Ackermann:2012uf, Ackermann:2012uf,Ando:2006cr,Ando:2013ff}.  A separate approach involves cross-correlating~\cite{Xia:2011ax,Ando:2014aoa,Ando:2013xwa,Xia:2015wka,Regis:2015zka,Cuoco:2015rfa,Ando:2016ang} the {\it Fermi} data with galaxy-count maps constructed from, \emph{e.g.}, the Two Micron All-Sky Survey (2MASS)~\cite{Jarrett:2000me,Bilicki:2013sza}.  A positive cross-correlation was detected with 2MASS galaxy counts~\cite{Xia:2015wka}, which could arise from annihilating DM with mass $\sim$$10$--$100$~GeV and a near-thermal annihilation rate~\cite{Regis:2015zka}.  However, other source classes, such as misaligned Active Galactic Nuclei, could also explain the signal~\cite{Cuoco:2015rfa}.
  
\begin{table*}[htb]
\footnotesize
\begin{tabular}{C{3.0cm}C{2.1cm}C{1.8cm}C{1.8cm}C{1.8cm}C{1.5cm}C{1.6cm}C{1.6cm}C{1.6cm}C{1.6m}}
\toprule
\Xhline{3\arrayrulewidth}
Name &   $\log_{10} J$  &  $\log_{10} M_\text{vir}$ &          $z \times 10^{3}$&        $\ell$ &        $b$ &  $\log_{10} c_\text{vir}$ &  $\theta_\text{s}$  &  $b_\text{sh}$   \\
 & {[GeV$^2$\,cm$^{-5}$\,sr]}& [$M_\odot$] &  & [deg] & [deg] & & [deg] &\\
\midrule
\hline
            NGC4472/Virgo &  19.11$\pm$0.35 &  14.6$\pm$0.14 &   3.58 &  283.94 &  74.52 &  0.80$\pm$0.18 &     1.15 &  4.53 \\
                  NGC0253 &  18.76$\pm$0.37 &  12.7$\pm$0.12 &   0.79 &   98.24 & -87.89 &  1.00$\pm$0.17 &     0.77 &  2.90 \\
                  NGC3031 &  18.58$\pm$0.36 &  12.6$\pm$0.12 &   0.83 &  141.88 &  40.87 &  1.02$\pm$0.17 &     0.64 &  2.76 \\
        NGC4696/Centaurus &  18.33$\pm$0.35 &  14.6$\pm$0.14 &   8.44 &  302.22 &  21.65 &  0.80$\pm$0.18 &     0.47 &  4.50 \\
                  NGC1399 &  18.30$\pm$0.37 &  13.8$\pm$0.13 &   4.11 &  236.62 & -53.88 &  0.89$\pm$0.17 &     0.45 &  3.87 \\
\bottomrule
\Xhline{3\arrayrulewidth}
\end{tabular}
\caption{The top five halos included in the analysis, as ranked by inferred $J$-factor, including the boost factor.  For each group, we show the brightest central galaxy and the common name, if one exists, as well as the virial mass, cosmological redshift, Galactic longitude $\ell$, Galactic latitude $b$, inferred virial concentration~\cite{Correa:2015dva}, angular extent, and boost factor~\cite{Bartels:2015uba}.  The angular extent is defined as $\theta_\text{s} \equiv \tan^{-1} (r_\text{s} / d_c[z])$, where $d_c[z]$ is the comoving distance and $r_\text{s}$ is the NFW scale radius.  A complete table of the galaxy groups used in this analysis, as well as their associated properties, are provided at \url{https://github.com/bsafdi/DMCat}.
}
\label{Jtab}
\end{table*}

An alternative to studying the full-sky imprint of extragalactic DM annihilation is to use individual galaxy clusters~\cite{Ackermann:2010rg, Ando:2012vu,Ackermann:2013iaq,Ackermann:2015fdi,Anderson:2015dpc,Rephaeli:2015nca,2016A&A...589A..33A,Liang:2016pvm,Adams:2016alz,Huang:2011xr}. Previous analyses along these lines have looked at  
  a small number of $\sim$$10^{14}$--$10^{15}$~M$_\odot$ X-ray--selected clusters. 
  Like the dSph searches, the cluster studies have the advantage that the expected signal is localized in the sky, which reduces the systematic uncertainties associated with modeling the foregrounds and unresolved extragalactic sources.  As we will show, however, the sensitivity to DM annihilation is enhanced---and is more robust---when a larger number of targets are included compared to previous studies.

 Our work aims to combine the best attributes of the cross-correlation and cluster studies to improve the search for extragalactic DM annihilation.  We use the galaxy group catalogs in Refs.~\cite{Tully:2015opa} and~\cite{2017ApJ...843...16K} (hereby T15 and T17, respectively), which contain accurate mass estimates for halos with mass greater than $\sim$$10^{12}$~M$_\odot$ and $z \lesssim 0.03$, to systematically determine the galaxy groups that are expected to yield the best limits on the annihilation rate.  The T15 catalog provides reliable redshift estimates in the range $0.01 \lesssim z \lesssim 0.03$, while the T17 catalog provides measured distances for nearby galaxies, $z \lesssim 0.01$, based on Ref.~\cite{Tully:2016ppz}. The T15 catalog was previously used for a gamma-ray line search~\cite{Adams:2016alz}, but our focus here is on the broader, and more challenging, class of continuum signatures.  We search for gamma-ray flux from these galaxy groups and interpret the null results as bounds on the annihilation cross section.   
 
 \noindent {\bf Galaxy Group Selection.}  The observed gamma-ray flux from DM annihilation in an extragalactic halo is proportional to both the particle physics properties of the DM, as well as its astrophysical distribution:
\es{particle}{
\frac{d\Phi}{dE_{\gamma}} &= \left.  J \, \times \frac{\langle\sigma v\rangle}{8 \pi m_{\chi}^{2}} \, \, \sum_i \text{Br}_{i}\, \frac{dN_{i}}{dE'_{\gamma}} \right|_{E_{\gamma}' = (1 +z) E_{\gamma}}   \,,
}
with units of $[{\rm counts} \,\,{\rm cm}^{-2} \, {\rm s}^{-1} \, {\rm GeV}^{-1}]$.  Here, $E_\gamma$ is the gamma-ray energy, $\langle \sigma v \rangle$ is the annihilation cross section, $m_\chi$ is the DM mass, $\text{Br}_{i}$ is the branching fraction to the $i^\text{th}$ annihilation channel, and $z$ is the cosmological redshift.  The energy spectrum for each channel is described the function $dN_{i}/dE_{\gamma}$, which is modeled using PPPC4DMID~\cite{Cirelli:2010xx}.  The $J$-factor that appears in~Eq.~\ref{particle} encodes the astrophysical properties of the halo.  It is proportional to the line-of-sight integral of the squared DM density distribution, $\rho_\text{DM}$, and is written in full as 
\begin{equation}
J = \left(1+b_\text{sh}[M_\text{vir}] \right)  \int ds\,d \Omega \,\rho^{2}_\text{DM}(s,\Omega) \, ,
\label{eq:Jfactor}
\end{equation}
where $b_\text{sh}[M_\text{vir}]$ is the boost factor, which accounts for the enhancement due to substructure.  For an extragalactic halo, where the comoving distance $d_c[z]$ is much greater than the virial radius $r_\text{vir}$, the integral in Eq.~\ref{eq:Jfactor} scales as $M_{\rm vir} c_{\rm vir}^3\rho_c/d_c^2[z]$ for the Navarro-Frenk-White (NFW) density profile~\cite{Navarro:1996gj}.  Here, $M_\text{vir}$ is the virial mass, $\rho_c$ is the critical density, and $c_\text{vir}=r_\text{vir}/r_s$ is the virial concentration, with $r_s$ the scale radius.  We infer $c_\text{vir}$ using the concentration-mass relation from Ref.~\cite{Correa:2015dva}, which we update with the Planck 2015 cosmology~\cite{Ade:2015xua}.  
For a given mass and redshift, the concentration is modeled as a log-normal distribution with mean given by the concentration-mass relation.  We estimate the dispersion by matching to that observed in the \texttt{DarkSky-400} simulation for an equivalent $M_\text{vir}$~\cite{Lehmann:2015ioa}.  Typical dispersions range from $\sim$$0.14$--$0.19$ over the halo masses considered. 

The halo mass and redshift also determine the boost factor enhancement that arises from annihilation in DM substructure.  Accurately modeling the boost factor is challenging as it involves extrapolating the halo-mass function and concentration to masses smaller than can be resolved with current simulations.  Some previous analyses of extragalactic DM annihilation have estimated boost factors $\sim$$10^2$--$10^3$ for cluster-size halos (see, for example, Ref.~\cite{Gao:2011rf}) based on phenomenological extrapolations of the subhalo mass and concentration relations.  However, more recent studies indicate that the concentration-mass relation likely flattens at low masses~\cite{Anderhalden:2013wd,Ludlow:2013vxa,Correa:2015dva}, suppressing the enhancement. We use the model of Ref.~\cite{Bartels:2015uba}---specifically, the ``self-consistent" model with $M_\text{min} = 10^{-6}$~M$_\odot$---which accounts for tidal stripping of bound subhalos and yields a modest boost $\sim$$5$ for $\sim$$10^{15}$~M$_\odot$ halos. Additionally, we model the boost factor as a multiplicative enhancement to the rate in our main analysis, though we consider the effect of possible spatial extension from the subhalo annihilation in the Supplementary Material. In particular, we find that modeling the boost component of the signal as tracing a subhalo population distributed as $\rho_\text{NFW}$ rather than $\rho^{2}_\text{NFW}$ degrades the upper limits obtained by almost an order of magnitude at higher masses $m_\chi \gtrsim 500$ GeV while strengthening the limit by a small $\mathcal O(1)$ factor at lower masses $m_\chi \lesssim 200$ GeV.

The halo masses and redshifts are taken from the galaxy group catalog T15~\cite{Tully:2015opa}, which is based on the 2MASS Redshift Survey (2MRS)~\cite{Crook:2006sw}, and T17~\cite{2017ApJ...843...16K}, which compiles an inventory of nearby galaxies and distances from several sources.  The catalogs provide group associations for these galaxies as well as mass estimates and uncertainties of the host halos, constructed from a luminosity-to-mass relation. The mass distribution is assumed to follow a log-normal distribution with uncertainty fixed at 1\% in log-space \cite{companion}, which translates to typical absolute uncertainties of 25-40\%.\footnote{To translate, approximately, between log- and linear-space uncertainties for the mass, we may write $x = \log_{10} M_\text{vir}$, which implies that the linear-space fractional uncertainties are $\delta M_\text{vir} / M_\text{vir} \sim (\delta x / x) \log M_\text{vir}$. } This is conservative compared to the 20\% uncertainty estimate given in T15 due to their inference procedure. The halo centers are assumed to coincide with the locations of the brightest galaxy in the group.  We infer the $J$-factor using Eq.~\ref{eq:Jfactor} and calculate its uncertainty by propagating the errors on $M_\text{vir}$ and $c_\text{vir}$, which we take to be uncorrelated.  Note that we neglect the distance uncertainties, which are expected to be $\sim$5\%~\cite{Tully:2016ppz,2017ApJ...843...16K}, as they are subdominant compared to the uncertainties on mass and concentration.  We compile an initial list of nearby targets using the T17 catalog, supplementing these with the T15 catalog.  We exclude from T15 all groups with Local Sheet velocity $V_\text{LS} < 3000$~km s$^{-1}$ ($z \lesssim 0.01$) and $V_\text{LS} > 10,000$~km s$^{-1}$ ($z \gtrsim 0.03$), the former because of peculiar velocity contamination and the latter because of large uncertainties in halo mass estimation due to less luminous satellites.  When groups overlap between the two catalogs, we preferentially choose distance and mass measurements from T17.

The galaxy groups are ranked by their inferred $J$-factors, excluding any groups that lie within $|b| \leq 20^\circ$ to mitigate contamination from Galactic diffuse emission.  We require that halos do not overlap to within $2^\circ$ of each other, which is approximately the scale radius of the largest halos.  The exclusion procedure is applied sequentially starting with a halo list ranked by $J$-factor.  We manually exclude Andromeda, the brightest halo in the catalog, because its large angular size is not ideally suited to our analysis pipeline and requires careful individual study~\cite{Ackermann:2017nya}.  
As discussed later in this Letter, halos are also excluded if they show large residuals that are inconsistent with DM annihilation in the other groups in the sample.  Starting with the top 1000 halos, we end up with 495 halos that pass all these requirements.  Of the excluded halos, 276 are removed because they fall too close to the Galactic plane, 134 are removed by the $2^\circ$ proximity requirement, and 95 are removed because of the cut on large residuals. 

Table~\ref{Jtab} lists the top five galaxy groups included in the analysis, labeled by their central galaxy or common name, if one exists.  We provide the inferred $J$-factor including the boost factor, the halo mass, redshift, position in Galactic coordinates, inferred concentration, and boost factor.  Additionally, we show $\theta_\text{s} \equiv \tan^{-1} (r_\text{s} / d_c[z])$ to indicate the spatial extension of the halo.  We find that $\theta_\text{s}$ is typically between the 68\% and 95\% containment radius for emission associated with annihilation in the halos, without accounting for spread from the point-spread function (PSF).  For reference, Andromeda has $\theta_\text{s} \sim 2.57^\circ$.  

\noindent {\bf Data Analysis.}  We analyze 413 weeks of Pass 8 {\it Fermi} data in the UltracleanVeto event class, from August 4, 2008 through July 7, 2016.  The data is binned in 26 logarithmically-spaced energy bins between 502~MeV and 251~GeV and spatially with a HEALPix pixelation~\cite{Gorski:2004by} with \texttt{nside}=128.\footnote{Our energy binning is constructed by taking 40 log-spaced bins between 200~MeV and 2~TeV and then removing the lowest four and highest ten bins, for reasons discussed in the companion paper~\cite{companion}. }  The recommended set of quality cuts are applied to the data corresponding to zenith angle less than $90^\circ$, $\texttt{LAT\_CONFIG}=1$, and $\texttt{DATA\_QUAL}>0$.\footnote{\url{https://fermi.gsfc.nasa.gov/ssc/data/analysis/documentation/Cicerone/Cicerone_Data_Exploration/Data_preparation.html}.}  We also mask known large-scale structures~\cite{companion}.

The template analysis that we perform using \texttt{NPTFit}~\cite{Mishra-Sharma:2016gis} is similar to that of previous dSph studies~\cite{Ackermann:2015zua,Fermi-LAT:2016uux}  and is detailed in our companion paper~\cite{companion}.  We summarize the relevant points here.  Each region-of-interest (ROI), defined as the $10^\circ$ area surrounding each halo center, has its own likelihood.  In each energy bin, this  likelihood is the product, over all pixels, of the Poisson probability for the observed photon counts per pixel.  This probability depends on the mean expected counts per pixel, which depends on contributions from known astrophysical emission as well as a potential DM signal. Note that the likelihood is also multiplied by the appropriate log-normal distribution for $J$, which we treat as a single nuisance parameter for each halo and account for through the profile likelihood method.  

\begin{figure*}[t]
\centering
\includegraphics[width=0.45\textwidth]{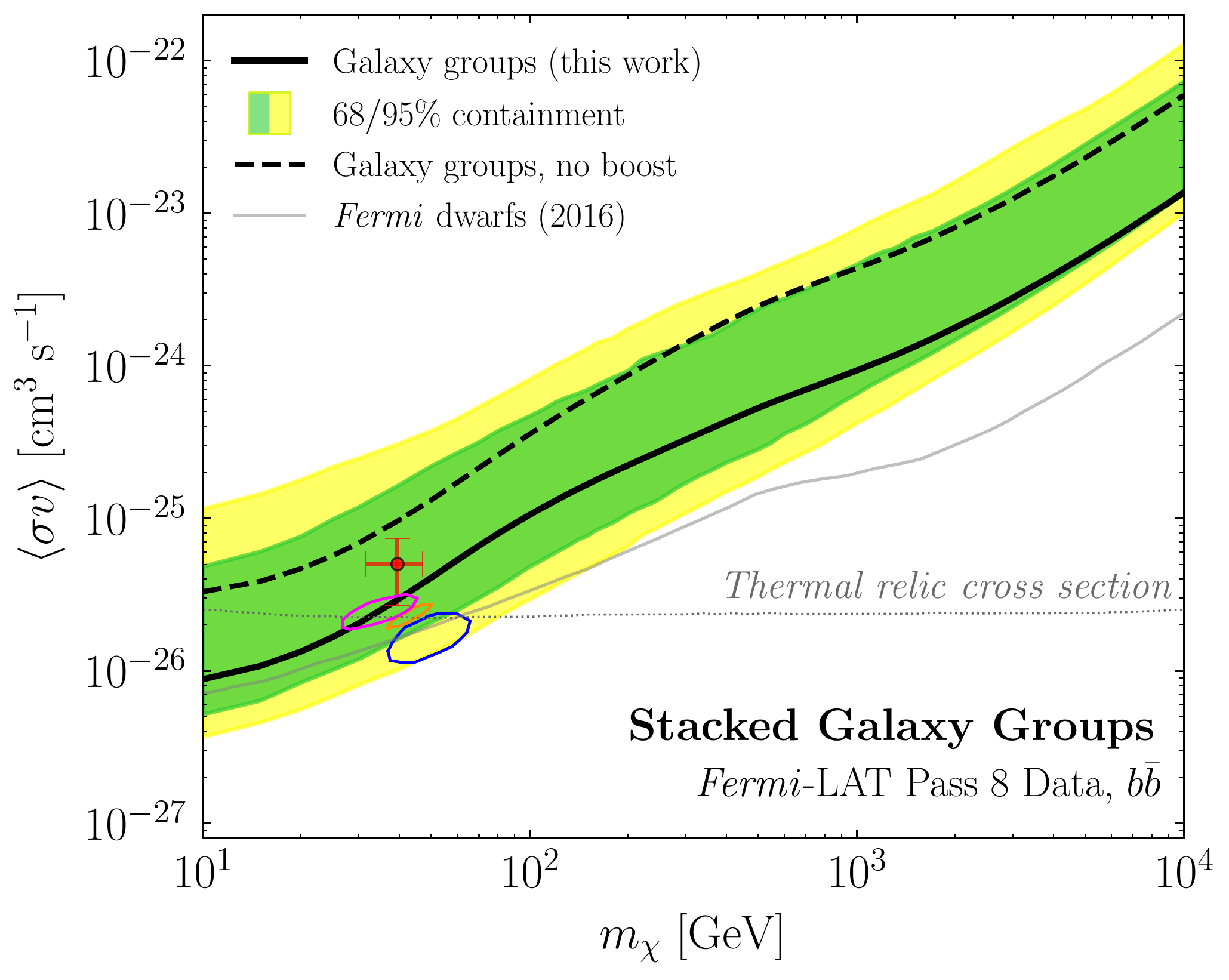} \hspace{4mm}
\includegraphics[width=0.45\textwidth]{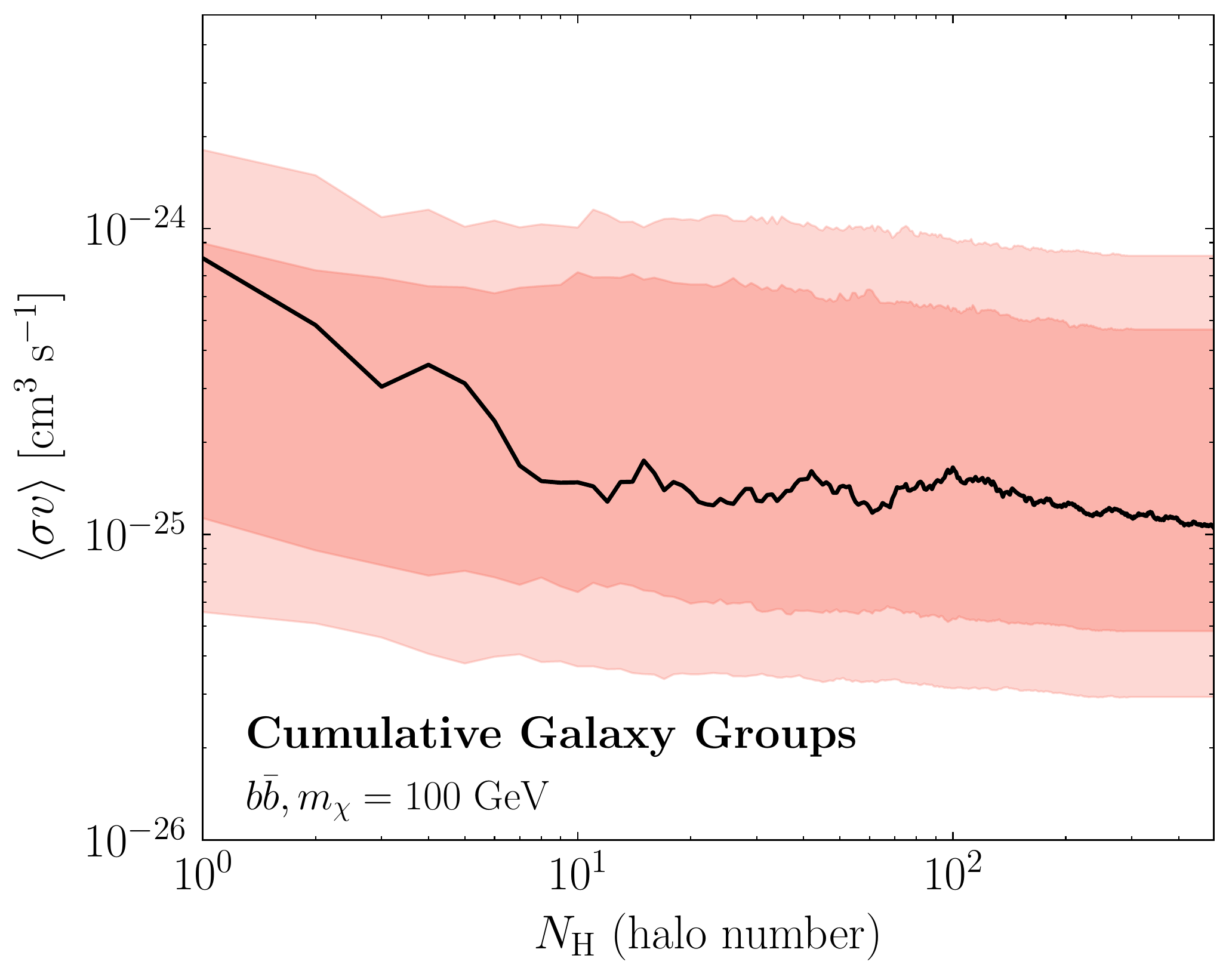}
\caption{(Left) The solid black line shows the 95\% confidence limit on the DM annihilation cross section, $\langle \sigma v \rangle$, as a function of the DM mass, $m_\chi$, for the $b \bar b$ final state, assuming the fiducial boost factor~\cite{Bartels:2015uba}. The containment regions are computed by performing the data analysis multiple times for random sky locations of the halos.  For comparison, the dashed black line shows the limit assuming no boost factor.  The {\it Fermi} dwarf limit is also shown, as well as the $2$$\sigma$ regions where DM may contribute to the Galactic Center Excess (see text for details).  The thermal relic cross section for a generic weakly interacting massive particle~\cite{Steigman:2012nb} is indicated by the thin dotted line.   (Right) The change in the limit for $m_\chi = 100$~GeV as a function of the number of halos that are included in the analysis, which are ranked in order of largest $J$-factor.  The result is compared to the expectation from random sky locations; the 68 and 95\% expectations from 200 random sky locations are indicated by the red bands.}
\label{fig:bounds}
\end{figure*}

To model the expected counts per pixel, we include several templates in the analysis that trace the emission associated with: (i) the projected NFW-squared profile modeling the putative DM signal, (ii) the diffuse background, as described by the  {\it Fermi} \texttt{gll\_iem\_v06 (p8r2)} model, (iii) isotropic emission, (iv) the {\it Fermi} bubbles~\cite{Su:2010qj}, (v) 3FGL sources within $10^\circ$ to $18^\circ$ of the halo center, floated together after fixing their individual fluxes to the values predicted by the 3FGL catalog~\cite{Acero:2015hja}, and (vi) all individual 3FGL point sources within $10^{\circ}$ of the halo center.  Note that we do not model the contributions from annihilation in the smooth Milky Way halo because the brightest groups have peak flux significantly (approximately an order of magnitude for the groups in Tab.~\ref{Jtab}) over the foreground emission from Galactic annihilation and because we expect Galactic annihilation to be subsumed by the isotropic component.   

We assume that the best-fit normalizations (\emph{i.e.}, profiled values) of the astrophysical components, which we treat as nuisance parameters, do not vary appreciably with DM template normalization. This allows us to obtain the likelihood profile in a given ROI and energy bin by profiling over them in the presence of the DM template, then fixing the normalizations of the background components to the best-fit values and scanning over the DM intensity. We then obtain the total likelihood by taking the product of the individual likelihoods from each energy bin. In order to avoid degeneracies at low energies due to the large PSF, we only include the DM template when obtaining the best-fit background normalizations at energies above $\sim$$1$~GeV. At the end of this procedure, the likelihood is only a function of the DM template intensity, which can then be mapped onto a mass and cross section for a given annihilation channel. We emphasize that the assumptions described above have been thoroughly vetted in our companion paper~\cite{companion}, where we show that this procedure is robust in the presence of a potential signal.

The final step of the analysis involves stacking the likelihoods from each ROI. The stacked log-likelihood, $\log \mathcal{L}$, is simply the sum of the log-likelihoods for each ROI.  It follows that the test statistic for data $d$ is defined as
 \begin{equation}\begin{aligned}
{\rm TS}(\mathcal{M}, \langle\sigma v\rangle, m_\chi) \equiv 2 &\left[ \log \mathcal{L}(d | \mathcal{M}, \langle\sigma v\rangle, m_\chi ) \right.\\
&\left.- \log \mathcal{L}(d | \mathcal{M}, \widehat{\langle\sigma v\rangle}, m_\chi ) \right]\,,
\label{eq:TSdef}
\end{aligned}\end{equation}
where $\widehat{\langle\sigma v\rangle}$ is the cross section that maximizes the likelihood for DM model $\mathcal{M}$.    The 95\% upper limit on the annihilation cross section is given by the value of $\langle\sigma v\rangle > \widehat{\langle \sigma v\rangle}$ where $\text{TS}=-2.71$.

Galaxy groups are expected to emit gamma-rays from standard cosmic-ray processes.  Using group catalogs to study gamma-ray emission from cosmic rays in these objects is an interesting study in its own right (see, {\it e.g.}, Ref.~\cite{Jeltema:2008vu,Huber:2013cia,Ackermann:2015fdi,Rephaeli:2015nca}), which we leave to future work.  For the purpose of the present analysis, however, we would like a way to remove groups with large residuals, likely arising from standard astrophysical processes in the clusters, to maintain maximum sensitivity to DM annihilation.  This requires care, however, as we must guarantee  that the procedure for removing halos does not remove a real signal, if one were present.  

We adopt the following algorithm to remove halos with large residuals that are inconsistent with DM annihilation in the other groups in the sample. A group is excluded if it meets two conditions. First, to ensure it is a statistically significant excess, we require twice the difference between the maximum log likelihood and the log likelihood with $\langle \sigma v \rangle = 0$ to be greater than 9 at any DM mass. This selects sources with large residuals at a given DM mass.  Second, the residuals must be strongly inconsistent with limits set by other galaxy groups. Specifically, the halo  must satisfy $\langle\sigma v\rangle_\text{best} > 10 \times \langle\sigma v\rangle^*_\text{lim}$, where $\langle\sigma v\rangle_\text{best}$ is the halo's best-fit cross section at \emph{any} mass and $\langle\sigma v\rangle^*_\text{lim}$ is the strongest limit out of all halos at the specified $m_\chi$. These conditions are designed to exclude galaxy groups where the gamma-ray emission is inconsistent with a DM origin.  This prescription has been extensively tested on mock data and, crucially, does not exclude injected signals~\cite{companion}.

\noindent  {\bf Results.}  The left panel of Fig.~\ref{fig:bounds} illustrates the main results of the stacked analysis.  The solid black line represents the limit obtained for DM annihilating to a $b \bar b$ final state using the fiducial boost factor model~\cite{Bartels:2015uba}, while the dashed line  shows the limit without the boost factor enhancement.  To estimate the expected limit under the null hypothesis, we repeat the analysis by randomizing the locations of the halos on the sky 200 times, though still requiring they pass the selection cuts described above.  
The colored bands indicate the 68 and 95\% containment regions for the expected limit.  
The limit is consistent with the expectation under the null hypothesis.

The right panel of Fig.~\ref{fig:bounds} illustrates how the limits evolve for the $b \bar b$ final state with $m_\chi = 100$~GeV as an increasing number of halos are stacked.  We also show the expected 68\% and 95\% containment regions, which are obtained from the random sky locations.  As can be seen, no single halo dominates the bounds.  For example, removing Virgo, the brightest halo in the catalog, from the stacking has no significant effect on the limit.  Indeed, the inclusion of all 495 halos buys one an additional order of magnitude in the sensitivity reach.

The limit derived in this work is complementary to the published dSph bound~\cite{Ackermann:2015zua,Fermi-LAT:2016uux}, shown as the solid gray line in the left panel of Fig.~\ref{fig:bounds}. Given the large systematic uncertainties associated with the dwarf analyses (see~\emph{e.g.}, Ref.~\cite{Geringer-Sameth:2014qqa}), we stress the importance of using complementary targets and detection strategies to probe the same region of parameter space. Our limit also probes the parameter space that may explain the Galactic Center excess (GCE); the best-fit models are marked by the orange cross~\cite{Abazajian:2014fta}, blue~\cite{Calore:2014xka}, red~\cite{Gordon:2013vta}, and orange~\cite{Daylan:2014rsa} $2$$\sigma$ regions.  The GCE is a spherically symmetric excess of $\sim$GeV gamma-rays observed to arise from the center of the Milky Way~\cite{Goodenough:2009gk,Hooper:2010mq,TheFermi-LAT:2015kwa,Karwin:2016tsw}.  The GCE has received a considerable amount of attention because it can be explained by annihilating DM.  However, it can also be explained by more standard astrophysical sources; indeed, recent analyses have shown that the distribution of photons in this region of sky is more consistent with a population of unresolved point sources, such as millisecond pulsars, compared to smooth emission from DM~\cite{Lee:2015fea, Bartels:2015aea,Linden:2016rcf, FermiLAT:2017yoi}.  Because systematic uncertainties can be significant and hard to quantify in indirect searches for DM, it is crucial to have independent probes of the parameter space where DM can explain the GCE.  While our null findings do not exclude the DM interpretation of the GCE, their consistency with the dwarf bounds put it further in tension.  This does not, however, account for the fact that the systematics on the modeling of the Milky Way's density distribution can potentially alleviate the tension by changing the best-fit cross section for the GCE.   

\noindent  {\bf Conclusions.} This Letter presents the results of the first systematic search for annihilating DM in nearby galaxy groups.  We introduced and validated  a prescription to infer properties of DM halos associated with  these groups, thereby allowing us to build a map of DM annihilation in the local Universe.  Using this map, we performed a stacked analysis of several hundred galaxy groups and obtained bounds that exclude thermal cross sections for DM  annihilating to $b \bar b$ with mass below $\sim$$30$~GeV, assuming a conservative boost factor model.  These limits are competitive with those obtained from the \emph{Fermi} dSph analyses and are in tension with the range of parameter space that can explain the GCE.  Moving forward, we plan to investigate the objects with gamma-ray excesses to see if they can be interpreted in the context of astrophysical emission.  In so doing, we can also develop more refined metrics for selecting the optimal galaxy groups for DM studies.    


\noindent{\textbf{Acknowledgements.}  We thank S.~Ando, N.~Bahcall, R.~Bartels, J.~Beacom, P.~Behroozi, F.~Calore, W.~Coulton, A.~Drlica-Wagner, D.~Hooper, S.~Horiuchi, A.~Kravtsov, T.~Linden, Y.~Mao, K.~Murase, L.~Necib, J.~Ostriker, A.~Peter, T.~Slatyer, B.~Tully, R.~Wechsler, C.~Weniger, and S.~Zimmer for helpful conversations. This research made use of the \texttt{Astropy}~\cite{2013A&A...558A..33A}, \texttt{IPython}~\cite{PER-GRA:2007}, \texttt{Minuit}~\cite{James:1975dr}, and \texttt{NPTFit}~\cite{Mishra-Sharma:2016gis} software packages.  
ML is supported by the DOE under contract DESC0007968, the Alfred P.~Sloan Foundation and the Cottrell Scholar Program through the Research Corporation for Science Advancement.  
NLR is supported by the DOE under contracts DESC00012567 and DESC0013999.
BRS is supported by a Pappalardo
Fellowship in Physics at MIT.  This work was performed in part at Aspen Center for Physics, which is supported by NSF grant PHY-1607611.

\twocolumngrid
\def\bibsection{} 
\bibliographystyle{apsrev}
\bibliography{fermi_darksky}

\begin{thebibliography}{87}
\expandafter\ifx\csname natexlab\endcsname\relax\def\natexlab#1{#1}\fi
\expandafter\ifx\csname bibnamefont\endcsname\relax
  \def\bibnamefont#1{#1}\fi
\expandafter\ifx\csname bibfnamefont\endcsname\relax
  \def\bibfnamefont#1{#1}\fi
\expandafter\ifx\csname citenamefont\endcsname\relax
  \def\citenamefont#1{#1}\fi
\expandafter\ifx\csname url\endcsname\relax
  \def\url#1{\texttt{#1}}\fi
\expandafter\ifx\csname urlprefix\endcsname\relax\def\urlprefix{URL }\fi
\providecommand{\bibinfo}[2]{#2}
\providecommand{\eprint}[2][]{\url{#2}}

\bibitem[{\citenamefont{Ackermann
  et~al.}(2015{\natexlab{a}})}]{Ackermann:2015zua}
\bibinfo{author}{\bibfnamefont{M.}~\bibnamefont{Ackermann}}
  \bibnamefont{et~al.} (\bibinfo{collaboration}{Fermi-LAT}),
  \bibinfo{journal}{Phys. Rev. Lett.} \textbf{\bibinfo{volume}{115}},
  \bibinfo{pages}{231301} (\bibinfo{year}{2015}{\natexlab{a}}),
  \eprint{1503.02641}.

\bibitem[{\citenamefont{Albert et~al.}(2017)}]{Fermi-LAT:2016uux}
\bibinfo{author}{\bibfnamefont{A.}~\bibnamefont{Albert}} \bibnamefont{et~al.}
  (\bibinfo{collaboration}{DES, Fermi-LAT}), \bibinfo{journal}{Astrophys. J.}
  \textbf{\bibinfo{volume}{834}}, \bibinfo{pages}{110} (\bibinfo{year}{2017}),
  \eprint{1611.03184}.

\bibitem[{\citenamefont{Lisanti et~al.}(2017)\citenamefont{Lisanti,
  Mishra-Sharma, Rodd, Safdi, and Wechsler}}]{companion}
\bibinfo{author}{\bibfnamefont{M.}~\bibnamefont{Lisanti}},
  \bibinfo{author}{\bibfnamefont{S.}~\bibnamefont{Mishra-Sharma}},
  \bibinfo{author}{\bibfnamefont{N.~L.} \bibnamefont{Rodd}},
  \bibinfo{author}{\bibfnamefont{B.~R.} \bibnamefont{Safdi}}, \bibnamefont{and}
  \bibinfo{author}{\bibfnamefont{R.~H.} \bibnamefont{Wechsler}}
  (\bibinfo{year}{2017}), \eprint{1709.00416}.

\bibitem[{\citenamefont{Ackermann
  et~al.}(2015{\natexlab{b}})}]{Ackermann:2015tah}
\bibinfo{author}{\bibfnamefont{M.}~\bibnamefont{Ackermann}}
  \bibnamefont{et~al.} (\bibinfo{collaboration}{Fermi-LAT}),
  \bibinfo{journal}{JCAP} \textbf{\bibinfo{volume}{1509}}, \bibinfo{pages}{008}
  (\bibinfo{year}{2015}{\natexlab{b}}), \eprint{1501.05464}.

\bibitem[{\citenamefont{Zechlin et~al.}(2016)\citenamefont{Zechlin, Cuoco,
  Donato, Fornengo, and Regis}}]{Zechlin:2016pme}
\bibinfo{author}{\bibfnamefont{H.-S.} \bibnamefont{Zechlin}},
  \bibinfo{author}{\bibfnamefont{A.}~\bibnamefont{Cuoco}},
  \bibinfo{author}{\bibfnamefont{F.}~\bibnamefont{Donato}},
  \bibinfo{author}{\bibfnamefont{N.}~\bibnamefont{Fornengo}}, \bibnamefont{and}
  \bibinfo{author}{\bibfnamefont{M.}~\bibnamefont{Regis}},
  \bibinfo{journal}{Astrophys. J.} \textbf{\bibinfo{volume}{826}},
  \bibinfo{pages}{L31} (\bibinfo{year}{2016}), \eprint{1605.04256}.

\bibitem[{\citenamefont{Lisanti et~al.}(2016)\citenamefont{Lisanti,
  Mishra-Sharma, Necib, and Safdi}}]{Lisanti:2016jub}
\bibinfo{author}{\bibfnamefont{M.}~\bibnamefont{Lisanti}},
  \bibinfo{author}{\bibfnamefont{S.}~\bibnamefont{Mishra-Sharma}},
  \bibinfo{author}{\bibfnamefont{L.}~\bibnamefont{Necib}}, \bibnamefont{and}
  \bibinfo{author}{\bibfnamefont{B.~R.} \bibnamefont{Safdi}}
  (\bibinfo{year}{2016}), \eprint{1606.04101}.

\bibitem[{\citenamefont{Ackermann et~al.}(2012)}]{Ackermann:2012uf}
\bibinfo{author}{\bibfnamefont{M.}~\bibnamefont{Ackermann}}
  \bibnamefont{et~al.} (\bibinfo{collaboration}{Fermi-LAT}),
  \bibinfo{journal}{Phys. Rev.} \textbf{\bibinfo{volume}{D85}},
  \bibinfo{pages}{083007} (\bibinfo{year}{2012}), \eprint{1202.2856}.

\bibitem[{\citenamefont{Ando et~al.}(2007)\citenamefont{Ando, Komatsu,
  Narumoto, and Totani}}]{Ando:2006cr}
\bibinfo{author}{\bibfnamefont{S.}~\bibnamefont{Ando}},
  \bibinfo{author}{\bibfnamefont{E.}~\bibnamefont{Komatsu}},
  \bibinfo{author}{\bibfnamefont{T.}~\bibnamefont{Narumoto}}, \bibnamefont{and}
  \bibinfo{author}{\bibfnamefont{T.}~\bibnamefont{Totani}},
  \bibinfo{journal}{Phys. Rev.} \textbf{\bibinfo{volume}{D75}},
  \bibinfo{pages}{063519} (\bibinfo{year}{2007}), \eprint{astro-ph/0612467}.

\bibitem[{\citenamefont{Ando and Komatsu}(2013)}]{Ando:2013ff}
\bibinfo{author}{\bibfnamefont{S.}~\bibnamefont{Ando}} \bibnamefont{and}
  \bibinfo{author}{\bibfnamefont{E.}~\bibnamefont{Komatsu}},
  \bibinfo{journal}{Phys. Rev.} \textbf{\bibinfo{volume}{D87}},
  \bibinfo{pages}{123539} (\bibinfo{year}{2013}), \eprint{1301.5901}.

\bibitem[{\citenamefont{Xia et~al.}(2011)\citenamefont{Xia, Cuoco, Branchini,
  Fornasa, and Viel}}]{Xia:2011ax}
\bibinfo{author}{\bibfnamefont{J.-Q.} \bibnamefont{Xia}},
  \bibinfo{author}{\bibfnamefont{A.}~\bibnamefont{Cuoco}},
  \bibinfo{author}{\bibfnamefont{E.}~\bibnamefont{Branchini}},
  \bibinfo{author}{\bibfnamefont{M.}~\bibnamefont{Fornasa}}, \bibnamefont{and}
  \bibinfo{author}{\bibfnamefont{M.}~\bibnamefont{Viel}},
  \bibinfo{journal}{Mon. Not. Roy. Astron. Soc.}
  \textbf{\bibinfo{volume}{416}}, \bibinfo{pages}{2247} (\bibinfo{year}{2011}),
  \eprint{1103.4861}.

\bibitem[{\citenamefont{Ando}(2014)}]{Ando:2014aoa}
\bibinfo{author}{\bibfnamefont{S.}~\bibnamefont{Ando}}, \bibinfo{journal}{JCAP}
  \textbf{\bibinfo{volume}{1410}}, \bibinfo{pages}{061} (\bibinfo{year}{2014}),
  \eprint{1407.8502}.

\bibitem[{\citenamefont{Ando et~al.}(2014)\citenamefont{Ando, Benoit-L{\'e}vy,
  and Komatsu}}]{Ando:2013xwa}
\bibinfo{author}{\bibfnamefont{S.}~\bibnamefont{Ando}},
  \bibinfo{author}{\bibfnamefont{A.}~\bibnamefont{Benoit-L{\'e}vy}},
  \bibnamefont{and} \bibinfo{author}{\bibfnamefont{E.}~\bibnamefont{Komatsu}},
  \bibinfo{journal}{Phys. Rev.} \textbf{\bibinfo{volume}{D90}},
  \bibinfo{pages}{023514} (\bibinfo{year}{2014}), \eprint{1312.4403}.

\bibitem[{\citenamefont{Xia et~al.}(2015)\citenamefont{Xia, Cuoco, Branchini,
  and Viel}}]{Xia:2015wka}
\bibinfo{author}{\bibfnamefont{J.-Q.} \bibnamefont{Xia}},
  \bibinfo{author}{\bibfnamefont{A.}~\bibnamefont{Cuoco}},
  \bibinfo{author}{\bibfnamefont{E.}~\bibnamefont{Branchini}},
  \bibnamefont{and} \bibinfo{author}{\bibfnamefont{M.}~\bibnamefont{Viel}},
  \bibinfo{journal}{Astrophys. J. Suppl.} \textbf{\bibinfo{volume}{217}},
  \bibinfo{pages}{15} (\bibinfo{year}{2015}), \eprint{1503.05918}.

\bibitem[{\citenamefont{Regis et~al.}(2015)\citenamefont{Regis, Xia, Cuoco,
  Branchini, Fornengo, and Viel}}]{Regis:2015zka}
\bibinfo{author}{\bibfnamefont{M.}~\bibnamefont{Regis}},
  \bibinfo{author}{\bibfnamefont{J.-Q.} \bibnamefont{Xia}},
  \bibinfo{author}{\bibfnamefont{A.}~\bibnamefont{Cuoco}},
  \bibinfo{author}{\bibfnamefont{E.}~\bibnamefont{Branchini}},
  \bibinfo{author}{\bibfnamefont{N.}~\bibnamefont{Fornengo}}, \bibnamefont{and}
  \bibinfo{author}{\bibfnamefont{M.}~\bibnamefont{Viel}},
  \bibinfo{journal}{Phys. Rev. Lett.} \textbf{\bibinfo{volume}{114}},
  \bibinfo{pages}{241301} (\bibinfo{year}{2015}), \eprint{1503.05922}.

\bibitem[{\citenamefont{Cuoco et~al.}(2015)\citenamefont{Cuoco, Xia, Regis,
  Branchini, Fornengo, and Viel}}]{Cuoco:2015rfa}
\bibinfo{author}{\bibfnamefont{A.}~\bibnamefont{Cuoco}},
  \bibinfo{author}{\bibfnamefont{J.-Q.} \bibnamefont{Xia}},
  \bibinfo{author}{\bibfnamefont{M.}~\bibnamefont{Regis}},
  \bibinfo{author}{\bibfnamefont{E.}~\bibnamefont{Branchini}},
  \bibinfo{author}{\bibfnamefont{N.}~\bibnamefont{Fornengo}}, \bibnamefont{and}
  \bibinfo{author}{\bibfnamefont{M.}~\bibnamefont{Viel}},
  \bibinfo{journal}{Astrophys. J. Suppl.} \textbf{\bibinfo{volume}{221}},
  \bibinfo{pages}{29} (\bibinfo{year}{2015}), \eprint{1506.01030}.

\bibitem[{\citenamefont{Ando and Ishiwata}(2016)}]{Ando:2016ang}
\bibinfo{author}{\bibfnamefont{S.}~\bibnamefont{Ando}} \bibnamefont{and}
  \bibinfo{author}{\bibfnamefont{K.}~\bibnamefont{Ishiwata}},
  \bibinfo{journal}{JCAP} \textbf{\bibinfo{volume}{1606}}, \bibinfo{pages}{045}
  (\bibinfo{year}{2016}), \eprint{1604.02263}.

\bibitem[{\citenamefont{Jarrett et~al.}(2000)\citenamefont{Jarrett, Chester,
  Cutri, Schneider, Skrutskie, and Huchra}}]{Jarrett:2000me}
\bibinfo{author}{\bibfnamefont{T.~H.} \bibnamefont{Jarrett}},
  \bibinfo{author}{\bibfnamefont{T.}~\bibnamefont{Chester}},
  \bibinfo{author}{\bibfnamefont{R.}~\bibnamefont{Cutri}},
  \bibinfo{author}{\bibfnamefont{S.}~\bibnamefont{Schneider}},
  \bibinfo{author}{\bibfnamefont{M.}~\bibnamefont{Skrutskie}},
  \bibnamefont{and} \bibinfo{author}{\bibfnamefont{J.~P.}
  \bibnamefont{Huchra}}, \bibinfo{journal}{Astron. J.}
  \textbf{\bibinfo{volume}{119}}, \bibinfo{pages}{2498} (\bibinfo{year}{2000}),
  \eprint{astro-ph/0004318}.

\bibitem[{\citenamefont{Bilicki et~al.}(2013)\citenamefont{Bilicki, Jarrett,
  Peacock, Cluver, and Steward}}]{Bilicki:2013sza}
\bibinfo{author}{\bibfnamefont{M.}~\bibnamefont{Bilicki}},
  \bibinfo{author}{\bibfnamefont{T.~H.} \bibnamefont{Jarrett}},
  \bibinfo{author}{\bibfnamefont{J.~A.} \bibnamefont{Peacock}},
  \bibinfo{author}{\bibfnamefont{M.~E.} \bibnamefont{Cluver}},
  \bibnamefont{and} \bibinfo{author}{\bibfnamefont{L.}~\bibnamefont{Steward}}
  (\bibinfo{year}{2013}), \bibinfo{note}{[Astrophys. J. Suppl.210,9(2014)]},
  \eprint{1311.5246}.

\bibitem[{\citenamefont{Correa et~al.}(2015)\citenamefont{Correa, Wyithe,
  Schaye, and Duffy}}]{Correa:2015dva}
\bibinfo{author}{\bibfnamefont{C.~A.} \bibnamefont{Correa}},
  \bibinfo{author}{\bibfnamefont{J.~S.~B.} \bibnamefont{Wyithe}},
  \bibinfo{author}{\bibfnamefont{J.}~\bibnamefont{Schaye}}, \bibnamefont{and}
  \bibinfo{author}{\bibfnamefont{A.~R.} \bibnamefont{Duffy}},
  \bibinfo{journal}{Mon. Not. Roy. Astron. Soc.}
  \textbf{\bibinfo{volume}{452}}, \bibinfo{pages}{1217} (\bibinfo{year}{2015}),
  \eprint{1502.00391}.

\bibitem[{\citenamefont{Bartels and Ando}(2015)}]{Bartels:2015uba}
\bibinfo{author}{\bibfnamefont{R.}~\bibnamefont{Bartels}} \bibnamefont{and}
  \bibinfo{author}{\bibfnamefont{S.}~\bibnamefont{Ando}},
  \bibinfo{journal}{Phys. Rev.} \textbf{\bibinfo{volume}{D92}},
  \bibinfo{pages}{123508} (\bibinfo{year}{2015}), \eprint{1507.08656}.

\bibitem[{\citenamefont{Ackermann et~al.}(2010)}]{Ackermann:2010rg}
\bibinfo{author}{\bibfnamefont{M.}~\bibnamefont{Ackermann}}
  \bibnamefont{et~al.}, \bibinfo{journal}{JCAP}
  \textbf{\bibinfo{volume}{1005}}, \bibinfo{pages}{025} (\bibinfo{year}{2010}),
  \eprint{1002.2239}.

\bibitem[{\citenamefont{Ando and Nagai}(2012)}]{Ando:2012vu}
\bibinfo{author}{\bibfnamefont{S.}~\bibnamefont{Ando}} \bibnamefont{and}
  \bibinfo{author}{\bibfnamefont{D.}~\bibnamefont{Nagai}},
  \bibinfo{journal}{JCAP} \textbf{\bibinfo{volume}{1207}}, \bibinfo{pages}{017}
  (\bibinfo{year}{2012}), \eprint{1201.0753}.

\bibitem[{\citenamefont{Ackermann et~al.}(2014)}]{Ackermann:2013iaq}
\bibinfo{author}{\bibfnamefont{M.}~\bibnamefont{Ackermann}}
  \bibnamefont{et~al.} (\bibinfo{collaboration}{Fermi-LAT}),
  \bibinfo{journal}{Astrophys. J.} \textbf{\bibinfo{volume}{787}},
  \bibinfo{pages}{18} (\bibinfo{year}{2014}), \eprint{1308.5654}.

\bibitem[{\citenamefont{Ackermann
  et~al.}(2015{\natexlab{c}})}]{Ackermann:2015fdi}
\bibinfo{author}{\bibfnamefont{M.}~\bibnamefont{Ackermann}}
  \bibnamefont{et~al.} (\bibinfo{collaboration}{Fermi-LAT}),
  \bibinfo{journal}{Astrophys. J.} \textbf{\bibinfo{volume}{812}},
  \bibinfo{pages}{159} (\bibinfo{year}{2015}{\natexlab{c}}),
  \eprint{1510.00004}.

\bibitem[{\citenamefont{Anderson et~al.}(2016)\citenamefont{Anderson, Zimmer,
  Conrad, Gustafsson, S{\'a}nchez-Conde, and Caputo}}]{Anderson:2015dpc}
\bibinfo{author}{\bibfnamefont{B.}~\bibnamefont{Anderson}},
  \bibinfo{author}{\bibfnamefont{S.}~\bibnamefont{Zimmer}},
  \bibinfo{author}{\bibfnamefont{J.}~\bibnamefont{Conrad}},
  \bibinfo{author}{\bibfnamefont{M.}~\bibnamefont{Gustafsson}},
  \bibinfo{author}{\bibfnamefont{M.}~\bibnamefont{S{\'a}nchez-Conde}},
  \bibnamefont{and} \bibinfo{author}{\bibfnamefont{R.}~\bibnamefont{Caputo}},
  \bibinfo{journal}{JCAP} \textbf{\bibinfo{volume}{1602}}, \bibinfo{pages}{026}
  (\bibinfo{year}{2016}), \eprint{1511.00014}.

\bibitem[{\citenamefont{Ackermann et~al.}(2016)}]{Rephaeli:2015nca}
\bibinfo{author}{\bibfnamefont{M.}~\bibnamefont{Ackermann}}
  \bibnamefont{et~al.} (\bibinfo{collaboration}{Fermi-LAT}),
  \bibinfo{journal}{Astrophys. J.} \textbf{\bibinfo{volume}{819}},
  \bibinfo{pages}{149} (\bibinfo{year}{2016}), \eprint{1507.08995}.

\bibitem[{\citenamefont{{Ahnen} et~al.}(2016)}]{2016A&A...589A..33A}
\bibinfo{author}{\bibfnamefont{M.~L.} \bibnamefont{{Ahnen}}}
  \bibnamefont{et~al.}, \bibinfo{journal}{{Astron. Astrophys.}}
  \textbf{\bibinfo{volume}{589}}, \bibinfo{eid}{A33} (\bibinfo{year}{2016}),
  \eprint{1602.03099}.

\bibitem[{\citenamefont{Liang et~al.}(2016)\citenamefont{Liang, Shen, Li, Fan,
  Huang, Lei, Feng, Liang, and Chang}}]{Liang:2016pvm}
\bibinfo{author}{\bibfnamefont{Y.-F.} \bibnamefont{Liang}},
  \bibinfo{author}{\bibfnamefont{Z.-Q.} \bibnamefont{Shen}},
  \bibinfo{author}{\bibfnamefont{X.}~\bibnamefont{Li}},
  \bibinfo{author}{\bibfnamefont{Y.-Z.} \bibnamefont{Fan}},
  \bibinfo{author}{\bibfnamefont{X.}~\bibnamefont{Huang}},
  \bibinfo{author}{\bibfnamefont{S.-J.} \bibnamefont{Lei}},
  \bibinfo{author}{\bibfnamefont{L.}~\bibnamefont{Feng}},
  \bibinfo{author}{\bibfnamefont{E.-W.} \bibnamefont{Liang}}, \bibnamefont{and}
  \bibinfo{author}{\bibfnamefont{J.}~\bibnamefont{Chang}},
  \bibinfo{journal}{Phys. Rev.} \textbf{\bibinfo{volume}{D93}},
  \bibinfo{pages}{103525} (\bibinfo{year}{2016}), \eprint{1602.06527}.

\bibitem[{\citenamefont{Adams et~al.}(2016)\citenamefont{Adams, Bergstrom, and
  Spolyar}}]{Adams:2016alz}
\bibinfo{author}{\bibfnamefont{D.~Q.} \bibnamefont{Adams}},
  \bibinfo{author}{\bibfnamefont{L.}~\bibnamefont{Bergstrom}},
  \bibnamefont{and} \bibinfo{author}{\bibfnamefont{D.}~\bibnamefont{Spolyar}}
  (\bibinfo{year}{2016}), \eprint{1606.09642}.

\bibitem[{\citenamefont{Huang et~al.}(2012)\citenamefont{Huang, Vertongen, and
  Weniger}}]{Huang:2011xr}
\bibinfo{author}{\bibfnamefont{X.}~\bibnamefont{Huang}},
  \bibinfo{author}{\bibfnamefont{G.}~\bibnamefont{Vertongen}},
  \bibnamefont{and} \bibinfo{author}{\bibfnamefont{C.}~\bibnamefont{Weniger}},
  \bibinfo{journal}{JCAP} \textbf{\bibinfo{volume}{1201}}, \bibinfo{pages}{042}
  (\bibinfo{year}{2012}), \eprint{1110.1529}.

\bibitem[{\citenamefont{Tully}(2015)}]{Tully:2015opa}
\bibinfo{author}{\bibfnamefont{R.~B.} \bibnamefont{Tully}},
  \bibinfo{journal}{Astron. J.} \textbf{\bibinfo{volume}{149}},
  \bibinfo{pages}{171} (\bibinfo{year}{2015}), \eprint{1503.03134}.

\bibitem[{\citenamefont{{Kourkchi} and {Tully}}(2017)}]{2017ApJ...843...16K}
\bibinfo{author}{\bibfnamefont{E.}~\bibnamefont{{Kourkchi}}} \bibnamefont{and}
  \bibinfo{author}{\bibfnamefont{R.~B.} \bibnamefont{{Tully}}},
  \bibinfo{journal}{Ap. J.} \textbf{\bibinfo{volume}{843}}, \bibinfo{eid}{16}
  (\bibinfo{year}{2017}), \eprint{1705.08068}.

\bibitem[{\citenamefont{Tully et~al.}(2016)\citenamefont{Tully, Courtois, and
  Sorce}}]{Tully:2016ppz}
\bibinfo{author}{\bibfnamefont{R.~B.} \bibnamefont{Tully}},
  \bibinfo{author}{\bibfnamefont{H.~M.} \bibnamefont{Courtois}},
  \bibnamefont{and} \bibinfo{author}{\bibfnamefont{J.~G.} \bibnamefont{Sorce}},
  \bibinfo{journal}{Astron. J.} \textbf{\bibinfo{volume}{152}},
  \bibinfo{pages}{50} (\bibinfo{year}{2016}), \eprint{1605.01765}.

\bibitem[{\citenamefont{Cirelli et~al.}(2011)\citenamefont{Cirelli, Corcella,
  Hektor, Hutsi, Kadastik, Panci, Raidal, Sala, and Strumia}}]{Cirelli:2010xx}
\bibinfo{author}{\bibfnamefont{M.}~\bibnamefont{Cirelli}},
  \bibinfo{author}{\bibfnamefont{G.}~\bibnamefont{Corcella}},
  \bibinfo{author}{\bibfnamefont{A.}~\bibnamefont{Hektor}},
  \bibinfo{author}{\bibfnamefont{G.}~\bibnamefont{Hutsi}},
  \bibinfo{author}{\bibfnamefont{M.}~\bibnamefont{Kadastik}},
  \bibinfo{author}{\bibfnamefont{P.}~\bibnamefont{Panci}},
  \bibinfo{author}{\bibfnamefont{M.}~\bibnamefont{Raidal}},
  \bibinfo{author}{\bibfnamefont{F.}~\bibnamefont{Sala}}, \bibnamefont{and}
  \bibinfo{author}{\bibfnamefont{A.}~\bibnamefont{Strumia}},
  \bibinfo{journal}{JCAP} \textbf{\bibinfo{volume}{1103}}, \bibinfo{pages}{051}
  (\bibinfo{year}{2011}), \bibinfo{note}{[Erratum: JCAP1210,E01(2012)]},
  \eprint{1012.4515}.

\bibitem[{\citenamefont{Navarro et~al.}(1997)\citenamefont{Navarro, Frenk, and
  White}}]{Navarro:1996gj}
\bibinfo{author}{\bibfnamefont{J.~F.} \bibnamefont{Navarro}},
  \bibinfo{author}{\bibfnamefont{C.~S.} \bibnamefont{Frenk}}, \bibnamefont{and}
  \bibinfo{author}{\bibfnamefont{S.~D.~M.} \bibnamefont{White}},
  \bibinfo{journal}{Astrophys. J.} \textbf{\bibinfo{volume}{490}},
  \bibinfo{pages}{493} (\bibinfo{year}{1997}), \eprint{astro-ph/9611107}.

\bibitem[{\citenamefont{Ade et~al.}(2016)}]{Ade:2015xua}
\bibinfo{author}{\bibfnamefont{P.~A.~R.} \bibnamefont{Ade}}
  \bibnamefont{et~al.} (\bibinfo{collaboration}{Planck}),
  \bibinfo{journal}{Astron. Astrophys.} \textbf{\bibinfo{volume}{594}},
  \bibinfo{pages}{A13} (\bibinfo{year}{2016}), \eprint{1502.01589}.

\bibitem[{\citenamefont{Lehmann et~al.}(2017)\citenamefont{Lehmann, Mao,
  Becker, Skillman, and Wechsler}}]{Lehmann:2015ioa}
\bibinfo{author}{\bibfnamefont{B.~V.} \bibnamefont{Lehmann}},
  \bibinfo{author}{\bibfnamefont{Y.-Y.} \bibnamefont{Mao}},
  \bibinfo{author}{\bibfnamefont{M.~R.} \bibnamefont{Becker}},
  \bibinfo{author}{\bibfnamefont{S.~W.} \bibnamefont{Skillman}},
  \bibnamefont{and} \bibinfo{author}{\bibfnamefont{R.~H.}
  \bibnamefont{Wechsler}}, \bibinfo{journal}{Astrophys. J.}
  \textbf{\bibinfo{volume}{834}}, \bibinfo{pages}{37} (\bibinfo{year}{2017}),
  \eprint{1510.05651}.

\bibitem[{\citenamefont{Gao et~al.}(2012)\citenamefont{Gao, Frenk, Jenkins,
  Springel, and White}}]{Gao:2011rf}
\bibinfo{author}{\bibfnamefont{L.}~\bibnamefont{Gao}},
  \bibinfo{author}{\bibfnamefont{C.~S.} \bibnamefont{Frenk}},
  \bibinfo{author}{\bibfnamefont{A.}~\bibnamefont{Jenkins}},
  \bibinfo{author}{\bibfnamefont{V.}~\bibnamefont{Springel}}, \bibnamefont{and}
  \bibinfo{author}{\bibfnamefont{S.~D.~M.} \bibnamefont{White}},
  \bibinfo{journal}{Mon. Not. Roy. Astron. Soc.}
  \textbf{\bibinfo{volume}{419}}, \bibinfo{pages}{1721} (\bibinfo{year}{2012}),
  \eprint{1107.1916}.

\bibitem[{\citenamefont{Anderhalden and Diemand}(2013)}]{Anderhalden:2013wd}
\bibinfo{author}{\bibfnamefont{D.}~\bibnamefont{Anderhalden}} \bibnamefont{and}
  \bibinfo{author}{\bibfnamefont{J.}~\bibnamefont{Diemand}},
  \bibinfo{journal}{JCAP} \textbf{\bibinfo{volume}{1304}}, \bibinfo{pages}{009}
  (\bibinfo{year}{2013}), \bibinfo{note}{[Erratum: JCAP1308,E02(2013)]},
  \eprint{1302.0003}.

\bibitem[{\citenamefont{Ludlow et~al.}(2014)\citenamefont{Ludlow, Navarro,
  Angulo, Boylan-Kolchin, Springel, Frenk, and White}}]{Ludlow:2013vxa}
\bibinfo{author}{\bibfnamefont{A.~D.} \bibnamefont{Ludlow}},
  \bibinfo{author}{\bibfnamefont{J.~F.} \bibnamefont{Navarro}},
  \bibinfo{author}{\bibfnamefont{R.~E.} \bibnamefont{Angulo}},
  \bibinfo{author}{\bibfnamefont{M.}~\bibnamefont{Boylan-Kolchin}},
  \bibinfo{author}{\bibfnamefont{V.}~\bibnamefont{Springel}},
  \bibinfo{author}{\bibfnamefont{C.}~\bibnamefont{Frenk}}, \bibnamefont{and}
  \bibinfo{author}{\bibfnamefont{S.~D.~M.} \bibnamefont{White}},
  \bibinfo{journal}{Mon. Not. Roy. Astron. Soc.}
  \textbf{\bibinfo{volume}{441}}, \bibinfo{pages}{378} (\bibinfo{year}{2014}),
  \eprint{1312.0945}.

\bibitem[{\citenamefont{Crook et~al.}(2007)\citenamefont{Crook, Huchra,
  Martimbeau, Masters, Jarrett, and Macri}}]{Crook:2006sw}
\bibinfo{author}{\bibfnamefont{A.~C.} \bibnamefont{Crook}},
  \bibinfo{author}{\bibfnamefont{J.~P.} \bibnamefont{Huchra}},
  \bibinfo{author}{\bibfnamefont{N.}~\bibnamefont{Martimbeau}},
  \bibinfo{author}{\bibfnamefont{K.~L.} \bibnamefont{Masters}},
  \bibinfo{author}{\bibfnamefont{T.}~\bibnamefont{Jarrett}}, \bibnamefont{and}
  \bibinfo{author}{\bibfnamefont{L.~M.} \bibnamefont{Macri}},
  \bibinfo{journal}{Astrophys. J.} \textbf{\bibinfo{volume}{655}},
  \bibinfo{pages}{790} (\bibinfo{year}{2007}), \eprint{astro-ph/0610732}.

\bibitem[{\citenamefont{Ackermann et~al.}(2017)}]{Ackermann:2017nya}
\bibinfo{author}{\bibfnamefont{M.}~\bibnamefont{Ackermann}}
  \bibnamefont{et~al.} (\bibinfo{collaboration}{Fermi-LAT}),
  \bibinfo{journal}{Astrophys. J.} \textbf{\bibinfo{volume}{836}},
  \bibinfo{pages}{208} (\bibinfo{year}{2017}), \eprint{1702.08602}.

\bibitem[{\citenamefont{Gorski et~al.}(2005)\citenamefont{Gorski, Hivon,
  Banday, Wandelt, Hansen, Reinecke, and Bartelman}}]{Gorski:2004by}
\bibinfo{author}{\bibfnamefont{K.~M.} \bibnamefont{Gorski}},
  \bibinfo{author}{\bibfnamefont{E.}~\bibnamefont{Hivon}},
  \bibinfo{author}{\bibfnamefont{A.~J.} \bibnamefont{Banday}},
  \bibinfo{author}{\bibfnamefont{B.~D.} \bibnamefont{Wandelt}},
  \bibinfo{author}{\bibfnamefont{F.~K.} \bibnamefont{Hansen}},
  \bibinfo{author}{\bibfnamefont{M.}~\bibnamefont{Reinecke}}, \bibnamefont{and}
  \bibinfo{author}{\bibfnamefont{M.}~\bibnamefont{Bartelman}},
  \bibinfo{journal}{Astrophys. J.} \textbf{\bibinfo{volume}{622}},
  \bibinfo{pages}{759} (\bibinfo{year}{2005}), \eprint{astro-ph/0409513}.

\bibitem[{\citenamefont{Mishra-Sharma et~al.}(2017)\citenamefont{Mishra-Sharma,
  Rodd, and Safdi}}]{Mishra-Sharma:2016gis}
\bibinfo{author}{\bibfnamefont{S.}~\bibnamefont{Mishra-Sharma}},
  \bibinfo{author}{\bibfnamefont{N.~L.} \bibnamefont{Rodd}}, \bibnamefont{and}
  \bibinfo{author}{\bibfnamefont{B.~R.} \bibnamefont{Safdi}},
  \bibinfo{journal}{Astron. J.} \textbf{\bibinfo{volume}{153}},
  \bibinfo{pages}{253} (\bibinfo{year}{2017}), \eprint{1612.03173}.

\bibitem[{\citenamefont{Steigman et~al.}(2012)\citenamefont{Steigman, Dasgupta,
  and Beacom}}]{Steigman:2012nb}
\bibinfo{author}{\bibfnamefont{G.}~\bibnamefont{Steigman}},
  \bibinfo{author}{\bibfnamefont{B.}~\bibnamefont{Dasgupta}}, \bibnamefont{and}
  \bibinfo{author}{\bibfnamefont{J.~F.} \bibnamefont{Beacom}},
  \bibinfo{journal}{Phys. Rev.} \textbf{\bibinfo{volume}{D86}},
  \bibinfo{pages}{023506} (\bibinfo{year}{2012}), \eprint{1204.3622}.

\bibitem[{\citenamefont{Su et~al.}(2010)\citenamefont{Su, Slatyer, and
  Finkbeiner}}]{Su:2010qj}
\bibinfo{author}{\bibfnamefont{M.}~\bibnamefont{Su}},
  \bibinfo{author}{\bibfnamefont{T.~R.} \bibnamefont{Slatyer}},
  \bibnamefont{and} \bibinfo{author}{\bibfnamefont{D.~P.}
  \bibnamefont{Finkbeiner}}, \bibinfo{journal}{Astrophys. J.}
  \textbf{\bibinfo{volume}{724}}, \bibinfo{pages}{1044} (\bibinfo{year}{2010}),
  \eprint{1005.5480}.

\bibitem[{\citenamefont{Acero et~al.}(2015)}]{Acero:2015hja}
\bibinfo{author}{\bibfnamefont{F.}~\bibnamefont{Acero}} \bibnamefont{et~al.}
  (\bibinfo{collaboration}{Fermi-LAT}), \bibinfo{journal}{Astrophys. J. Suppl.}
  \textbf{\bibinfo{volume}{218}}, \bibinfo{pages}{23} (\bibinfo{year}{2015}),
  \eprint{1501.02003}.

\bibitem[{\citenamefont{Jeltema et~al.}(2009)\citenamefont{Jeltema, Kehayias,
  and Profumo}}]{Jeltema:2008vu}
\bibinfo{author}{\bibfnamefont{T.~E.} \bibnamefont{Jeltema}},
  \bibinfo{author}{\bibfnamefont{J.}~\bibnamefont{Kehayias}}, \bibnamefont{and}
  \bibinfo{author}{\bibfnamefont{S.}~\bibnamefont{Profumo}},
  \bibinfo{journal}{Phys. Rev.} \textbf{\bibinfo{volume}{D80}},
  \bibinfo{pages}{023005} (\bibinfo{year}{2009}), \eprint{0812.0597}.

\bibitem[{\citenamefont{Huber et~al.}(2013)\citenamefont{Huber, Tchernin,
  Eckert, Farnier, Manalaysay, Straumann, and Walter}}]{Huber:2013cia}
\bibinfo{author}{\bibfnamefont{B.}~\bibnamefont{Huber}},
  \bibinfo{author}{\bibfnamefont{C.}~\bibnamefont{Tchernin}},
  \bibinfo{author}{\bibfnamefont{D.}~\bibnamefont{Eckert}},
  \bibinfo{author}{\bibfnamefont{C.}~\bibnamefont{Farnier}},
  \bibinfo{author}{\bibfnamefont{A.}~\bibnamefont{Manalaysay}},
  \bibinfo{author}{\bibfnamefont{U.}~\bibnamefont{Straumann}},
  \bibnamefont{and} \bibinfo{author}{\bibfnamefont{R.}~\bibnamefont{Walter}},
  \bibinfo{journal}{Astron. Astrophys.} \textbf{\bibinfo{volume}{560}},
  \bibinfo{pages}{A64} (\bibinfo{year}{2013}), \eprint{1308.6278}.

\bibitem[{\citenamefont{Geringer-Sameth
  et~al.}(2015)\citenamefont{Geringer-Sameth, Koushiappas, and
  Walker}}]{Geringer-Sameth:2014qqa}
\bibinfo{author}{\bibfnamefont{A.}~\bibnamefont{Geringer-Sameth}},
  \bibinfo{author}{\bibfnamefont{S.~M.} \bibnamefont{Koushiappas}},
  \bibnamefont{and} \bibinfo{author}{\bibfnamefont{M.~G.}
  \bibnamefont{Walker}}, \bibinfo{journal}{Phys. Rev.}
  \textbf{\bibinfo{volume}{D91}}, \bibinfo{pages}{083535}
  (\bibinfo{year}{2015}), \eprint{1410.2242}.

\bibitem[{\citenamefont{Abazajian et~al.}(2014)\citenamefont{Abazajian, Canac,
  Horiuchi, and Kaplinghat}}]{Abazajian:2014fta}
\bibinfo{author}{\bibfnamefont{K.~N.} \bibnamefont{Abazajian}},
  \bibinfo{author}{\bibfnamefont{N.}~\bibnamefont{Canac}},
  \bibinfo{author}{\bibfnamefont{S.}~\bibnamefont{Horiuchi}}, \bibnamefont{and}
  \bibinfo{author}{\bibfnamefont{M.}~\bibnamefont{Kaplinghat}},
  \bibinfo{journal}{Phys. Rev.} \textbf{\bibinfo{volume}{D90}},
  \bibinfo{pages}{023526} (\bibinfo{year}{2014}), \eprint{1402.4090}.

\bibitem[{\citenamefont{Calore et~al.}(2015)\citenamefont{Calore, Cholis, and
  Weniger}}]{Calore:2014xka}
\bibinfo{author}{\bibfnamefont{F.}~\bibnamefont{Calore}},
  \bibinfo{author}{\bibfnamefont{I.}~\bibnamefont{Cholis}}, \bibnamefont{and}
  \bibinfo{author}{\bibfnamefont{C.}~\bibnamefont{Weniger}},
  \bibinfo{journal}{JCAP} \textbf{\bibinfo{volume}{1503}}, \bibinfo{pages}{038}
  (\bibinfo{year}{2015}), \eprint{1409.0042}.

\bibitem[{\citenamefont{Gordon and Macias}(2013)}]{Gordon:2013vta}
\bibinfo{author}{\bibfnamefont{C.}~\bibnamefont{Gordon}} \bibnamefont{and}
  \bibinfo{author}{\bibfnamefont{O.}~\bibnamefont{Macias}},
  \bibinfo{journal}{Phys. Rev.} \textbf{\bibinfo{volume}{D88}},
  \bibinfo{pages}{083521} (\bibinfo{year}{2013}), \bibinfo{note}{[Erratum:
  Phys. Rev.D89,no.4,049901(2014)]}, \eprint{1306.5725}.

\bibitem[{\citenamefont{Daylan et~al.}(2016)\citenamefont{Daylan, Finkbeiner,
  Hooper, Linden, Portillo, Rodd, and Slatyer}}]{Daylan:2014rsa}
\bibinfo{author}{\bibfnamefont{T.}~\bibnamefont{Daylan}},
  \bibinfo{author}{\bibfnamefont{D.~P.} \bibnamefont{Finkbeiner}},
  \bibinfo{author}{\bibfnamefont{D.}~\bibnamefont{Hooper}},
  \bibinfo{author}{\bibfnamefont{T.}~\bibnamefont{Linden}},
  \bibinfo{author}{\bibfnamefont{S.~K.~N.} \bibnamefont{Portillo}},
  \bibinfo{author}{\bibfnamefont{N.~L.} \bibnamefont{Rodd}}, \bibnamefont{and}
  \bibinfo{author}{\bibfnamefont{T.~R.} \bibnamefont{Slatyer}},
  \bibinfo{journal}{Phys. Dark Univ.} \textbf{\bibinfo{volume}{12}},
  \bibinfo{pages}{1} (\bibinfo{year}{2016}), \eprint{1402.6703}.

\bibitem[{\citenamefont{Goodenough and Hooper}(2009)}]{Goodenough:2009gk}
\bibinfo{author}{\bibfnamefont{L.}~\bibnamefont{Goodenough}} \bibnamefont{and}
  \bibinfo{author}{\bibfnamefont{D.}~\bibnamefont{Hooper}}
  (\bibinfo{year}{2009}), \eprint{0910.2998}.

\bibitem[{\citenamefont{Hooper and Goodenough}(2011)}]{Hooper:2010mq}
\bibinfo{author}{\bibfnamefont{D.}~\bibnamefont{Hooper}} \bibnamefont{and}
  \bibinfo{author}{\bibfnamefont{L.}~\bibnamefont{Goodenough}},
  \bibinfo{journal}{Phys. Lett.} \textbf{\bibinfo{volume}{B697}},
  \bibinfo{pages}{412} (\bibinfo{year}{2011}), \eprint{1010.2752}.

\bibitem[{\citenamefont{Ajello et~al.}(2016)}]{TheFermi-LAT:2015kwa}
\bibinfo{author}{\bibfnamefont{M.}~\bibnamefont{Ajello}} \bibnamefont{et~al.}
  (\bibinfo{collaboration}{Fermi-LAT}), \bibinfo{journal}{Astrophys. J.}
  \textbf{\bibinfo{volume}{819}}, \bibinfo{pages}{44} (\bibinfo{year}{2016}),
  \eprint{1511.02938}.

\bibitem[{\citenamefont{Karwin et~al.}(2017)\citenamefont{Karwin, Murgia, Tait,
  Porter, and Tanedo}}]{Karwin:2016tsw}
\bibinfo{author}{\bibfnamefont{C.}~\bibnamefont{Karwin}},
  \bibinfo{author}{\bibfnamefont{S.}~\bibnamefont{Murgia}},
  \bibinfo{author}{\bibfnamefont{T.~M.~P.} \bibnamefont{Tait}},
  \bibinfo{author}{\bibfnamefont{T.~A.} \bibnamefont{Porter}},
  \bibnamefont{and} \bibinfo{author}{\bibfnamefont{P.}~\bibnamefont{Tanedo}},
  \bibinfo{journal}{Phys. Rev.} \textbf{\bibinfo{volume}{D95}},
  \bibinfo{pages}{103005} (\bibinfo{year}{2017}), \eprint{1612.05687}.

\bibitem[{\citenamefont{Lee et~al.}(2016)\citenamefont{Lee, Lisanti, Safdi,
  Slatyer, and Xue}}]{Lee:2015fea}
\bibinfo{author}{\bibfnamefont{S.~K.} \bibnamefont{Lee}},
  \bibinfo{author}{\bibfnamefont{M.}~\bibnamefont{Lisanti}},
  \bibinfo{author}{\bibfnamefont{B.~R.} \bibnamefont{Safdi}},
  \bibinfo{author}{\bibfnamefont{T.~R.} \bibnamefont{Slatyer}},
  \bibnamefont{and} \bibinfo{author}{\bibfnamefont{W.}~\bibnamefont{Xue}},
  \bibinfo{journal}{Phys. Rev. Lett.} \textbf{\bibinfo{volume}{116}},
  \bibinfo{pages}{051103} (\bibinfo{year}{2016}), \eprint{1506.05124}.

\bibitem[{\citenamefont{Bartels et~al.}(2016)\citenamefont{Bartels,
  Krishnamurthy, and Weniger}}]{Bartels:2015aea}
\bibinfo{author}{\bibfnamefont{R.}~\bibnamefont{Bartels}},
  \bibinfo{author}{\bibfnamefont{S.}~\bibnamefont{Krishnamurthy}},
  \bibnamefont{and} \bibinfo{author}{\bibfnamefont{C.}~\bibnamefont{Weniger}},
  \bibinfo{journal}{Phys. Rev. Lett.} \textbf{\bibinfo{volume}{116}},
  \bibinfo{pages}{051102} (\bibinfo{year}{2016}), \eprint{1506.05104}.

\bibitem[{\citenamefont{Linden et~al.}(2016)\citenamefont{Linden, Rodd, Safdi,
  and Slatyer}}]{Linden:2016rcf}
\bibinfo{author}{\bibfnamefont{T.}~\bibnamefont{Linden}},
  \bibinfo{author}{\bibfnamefont{N.~L.} \bibnamefont{Rodd}},
  \bibinfo{author}{\bibfnamefont{B.~R.} \bibnamefont{Safdi}}, \bibnamefont{and}
  \bibinfo{author}{\bibfnamefont{T.~R.} \bibnamefont{Slatyer}},
  \bibinfo{journal}{Phys. Rev.} \textbf{\bibinfo{volume}{D94}},
  \bibinfo{pages}{103013} (\bibinfo{year}{2016}), \eprint{1604.01026}.

\bibitem[{\citenamefont{Ajello et~al.}(2017)}]{FermiLAT:2017yoi}
\bibinfo{author}{\bibfnamefont{M.}~\bibnamefont{Ajello}} \bibnamefont{et~al.}
  (\bibinfo{collaboration}{Fermi-LAT}), \bibinfo{journal}{Submitted to:
  Astrophys. J.}  (\bibinfo{year}{2017}), \eprint{1705.00009}.

\bibitem[{\citenamefont{{Astropy Collaboration}
  et~al.}(2013)\citenamefont{{Astropy Collaboration}, {Robitaille}, {Tollerud},
  {Greenfield}, {Droettboom}, {Bray}, {Aldcroft}, {Davis}, {Ginsburg},
  {Price-Whelan} et~al.}}]{2013A&A...558A..33A}
\bibinfo{author}{\bibnamefont{{Astropy Collaboration}}},
  \bibinfo{author}{\bibfnamefont{T.~P.} \bibnamefont{{Robitaille}}},
  \bibinfo{author}{\bibfnamefont{E.~J.} \bibnamefont{{Tollerud}}},
  \bibinfo{author}{\bibfnamefont{P.}~\bibnamefont{{Greenfield}}},
  \bibinfo{author}{\bibfnamefont{M.}~\bibnamefont{{Droettboom}}},
  \bibinfo{author}{\bibfnamefont{E.}~\bibnamefont{{Bray}}},
  \bibinfo{author}{\bibfnamefont{T.}~\bibnamefont{{Aldcroft}}},
  \bibinfo{author}{\bibfnamefont{M.}~\bibnamefont{{Davis}}},
  \bibinfo{author}{\bibfnamefont{A.}~\bibnamefont{{Ginsburg}}},
  \bibinfo{author}{\bibfnamefont{A.~M.} \bibnamefont{{Price-Whelan}}},
  \bibnamefont{et~al.}, \bibinfo{journal}{AAP} \textbf{\bibinfo{volume}{558}},
  \bibinfo{eid}{A33} (\bibinfo{year}{2013}), \eprint{1307.6212}.

\bibitem[{\citenamefont{P\'erez and Granger}(2007)}]{PER-GRA:2007}
\bibinfo{author}{\bibfnamefont{F.}~\bibnamefont{P\'erez}} \bibnamefont{and}
  \bibinfo{author}{\bibfnamefont{B.~E.} \bibnamefont{Granger}},
  \bibinfo{journal}{Computing in Science and Engineering}
  \textbf{\bibinfo{volume}{9}}, \bibinfo{pages}{21} (\bibinfo{year}{2007}),
  ISSN \bibinfo{issn}{1521-9615}, \urlprefix\url{http://ipython.org}.

\bibitem[{\citenamefont{James and Roos}(1975)}]{James:1975dr}
\bibinfo{author}{\bibfnamefont{F.}~\bibnamefont{James}} \bibnamefont{and}
  \bibinfo{author}{\bibfnamefont{M.}~\bibnamefont{Roos}},
  \bibinfo{journal}{Comput. Phys. Commun.} \textbf{\bibinfo{volume}{10}},
  \bibinfo{pages}{343} (\bibinfo{year}{1975}).

\bibitem[{\citenamefont{Elor et~al.}(2015)\citenamefont{Elor, Rodd, and
  Slatyer}}]{Elor:2015tva}
\bibinfo{author}{\bibfnamefont{G.}~\bibnamefont{Elor}},
  \bibinfo{author}{\bibfnamefont{N.~L.} \bibnamefont{Rodd}}, \bibnamefont{and}
  \bibinfo{author}{\bibfnamefont{T.~R.} \bibnamefont{Slatyer}},
  \bibinfo{journal}{Phys. Rev.} \textbf{\bibinfo{volume}{D91}},
  \bibinfo{pages}{103531} (\bibinfo{year}{2015}), \eprint{1503.01773}.

\bibitem[{\citenamefont{Elor et~al.}(2016)\citenamefont{Elor, Rodd, Slatyer,
  and Xue}}]{Elor:2015bho}
\bibinfo{author}{\bibfnamefont{G.}~\bibnamefont{Elor}},
  \bibinfo{author}{\bibfnamefont{N.~L.} \bibnamefont{Rodd}},
  \bibinfo{author}{\bibfnamefont{T.~R.} \bibnamefont{Slatyer}},
  \bibnamefont{and} \bibinfo{author}{\bibfnamefont{W.}~\bibnamefont{Xue}},
  \bibinfo{journal}{JCAP} \textbf{\bibinfo{volume}{1606}}, \bibinfo{pages}{024}
  (\bibinfo{year}{2016}), \eprint{1511.08787}.

\bibitem[{\citenamefont{Cirelli et~al.}(2010)\citenamefont{Cirelli, Panci, and
  Serpico}}]{Cirelli:2009dv}
\bibinfo{author}{\bibfnamefont{M.}~\bibnamefont{Cirelli}},
  \bibinfo{author}{\bibfnamefont{P.}~\bibnamefont{Panci}}, \bibnamefont{and}
  \bibinfo{author}{\bibfnamefont{P.~D.} \bibnamefont{Serpico}},
  \bibinfo{journal}{Nucl. Phys.} \textbf{\bibinfo{volume}{B840}},
  \bibinfo{pages}{284} (\bibinfo{year}{2010}), \eprint{0912.0663}.

\bibitem[{\citenamefont{Murase and Beacom}(2012)}]{Murase:2012xs}
\bibinfo{author}{\bibfnamefont{K.}~\bibnamefont{Murase}} \bibnamefont{and}
  \bibinfo{author}{\bibfnamefont{J.~F.} \bibnamefont{Beacom}},
  \bibinfo{journal}{JCAP} \textbf{\bibinfo{volume}{1210}}, \bibinfo{pages}{043}
  (\bibinfo{year}{2012}), \eprint{1206.2595}.

\bibitem[{\citenamefont{Cowan et~al.}(2011)\citenamefont{Cowan, Cranmer, Gross,
  and Vitells}}]{Cowan:2010js}
\bibinfo{author}{\bibfnamefont{G.}~\bibnamefont{Cowan}},
  \bibinfo{author}{\bibfnamefont{K.}~\bibnamefont{Cranmer}},
  \bibinfo{author}{\bibfnamefont{E.}~\bibnamefont{Gross}}, \bibnamefont{and}
  \bibinfo{author}{\bibfnamefont{O.}~\bibnamefont{Vitells}},
  \bibinfo{journal}{Eur. Phys. J.} \textbf{\bibinfo{volume}{C71}},
  \bibinfo{pages}{1554} (\bibinfo{year}{2011}), \bibinfo{note}{[Erratum: Eur.
  Phys. J.C73,2501(2013)]}, \eprint{1007.1727}.

\bibitem[{\citenamefont{Edwards and Weniger}(2017)}]{Edwards:2017mnf}
\bibinfo{author}{\bibfnamefont{T.~D.~P.} \bibnamefont{Edwards}}
  \bibnamefont{and} \bibinfo{author}{\bibfnamefont{C.}~\bibnamefont{Weniger}}
  (\bibinfo{year}{2017}), \eprint{1704.05458}.

\bibitem[{\citenamefont{Burkert}(1996)}]{Burkert:1995yz}
\bibinfo{author}{\bibfnamefont{A.}~\bibnamefont{Burkert}},
  \bibinfo{journal}{IAU Symp.} \textbf{\bibinfo{volume}{171}},
  \bibinfo{pages}{175} (\bibinfo{year}{1996}), \bibinfo{note}{[Astrophys.
  J.447,L25(1995)]}, \eprint{astro-ph/9504041}.

\bibitem[{\citenamefont{Diemer and Kravtsov}(2015)}]{Diemer:2014gba}
\bibinfo{author}{\bibfnamefont{B.}~\bibnamefont{Diemer}} \bibnamefont{and}
  \bibinfo{author}{\bibfnamefont{A.~V.} \bibnamefont{Kravtsov}},
  \bibinfo{journal}{Astrophys. J.} \textbf{\bibinfo{volume}{799}},
  \bibinfo{pages}{108} (\bibinfo{year}{2015}), \eprint{1407.4730}.

\bibitem[{\citenamefont{Strong et~al.}(2007)\citenamefont{Strong, Moskalenko,
  and Ptuskin}}]{Strong:2007nh}
\bibinfo{author}{\bibfnamefont{A.~W.} \bibnamefont{Strong}},
  \bibinfo{author}{\bibfnamefont{I.~V.} \bibnamefont{Moskalenko}},
  \bibnamefont{and} \bibinfo{author}{\bibfnamefont{V.~S.}
  \bibnamefont{Ptuskin}}, \bibinfo{journal}{Ann. Rev. Nucl. Part. Sci.}
  \textbf{\bibinfo{volume}{57}}, \bibinfo{pages}{285} (\bibinfo{year}{2007}),
  \eprint{astro-ph/0701517}.

\bibitem[{\citenamefont{Acero et~al.}(2016)}]{Acero:2016qlg}
\bibinfo{author}{\bibfnamefont{F.}~\bibnamefont{Acero}} \bibnamefont{et~al.}
  (\bibinfo{collaboration}{Fermi-LAT}), \bibinfo{journal}{Astrophys. J. Suppl.}
  \textbf{\bibinfo{volume}{223}}, \bibinfo{pages}{26} (\bibinfo{year}{2016}),
  \eprint{1602.07246}.

\bibitem[{\citenamefont{Narayanan and Slatyer}(2017)}]{Narayanan:2016nzy}
\bibinfo{author}{\bibfnamefont{S.~A.} \bibnamefont{Narayanan}}
  \bibnamefont{and} \bibinfo{author}{\bibfnamefont{T.~R.}
  \bibnamefont{Slatyer}}, \bibinfo{journal}{Mon. Not. Roy. Astron. Soc.}
  \textbf{\bibinfo{volume}{468}}, \bibinfo{pages}{3051} (\bibinfo{year}{2017}),
  \eprint{1603.06582}.

\bibitem[{\citenamefont{Cohen et~al.}(2017)\citenamefont{Cohen, Murase, Rodd,
  Safdi, and Soreq}}]{Cohen:2016uyg}
\bibinfo{author}{\bibfnamefont{T.}~\bibnamefont{Cohen}},
  \bibinfo{author}{\bibfnamefont{K.}~\bibnamefont{Murase}},
  \bibinfo{author}{\bibfnamefont{N.~L.} \bibnamefont{Rodd}},
  \bibinfo{author}{\bibfnamefont{B.~R.} \bibnamefont{Safdi}}, \bibnamefont{and}
  \bibinfo{author}{\bibfnamefont{Y.}~\bibnamefont{Soreq}},
  \bibinfo{journal}{Phys. Rev. Lett.} \textbf{\bibinfo{volume}{119}},
  \bibinfo{pages}{021102} (\bibinfo{year}{2017}), \eprint{1612.05638}.

\bibitem[{\citenamefont{Springel et~al.}(2008)\citenamefont{Springel, Wang,
  Vogelsberger, Ludlow, Jenkins, Helmi, Navarro, Frenk, and
  White}}]{Springel:2008cc}
\bibinfo{author}{\bibfnamefont{V.}~\bibnamefont{Springel}},
  \bibinfo{author}{\bibfnamefont{J.}~\bibnamefont{Wang}},
  \bibinfo{author}{\bibfnamefont{M.}~\bibnamefont{Vogelsberger}},
  \bibinfo{author}{\bibfnamefont{A.}~\bibnamefont{Ludlow}},
  \bibinfo{author}{\bibfnamefont{A.}~\bibnamefont{Jenkins}},
  \bibinfo{author}{\bibfnamefont{A.}~\bibnamefont{Helmi}},
  \bibinfo{author}{\bibfnamefont{J.~F.} \bibnamefont{Navarro}},
  \bibinfo{author}{\bibfnamefont{C.~S.} \bibnamefont{Frenk}}, \bibnamefont{and}
  \bibinfo{author}{\bibfnamefont{S.~D.~M.} \bibnamefont{White}},
  \bibinfo{journal}{Mon. Not. Roy. Astron. Soc.}
  \textbf{\bibinfo{volume}{391}}, \bibinfo{pages}{1685} (\bibinfo{year}{2008}),
  \eprint{0809.0898}.

\bibitem[{\citenamefont{Schaller et~al.}(2015)\citenamefont{Schaller, Frenk,
  Bower, Theuns, Jenkins, Schaye, Crain, Furlong, Vecchia, and
  McCarthy}}]{Schaller:2014uwa}
\bibinfo{author}{\bibfnamefont{M.}~\bibnamefont{Schaller}},
  \bibinfo{author}{\bibfnamefont{C.~S.} \bibnamefont{Frenk}},
  \bibinfo{author}{\bibfnamefont{R.~G.} \bibnamefont{Bower}},
  \bibinfo{author}{\bibfnamefont{T.}~\bibnamefont{Theuns}},
  \bibinfo{author}{\bibfnamefont{A.}~\bibnamefont{Jenkins}},
  \bibinfo{author}{\bibfnamefont{J.}~\bibnamefont{Schaye}},
  \bibinfo{author}{\bibfnamefont{R.~A.} \bibnamefont{Crain}},
  \bibinfo{author}{\bibfnamefont{M.}~\bibnamefont{Furlong}},
  \bibinfo{author}{\bibfnamefont{C.~D.} \bibnamefont{Vecchia}},
  \bibnamefont{and} \bibinfo{author}{\bibfnamefont{I.~G.}
  \bibnamefont{McCarthy}}, \bibinfo{journal}{Mon. Not. Roy. Astron. Soc.}
  \textbf{\bibinfo{volume}{451}}, \bibinfo{pages}{1247} (\bibinfo{year}{2015}),
  \eprint{1409.8617}.

\bibitem[{\citenamefont{Bullock et~al.}(2001)\citenamefont{Bullock, Kolatt,
  Sigad, Somerville, Kravtsov, Klypin, Primack, and Dekel}}]{Bullock:1999he}
\bibinfo{author}{\bibfnamefont{J.~S.} \bibnamefont{Bullock}},
  \bibinfo{author}{\bibfnamefont{T.~S.} \bibnamefont{Kolatt}},
  \bibinfo{author}{\bibfnamefont{Y.}~\bibnamefont{Sigad}},
  \bibinfo{author}{\bibfnamefont{R.~S.} \bibnamefont{Somerville}},
  \bibinfo{author}{\bibfnamefont{A.~V.} \bibnamefont{Kravtsov}},
  \bibinfo{author}{\bibfnamefont{A.~A.} \bibnamefont{Klypin}},
  \bibinfo{author}{\bibfnamefont{J.~R.} \bibnamefont{Primack}},
  \bibnamefont{and} \bibinfo{author}{\bibfnamefont{A.}~\bibnamefont{Dekel}},
  \bibinfo{journal}{Mon. Not. Roy. Astron. Soc.}
  \textbf{\bibinfo{volume}{321}}, \bibinfo{pages}{559} (\bibinfo{year}{2001}),
  \eprint{astro-ph/9908159}.

\bibitem[{\citenamefont{Molin{\'e} et~al.}(2017)\citenamefont{Molin{\'e},
  S{\'a}nchez-Conde, Palomares-Ruiz, and Prada}}]{Moline:2016pbm}
\bibinfo{author}{\bibfnamefont{{\'A}.}~\bibnamefont{Molin{\'e}}},
  \bibinfo{author}{\bibfnamefont{M.~A.} \bibnamefont{S{\'a}nchez-Conde}},
  \bibinfo{author}{\bibfnamefont{S.}~\bibnamefont{Palomares-Ruiz}},
  \bibnamefont{and} \bibinfo{author}{\bibfnamefont{F.}~\bibnamefont{Prada}},
  \bibinfo{journal}{Mon. Not. Roy. Astron. Soc.}
  \textbf{\bibinfo{volume}{466}}, \bibinfo{pages}{4974} (\bibinfo{year}{2017}),
  \eprint{1603.04057}.

\bibitem[{\citenamefont{Nezri et~al.}(2012)\citenamefont{Nezri, White, Combet,
  Maurin, Pointecouteau, and Hinton}}]{Nezri:2012tu}
\bibinfo{author}{\bibfnamefont{E.}~\bibnamefont{Nezri}},
  \bibinfo{author}{\bibfnamefont{R.}~\bibnamefont{White}},
  \bibinfo{author}{\bibfnamefont{C.}~\bibnamefont{Combet}},
  \bibinfo{author}{\bibfnamefont{D.}~\bibnamefont{Maurin}},
  \bibinfo{author}{\bibfnamefont{E.}~\bibnamefont{Pointecouteau}},
  \bibnamefont{and} \bibinfo{author}{\bibfnamefont{J.~A.}
  \bibnamefont{Hinton}}, \bibinfo{journal}{Mon. Not. Roy. Astron. Soc.}
  \textbf{\bibinfo{volume}{425}}, \bibinfo{pages}{477} (\bibinfo{year}{2012}),
  \eprint{1203.1165}.

\bibitem[{\citenamefont{S{\'a}nchez-Conde and
  Prada}(2014)}]{Sanchez-Conde:2013yxa}
\bibinfo{author}{\bibfnamefont{M.~A.} \bibnamefont{S{\'a}nchez-Conde}}
  \bibnamefont{and} \bibinfo{author}{\bibfnamefont{F.}~\bibnamefont{Prada}},
  \bibinfo{journal}{Mon. Not. Roy. Astron. Soc.}
  \textbf{\bibinfo{volume}{442}}, \bibinfo{pages}{2271} (\bibinfo{year}{2014}),
  \eprint{1312.1729}.

\bibitem[{\citenamefont{Lu et~al.}(2016)\citenamefont{Lu, Yang, Shi, Mo, Tweed,
  Wang, Zhang, Li, and Lim}}]{Lu:2016vmu}
\bibinfo{author}{\bibfnamefont{Y.}~\bibnamefont{Lu}},
  \bibinfo{author}{\bibfnamefont{X.}~\bibnamefont{Yang}},
  \bibinfo{author}{\bibfnamefont{F.}~\bibnamefont{Shi}},
  \bibinfo{author}{\bibfnamefont{H.~J.} \bibnamefont{Mo}},
  \bibinfo{author}{\bibfnamefont{D.}~\bibnamefont{Tweed}},
  \bibinfo{author}{\bibfnamefont{H.}~\bibnamefont{Wang}},
  \bibinfo{author}{\bibfnamefont{Y.}~\bibnamefont{Zhang}},
  \bibinfo{author}{\bibfnamefont{S.}~\bibnamefont{Li}}, \bibnamefont{and}
  \bibinfo{author}{\bibfnamefont{S.~H.} \bibnamefont{Lim}},
  \bibinfo{journal}{Astrophys. J.} \textbf{\bibinfo{volume}{832}},
  \bibinfo{pages}{39} (\bibinfo{year}{2016}), \eprint{1607.03982}.

\bibitem[{\citenamefont{{Karachentsev} and
  {Makarov}}(1996)}]{1996AJ....111..794K}
\bibinfo{author}{\bibfnamefont{I.~D.} \bibnamefont{{Karachentsev}}}
  \bibnamefont{and} \bibinfo{author}{\bibfnamefont{D.~A.}
  \bibnamefont{{Makarov}}}, \bibinfo{journal}{AJ}
  \textbf{\bibinfo{volume}{111}}, \bibinfo{pages}{794} (\bibinfo{year}{1996}).

\bibitem[{\citenamefont{{Karachentsev}
  et~al.}(2014)\citenamefont{{Karachentsev}, {Tully}, {Wu}, {Shaya}, and
  {Dolphin}}}]{2014ApJ...782....4K}
\bibinfo{author}{\bibfnamefont{I.~D.} \bibnamefont{{Karachentsev}}},
  \bibinfo{author}{\bibfnamefont{R.~B.} \bibnamefont{{Tully}}},
  \bibinfo{author}{\bibfnamefont{P.-F.} \bibnamefont{{Wu}}},
  \bibinfo{author}{\bibfnamefont{E.~J.} \bibnamefont{{Shaya}}},
  \bibnamefont{and} \bibinfo{author}{\bibfnamefont{A.~E.}
  \bibnamefont{{Dolphin}}}, \bibinfo{journal}{\apj}
  \textbf{\bibinfo{volume}{782}}, \bibinfo{eid}{4} (\bibinfo{year}{2014}),
  \eprint{1312.6769}.

\bibitem[{\citenamefont{{Diaferio} et~al.}(1993)\citenamefont{{Diaferio},
  {Ramella}, {Geller}, and {Ferrari}}}]{1993AJ....105.2035D}
\bibinfo{author}{\bibfnamefont{A.}~\bibnamefont{{Diaferio}}},
  \bibinfo{author}{\bibfnamefont{M.}~\bibnamefont{{Ramella}}},
  \bibinfo{author}{\bibfnamefont{M.~J.} \bibnamefont{{Geller}}},
  \bibnamefont{and}
  \bibinfo{author}{\bibfnamefont{A.}~\bibnamefont{{Ferrari}}},
  \bibinfo{journal}{Astrophys. J.} \textbf{\bibinfo{volume}{105}},
  \bibinfo{pages}{2035} (\bibinfo{year}{1993}).

\end{thebibliography}

\clearpage
\newpage
\maketitle
\onecolumngrid
\begin{center}
\textbf{\large A Search for Dark Matter Annihilation in Galaxy Groups} \\ 
\vspace{0.05in}
{ \it \large Supplementary Material}\\ 
\vspace{0.05in}
{ Mariangela Lisanti, Siddharth Mishra-Sharma, Nicholas L. Rodd, and Benjamin R. Safdi}
\end{center}
\setcounter{equation}{0}
\setcounter{figure}{0}
\setcounter{table}{0}
\setcounter{section}{0}
\makeatletter
\renewcommand{\theequation}{S\arabic{equation}}
\renewcommand{\thefigure}{S\arabic{figure}}
\renewcommand{\thetable}{S\arabic{table}}
\newcommand\ptwiddle[1]{\mathord{\mathop{#1}\limits^{\scriptscriptstyle(\sim)}}}

This Supplementary Material is organized as follows. First, we provide an extended description of the main analysis results presented in this Letter, including limits for different annihilation channels, injected signal tests, individual bounds on the top ten galaxy groups studied, and sky maps of the extragalactic DM halos. Secondly, we show how the results are affected by variations in the analysis procedure, focusing specifically on the halo selection criteria, data set type, foreground models, halo density and concentration, substructure boost, and the galaxy group catalog.  

\section{Extended results}
\label{sec:extended}

\noindent  {\bf The $\mathbf{b\bar{b}}$ Channel.}  
In the main Letter, the right panel of Fig.~\ref{fig:bounds}  demonstrates how the limit on the  $b\bar b$ annihilation cross section  depends on the number of halos included in the stacking, for the case where  $m_\chi = 100$~GeV. In Fig.~\ref{fig:moreelephants}, we show the corresponding plot for $m_\chi = 10$~GeV (left) and $10$~TeV (right).  As in the 100~GeV case, we see that no single halo dominates the bound and that stacking a large number of halos considerably improves the sensitivity.

The left panel of Fig.~\ref{fig:other_lims} shows the maximum test statistic, TS$_\text{max}$, recovered for the stacked analysis in the $b\bar{b}$ channel.  For a given data set $d$, we define the maximum test-statistic in preference for the DM model, relative to the null hypothesis without DM, as 
\es{maxTS}{
{\rm TS}_\text{max}(\mathcal{M}, m_{\rm \chi}) \equiv & \,2 \left[ \log \mathcal{L}(d | \mathcal{M}, \widehat{\langle\sigma v\rangle}, m_\chi ) - \log \mathcal{L}(d | \mathcal{M}, \langle\sigma v\rangle =0, m_\chi ) \right] \, ,
}
where $\widehat{\langle\sigma v\rangle}$ is the cross section that maximizes the likelihood for DM model $\mathcal{M}$.  The observed TS$_\text{max}$ is negligible at all masses and well-within the null expectation (green/yellow bands), consistent with the conclusion that we find no evidence for DM annihilation.  \vspace{0.1in}

\begin{figure*}[b]
  \centering
	\includegraphics[width=.45\textwidth]{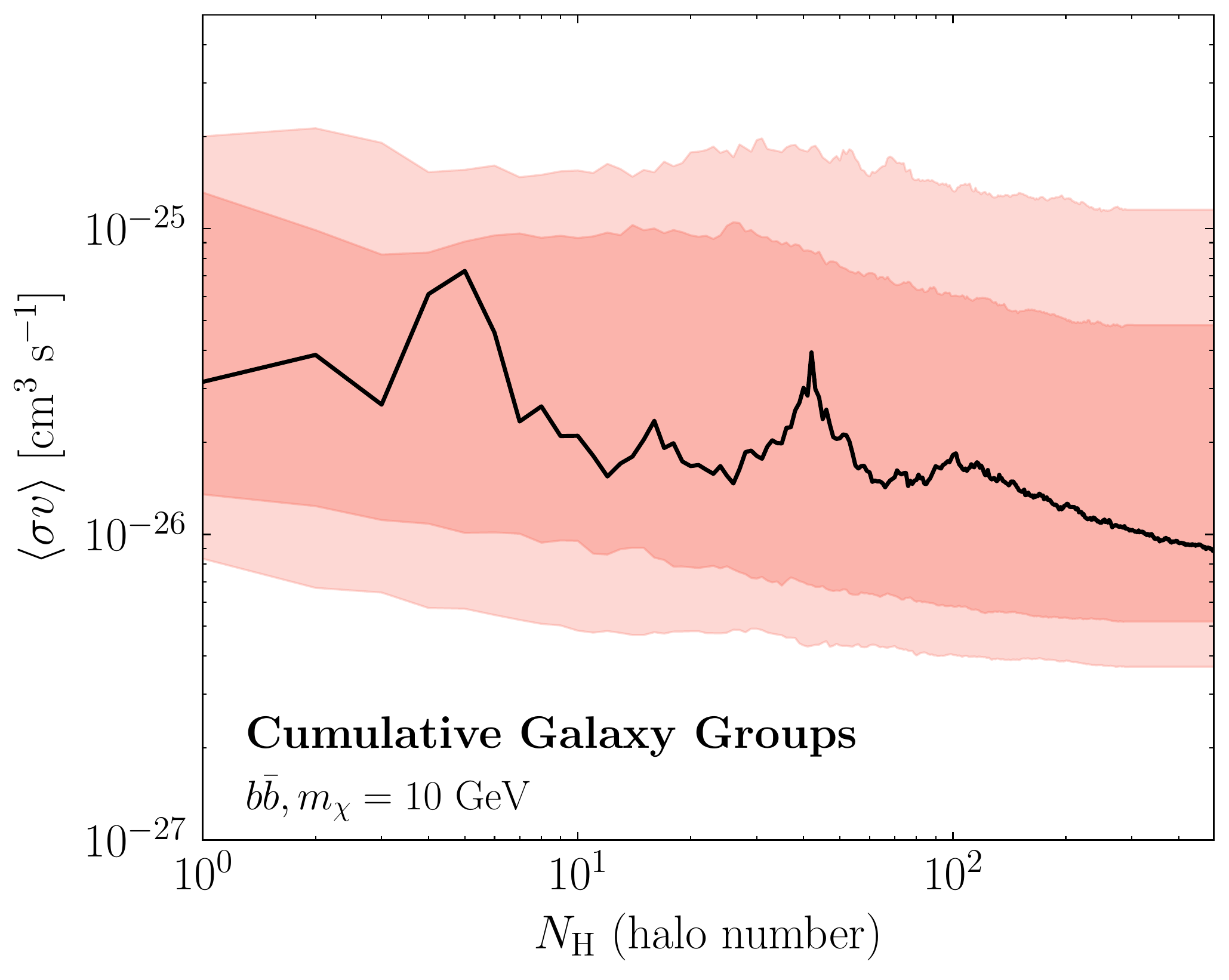} 
	\includegraphics[width=.45\textwidth]{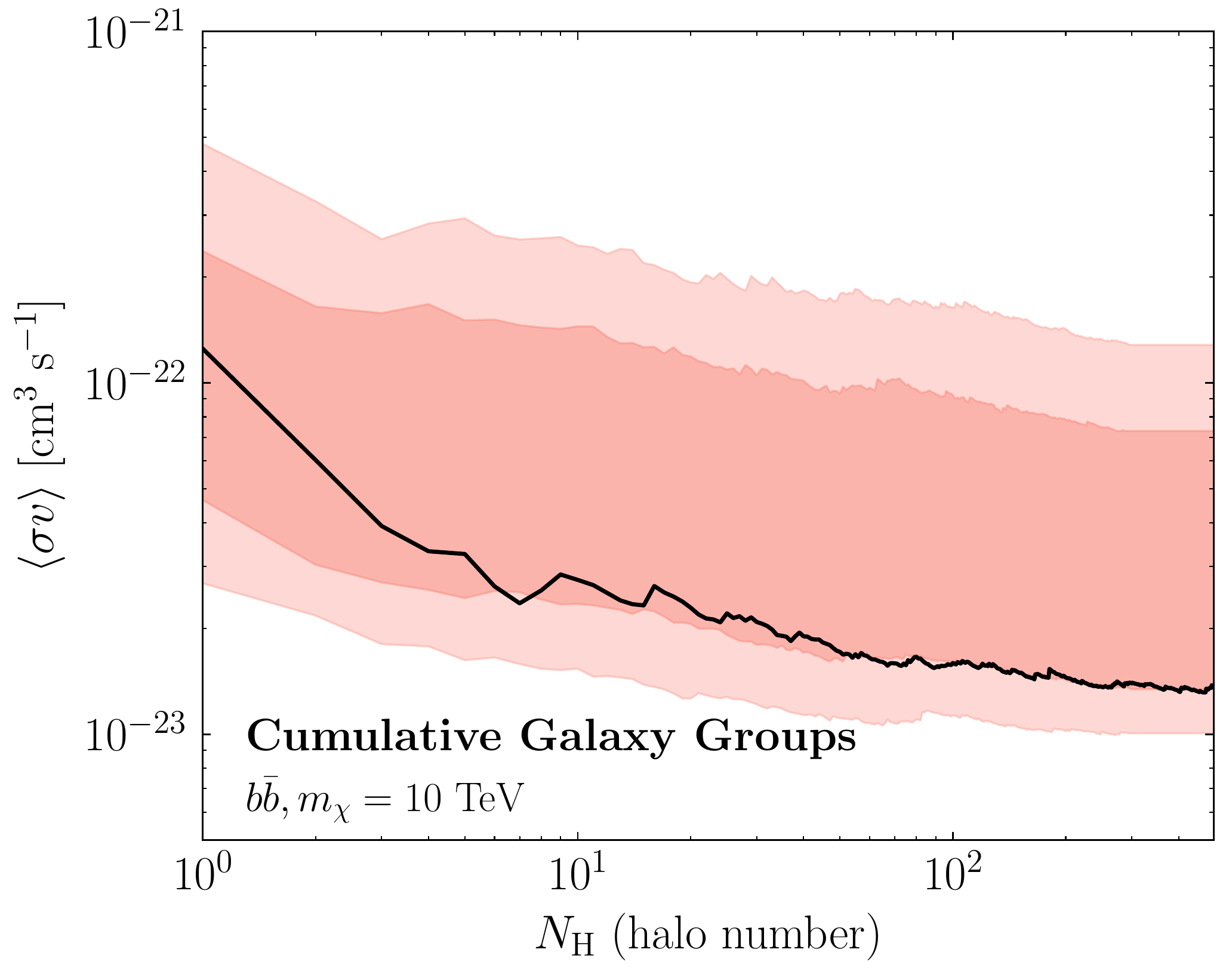} 
  \caption{The change in the limit on the $b\bar{b}$ annihilation channel as a function of the number of halos included in the stacking, for $m_\chi= 10$~GeV (left) and 10~TeV (right). The 68 and 95\% expectations from 200 random sky locations are indicated by the red bands.}
  \label{fig:moreelephants}
\end{figure*}

\noindent  {\bf Other Annihilation Channels.}  
In general, DM may annihilate to a variety of Standard Model final states.  Figure~\ref{fig:other_lims} (right) interprets the results of the analysis in terms of limits on additional final states that also lead to continuum gamma-ray emission.  Final states that predominantly decay hadronically  ($W^+ W^-$, $ZZ$, $q \bar{q}$, $c \bar c$, $b \bar b$, $t \bar t$) give similar limits because their energy spectra are mostly set by boosted pion decay.  The leptonic channels ($e^+ e^-$, $\mu^+ \mu^-$) give weaker limits because gamma-rays predominantly arise from final-state radiation or, in the case of the muon, radiative decays.  The $\tau^+ \tau^-$ limit is intermediate because roughly 35\% of the $\tau$ decays are leptonic, while the remaining are hadronic.   Of course, the DM could annihilate into even more complicated final states than the two-body cases considered here and the results can be extended to these cases~\cite{Elor:2015tva,Elor:2015bho}.  Note that the limits we present for the leptonic final states are conservative, as they neglect Inverse Compton (IC) emission and electromagnetic cascades, which are likely important at high DM masses---see~\emph{e.g.}, Ref.~\cite{Cirelli:2009dv, Murase:2012xs}.  A more careful treatment of these final states requires modeling the magnetic field strength and energy loss mechanisms within the galaxy groups. 
\vspace{0.1in}

\begin{figure*}[t]
  \centering
   \includegraphics[width=0.45\textwidth]{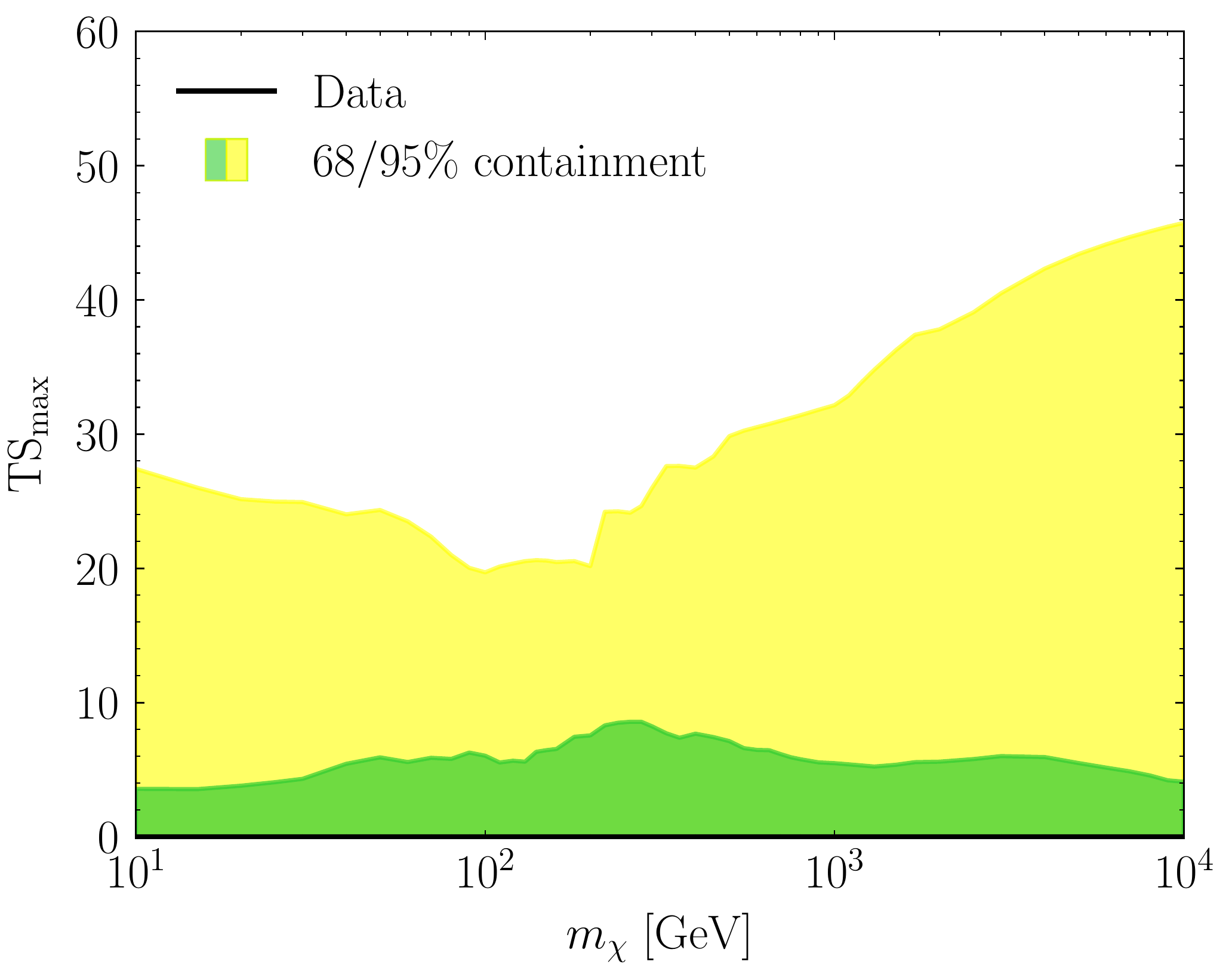}
  \includegraphics[width=.45\textwidth]{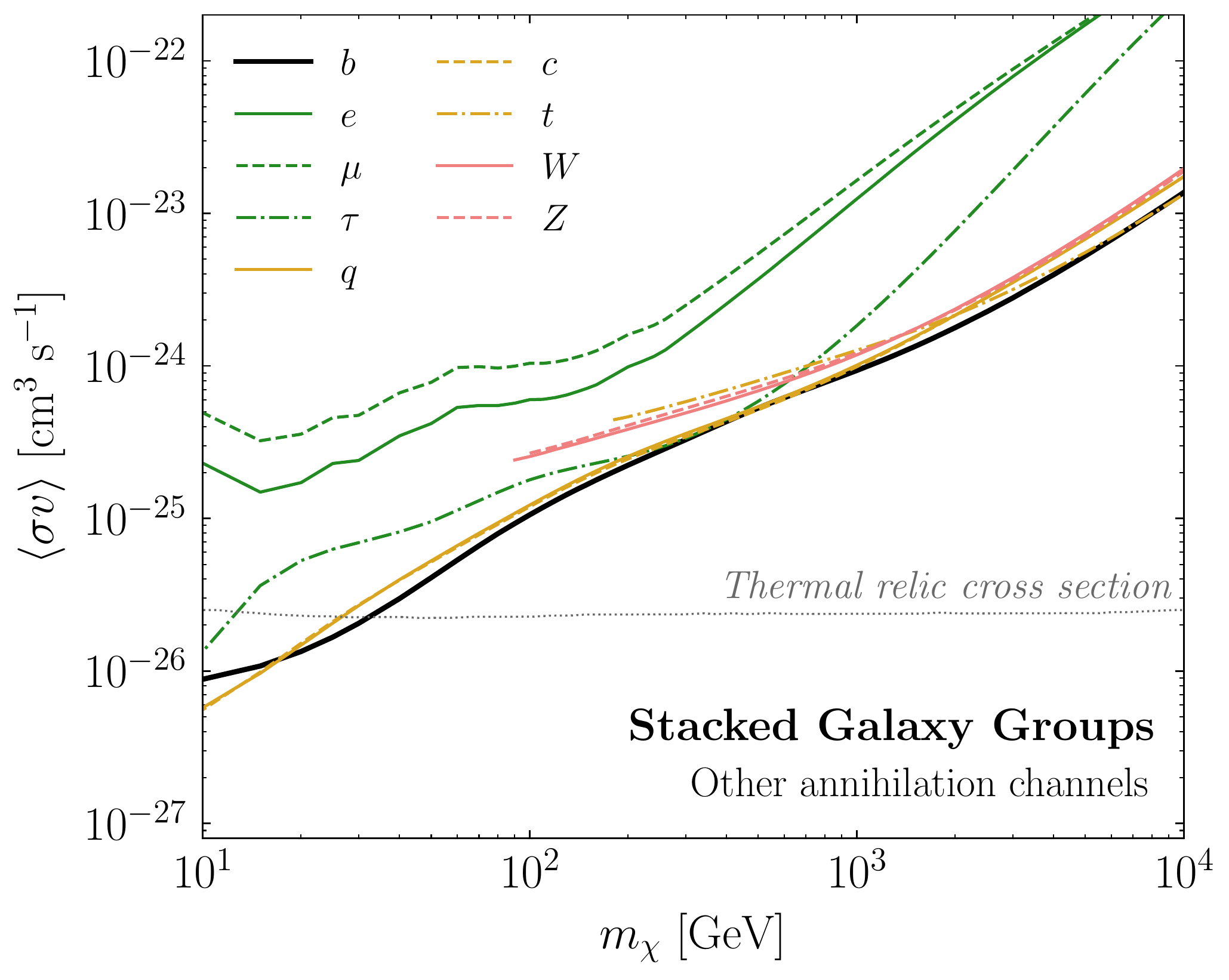}
  \caption{(Left) Maximum test statistic, TS$_\text{max}$, for the stacked analysis comparing the model with and without DM annihilating to $b \bar b$.  The green~(yellow) bands show the 68\%~(95\%) containment over multiple random sky locations.  (Right) The 95\% confidence limits on the DM annihilation cross section, as a function of the DM mass, for the Standard Model final states indicated in the legend.  These limits assume the fiducial boost factor taken from Ref.~\cite{Bartels:2015uba}.  Note that we neglect Inverse Compton emission and electromagnetic cascades, which can be relevant for the leptonic decay channels at high energies.}
  \label{fig:other_lims}
\end{figure*}

\noindent  {\bf Injected Signal.} An important consistency requirement is to ensure that the limit-setting procedure does not exclude a putative DM signal. The likelihood procedure employed here was extensively vetted in our companion paper~\cite{companion}, where we demonstrated that the limit never excludes an injected signal.  In Fig.~\ref{fig:injsig}, we demonstrate a data-driven version of this test. In detail, we inject a DM signal on top of the actual data set used in the main analysis, focusing on the case of DM annihilation to $b \bar{b}$ for a variety of cross sections and masses. We then apply the analysis pipeline to these maps.  The top panel of Fig.~\ref{fig:injsig} shows the recovered cross sections, as a function of the injected values.  The green line corresponds to the 95\% cross section limit, while the blue line shows the best-fit cross section.  Note that statistical uncertainties arising from DM annihilation photon counts are not significant here, as the dominant source of counts arises from the data itself. 
The columns correspond to 10, 100, and 10$^4$~GeV DM annihilating to $b \bar b$ (left, center, right, respectively).  The bottom row shows the maximum test statistic in favor of the model with DM as a function of the injected cross section.  The best-fit cross sections are only meaningful when the maximum test statistic is $\gtrsim 1$, implying evidence for DM annihilation.     
We see that across all masses, the cross section limit  (green line) is always weaker than the injected value.  Additionally, the recovered cross section (blue line) closely approaches that of the injected signal as the significance of the DM excess  increases.   
\vspace{0.1in}

\begin{figure*}[t]
  \centering
	\includegraphics[width=.32\textwidth]{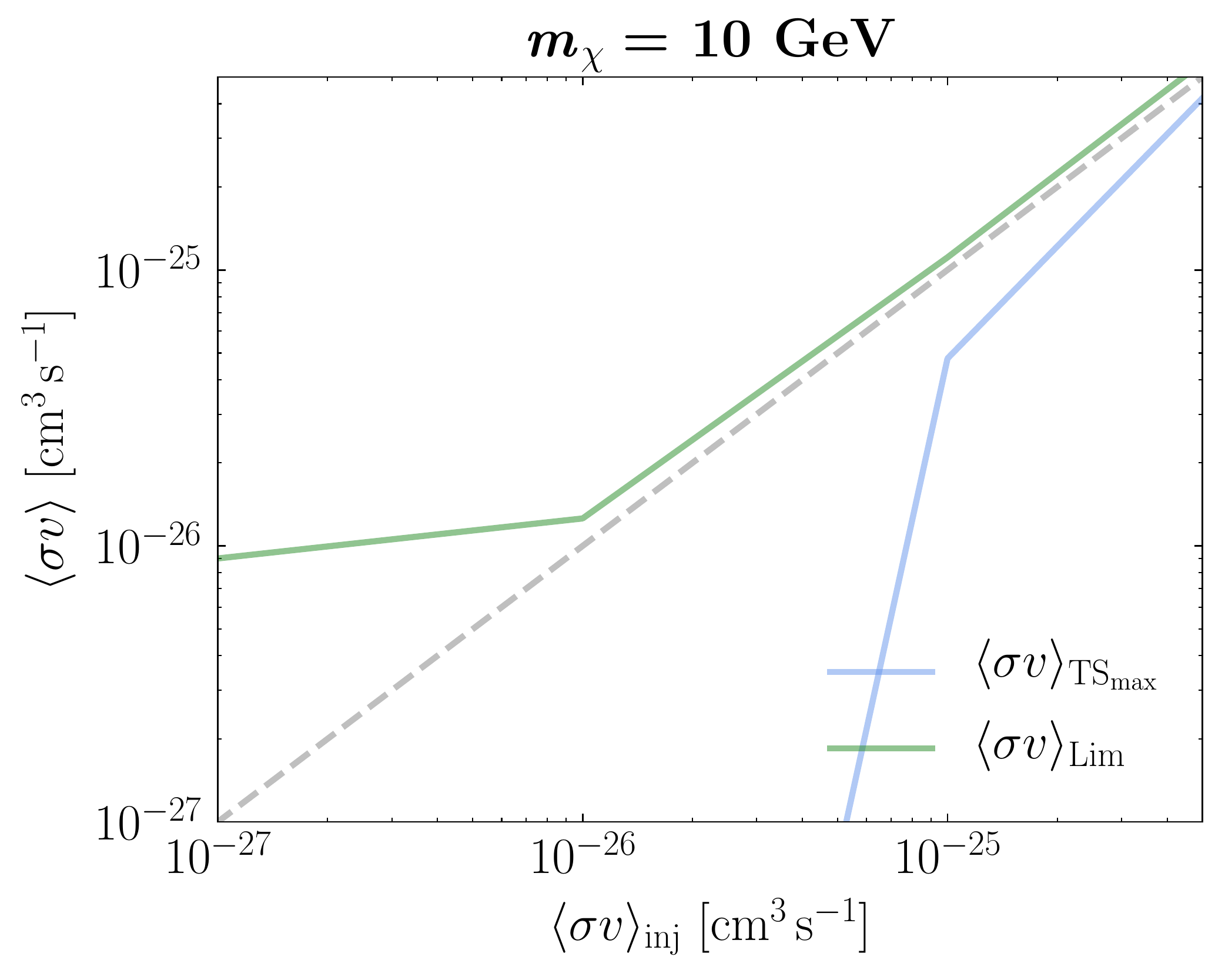}
	\includegraphics[width=.32\textwidth]{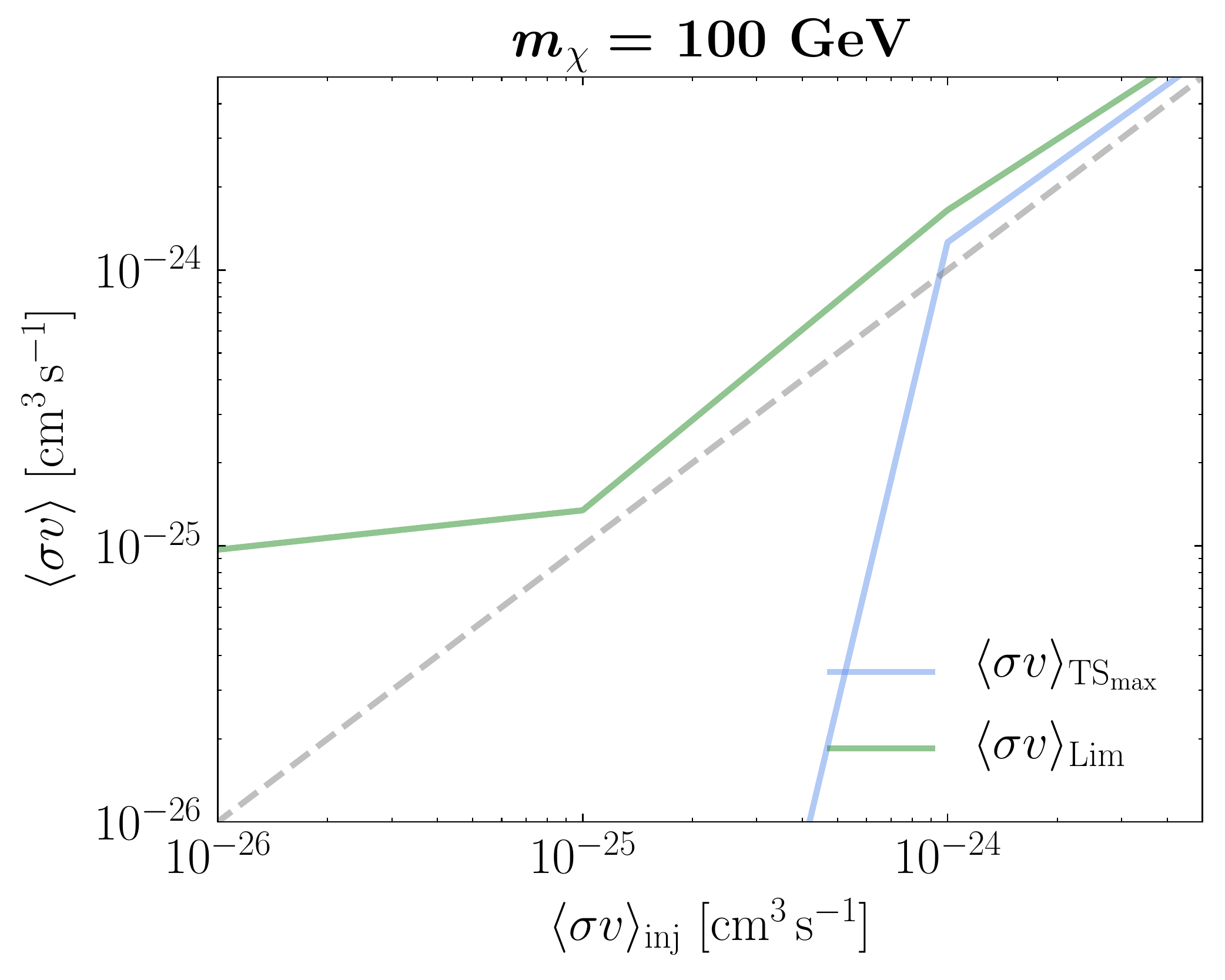}
	\includegraphics[width=.32\textwidth]{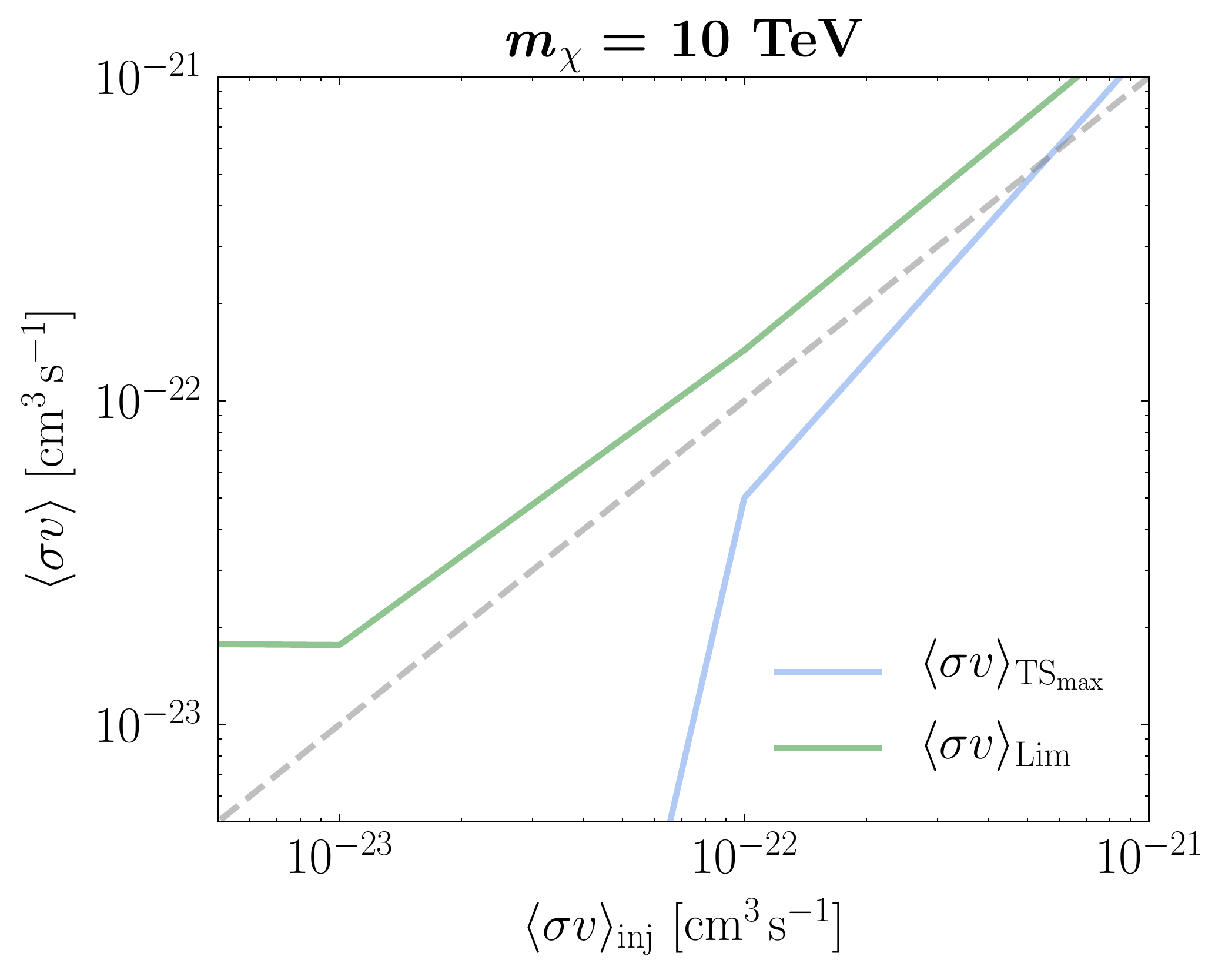} \\
	\includegraphics[width=.32\textwidth]{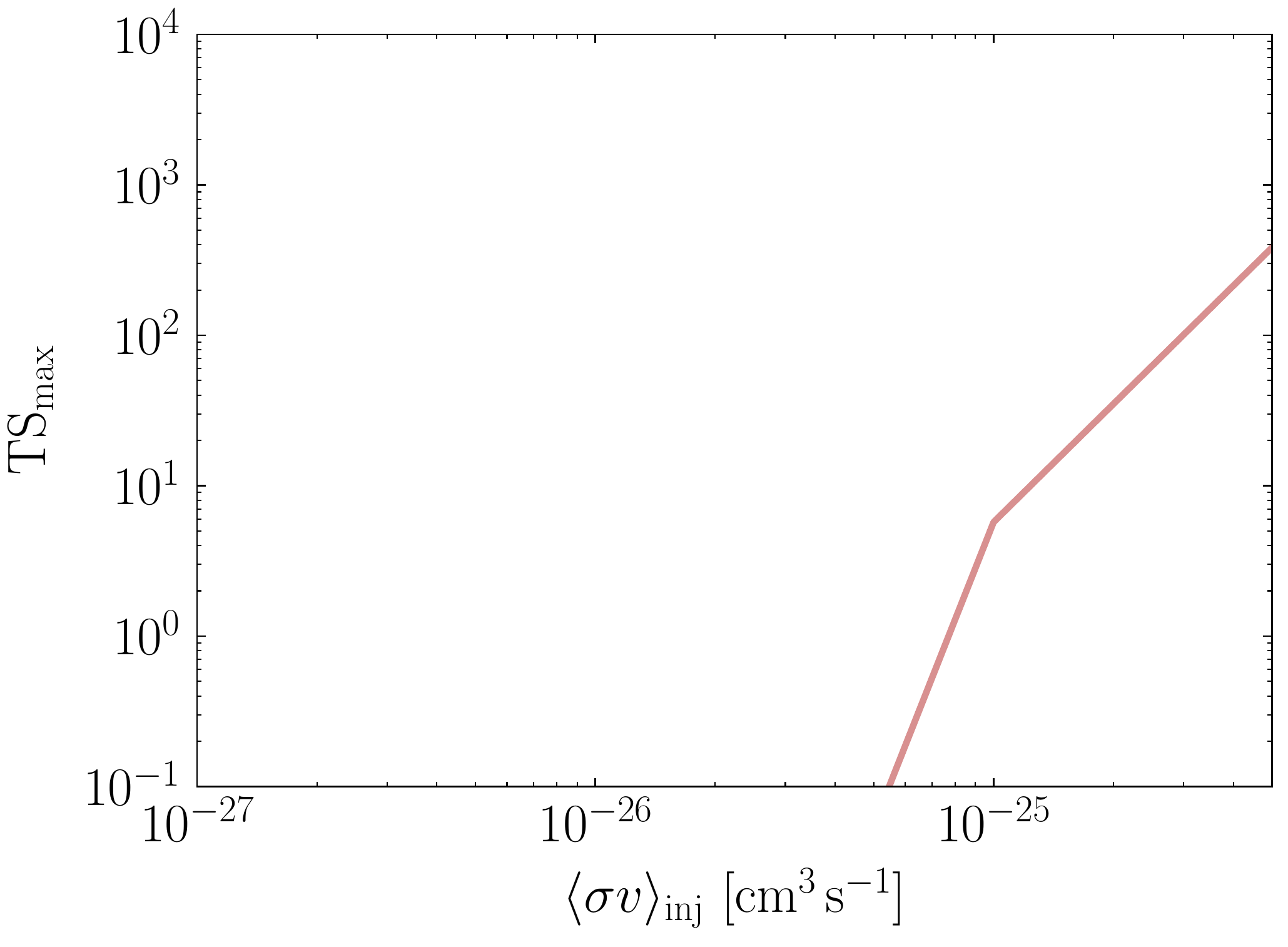}
	\includegraphics[width=.32\textwidth]{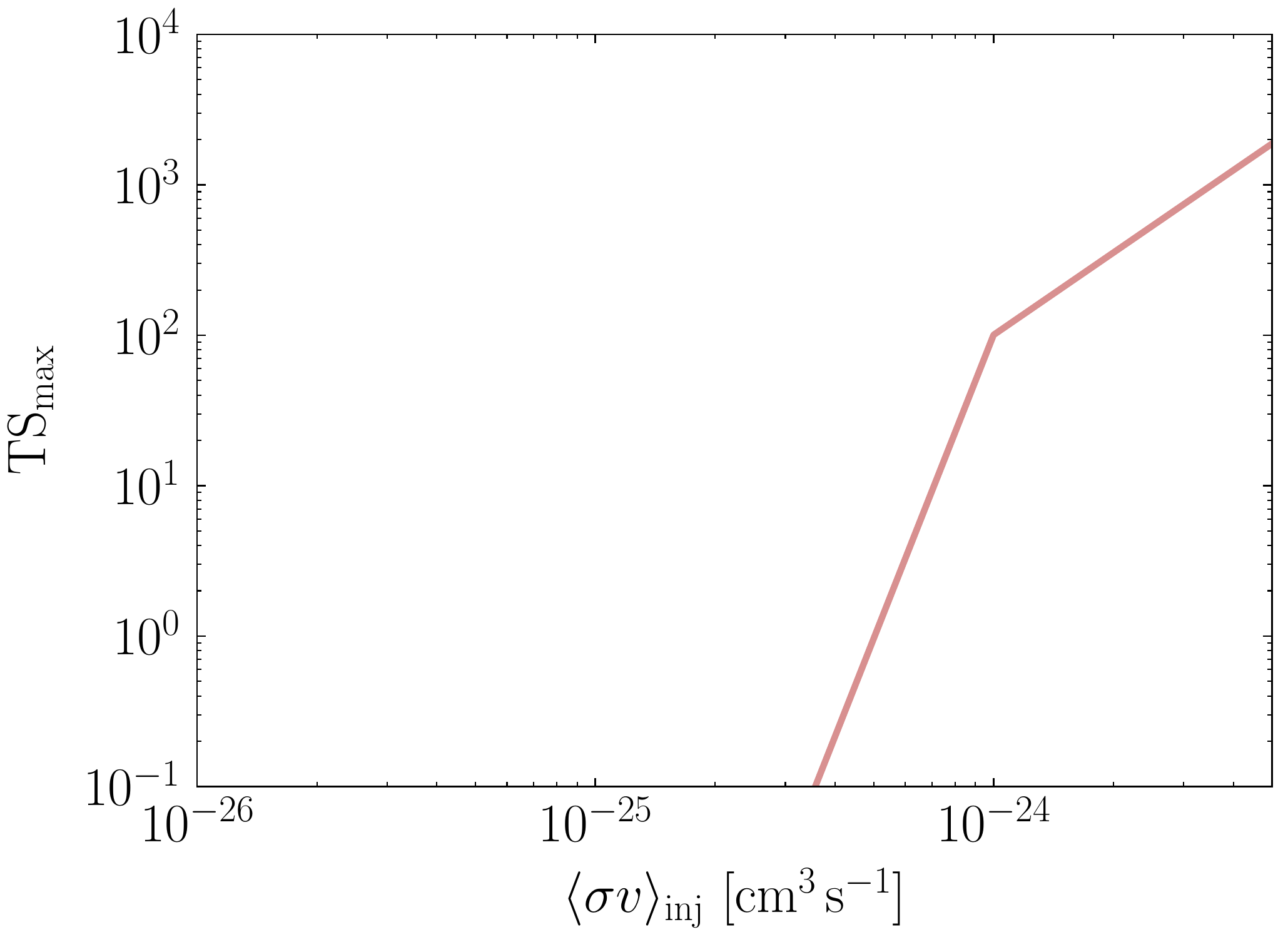}
	\includegraphics[width=.32\textwidth]{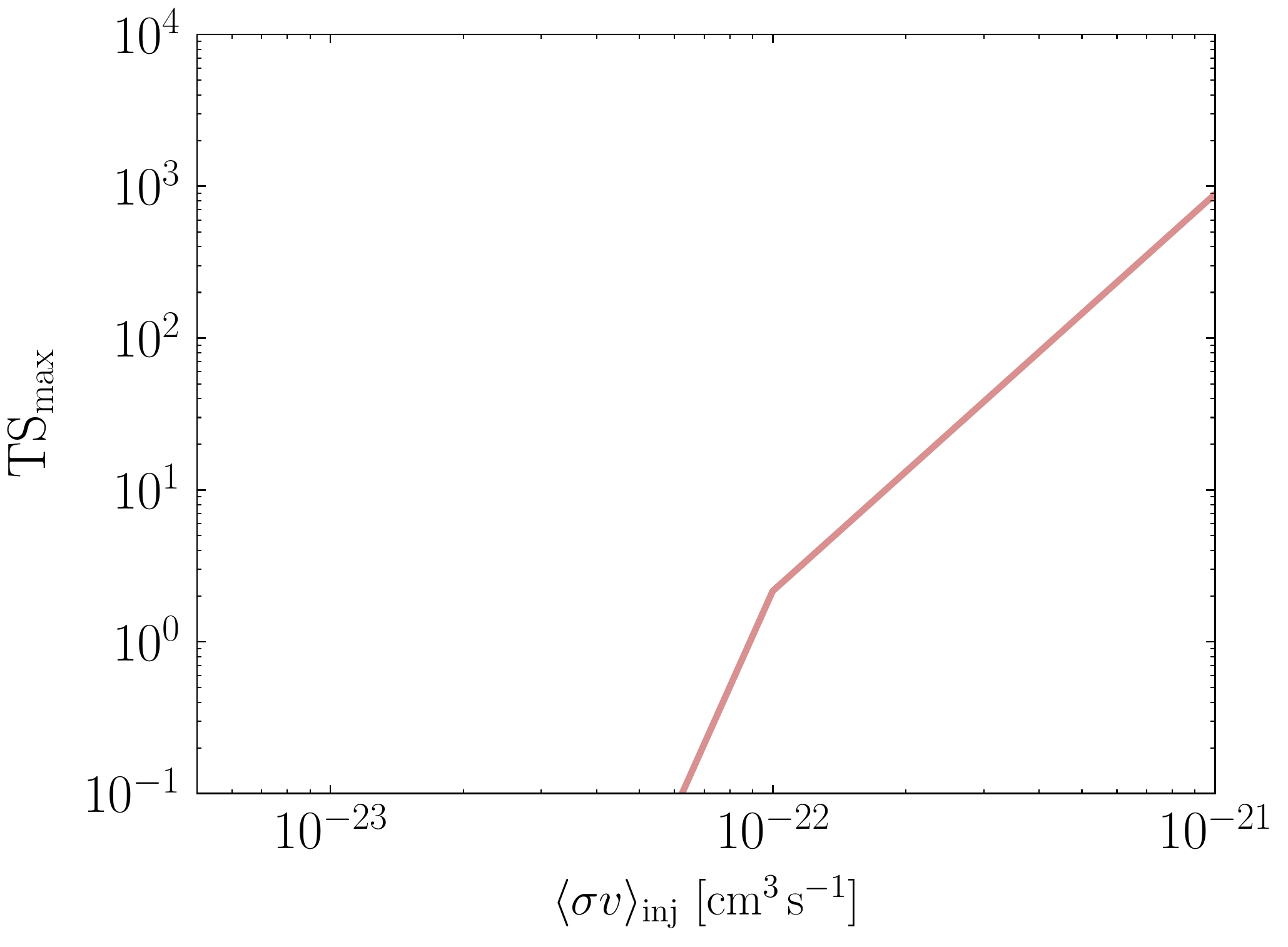}
  \caption{(Top) Recovered cross section at maxiumum test statistic, TS$_\text{max}$, (blue line) and limit (green line) obtained for various signals injected on top of the data. (Bottom) The maximum test statistic obtained at various injected cross section values. }
  \label{fig:injsig}
\end{figure*}

\noindent  {\bf Results for Individual Halos.}  Here, we explore the properties of the individual galaxy groups that are included in the stacked analysis.  These galaxy groups are taken from the catalogs in Ref.~\cite{Tully:2015opa} and~\cite{2017ApJ...843...16K}, which we refer to as T15 and T17, respectively.  Table~\ref{tab:tully_extended} lists the top thirty galaxy groups, ordered by the relative brightness of their inferred $J$-factor.  If a group in the table is not labeled with a checkmark, then it is not included in the stacking because one of the following conditions is met:
\es{eq:selection}{
\begin{cases} 
   |b| \leq 20^\circ \, , \\
   \text{overlaps another halo to within}~2^\circ~\text{of its center} \, ,\\
   \text{TS}_\text{max} > 9\, \text{ and } (\sigma v)_\text{best} > 10 \times (\sigma v)^*_\text{lim} \, .
\end{cases}
}
Note that the overlap criteria is applied sequentially in order of increasing $J$-factor.
These selection criteria have been extensively studied on mock data in our companion paper~\cite{companion} and have been verified to not exclude a potential DM signal, even on data as discussed above. Of the five halos with the largest $J$-factors that are excluded, Andromeda is removed because of its large angular extent, and the rest fail the latitude cut. 

The exclusion of Andromeda is not a result of the criteria in Eq.~\ref{eq:selection}, so some more justification is warranted. As can be seen in Table~\ref{tab:tully_extended}, the angular extent of Andromeda's  scale radius, $\theta_{s}$, is significantly larger than that of any other halo.  To justify $\theta_{s}$ as a proxy for angular extent of the emission, we calculate the 68\% (95\%) containment angle of the expected DM annihilation flux, without accounting for the PSF, and find 1.2$^{\circ}$ (4.4$^{\circ}$). This can be contrasted with the equivalent numbers for the next most important halo, Virgo, where the corresponding  68\% (95\%) containment angles are 0.5$^{\circ}$ (2.0$^{\circ}$). 
Because Andromeda is noticeably more extended beyond the \textit{Fermi} PSF, one must carefully model the spatial distribution of both the smooth DM component and the substructure.  Such a dedicated analysis of Andromeda was recently performed by the \emph{Fermi} collaboration~\cite{Ackermann:2017nya}.  Out of an abundance of caution, we remove Andromeda from the main joint analysis, but we do show how the limits change when Andromeda is included further below.

Figure~\ref{fig:individual_lims} shows the individual limits on the $b\bar{b}$ annihilation cross section for the top ten halos that pass the selection cuts and Fig.~\ref{fig:individual_maxts}  shows the maximum test statistic (TS$_\text{max}$), as a function of $m_\chi$, for these same halos. The green and yellow bands in Fig.~\ref{fig:individual_lims} and~\ref{fig:individual_maxts} represent the 68\% and 95\% containment regions obtained by randomly changing the sky location of each individual halo 200 times (subject to the selection criteria listed above). 
As is evident, the individual limits for the halos  are consistent with expectation under the null hypothesis---\emph{i.e.}, the black line falls within the green/yellow bands for each of these halos.  Some of these groups have been analyzed in previous cluster studies.  For example, the \emph{Fermi} Collaboration provided DM bounds for Virgo~\cite{Ackermann:2015fdi}; our limit is roughly consistent with theirs, and possibly a bit stronger, though an exact comparison is difficult to make due to differences in the data set and DM model assumptions.\footnote{Note that the $J$-factor in Ref.~\cite{Ackermann:2015fdi} is a factor of $4\pi$ too small.}

Figure~\ref{fig:individual_flux} provides the 95\% upper limits on the gamma-ray flux associated with the DM template for each of the top ten halos.  The upper limits are provided for 26 energy bins and compared to the expectations under the null hypothesis.  The upper limits are generally consistent with the expectations under the null hypothesis, though small systematic discrepancies do exist for a few halos, such as NGC3031, at high energies.  This could be due to subtle differences in the sky locations and angular extents between the objects of interest and the set of representative halos used to create the null hypothesis expectations. 

To demonstrate the case of a galaxy group with an excess, we show the TS$_\text{max}$ distribution and the limit for NGC6822 in Fig.~\ref{fig:maxTSoneobject}.  This object fails the selection criteria because it is too close to the Galactic plane. However, it also exhibits a TS$_\text{max}$ excess and, as expected, the limit is weaker than the expectation under the null hypothesis. \vspace{0.1in}

\noindent  {\bf Sky maps.} Fig.~\ref{fig:jfactor_maps} shows a Mollweide projection of all the $J$-factors inferred using the T15 and T17 catalogs,  smoothed at $2^\circ$ with a Gaussian kernel. The map is shown in Galactic coordinates with the Galactic Center at the origin. Looking beyond astrophysical sources, this is how an extragalactic DM signal might show up in the sky. Although this map has no masks added to it, a clear extinction is still visible along the Galactic plane. This originates from the incompleteness of the catalogs along the Galactic plane. 

In Fig.~\ref{fig:individual_skyrois}, we show the counts map in $20^\circ \times 20^\circ$ square regions around each of the top nine halos that pass the selection cuts.  For each map, we show all photons with energies above $\sim$500 MeV, indicate all {\it Fermi} 3FGL point sources with orange stars, and show the extent of $\theta_s$ with a dashed orange circle.  Given a DM signal, we would expect to see emission extend out to $\theta_s$ at the center of these images.

\begin{table*}[htb]
\footnotesize
\begin{tabular}{C{3cm}C{2.1cm}C{1.5cm}C{1.5cm}C{1.2cm}C{1.2cm}C{1.3cm}C{1.2cm}C{1.2cm}C{1.2cm}C{0.8cm}}
\toprule
\Xhline{3\arrayrulewidth}
Name &   $\log_{10} J$  &  $\log_{10} M_\text{vir}$ &          $z \times 10^{3}$&        $\ell$ &        $b$ &  $\log_{10} c_\text{vir}$  & $\theta_\text{s}$ &   $b_\text{sh}$ & TS$_\text{max}$ &  Incl. \\
& {[GeV$^2$ cm$^{-5}$ sr]} & [$M_\odot$] &  & [deg] & [deg] & & [deg] & &\\
\midrule
\hline
             Andromeda &  19.79$\pm$0.36 &  12.4$\pm$0.12 &   0.17 &  121.51 & -21.79 &  1.04$\pm$0.17 &     2.57 &  2.64 &   2.92 &             \\
         NGC4472/Virgo &  19.11$\pm$0.35 &  14.6$\pm$0.14 &   3.58 &  283.94 &  74.52 &  0.80$\pm$0.18 &     1.15 &  4.53 &   1.04 &  \checkmark \\
               NGC5128 &  18.89$\pm$0.37 &  12.9$\pm$0.12 &   0.82 &  307.88 &  17.08 &  0.99$\pm$0.17 &     0.88 &  3.14 &   0.00 &             \\
               NGC0253 &  18.76$\pm$0.37 &  12.7$\pm$0.12 &   0.79 &   98.24 & -87.89 &  1.00$\pm$0.17 &     0.77 &  2.90 &   0.63 &  \checkmark \\
              Maffei 1 &  18.68$\pm$0.37 &  12.6$\pm$0.12 &   0.78 &  136.23 &  -0.44 &  1.01$\pm$0.17 &     0.71 &  2.81 &   7.26 &             \\
              NGC6822 &  18.59$\pm$0.37 &  10.7$\pm$0.10 &   0.11 &   25.34 & -18.40 &  1.17$\pm$0.17 &     0.77 &  1.70 &  16.65 &             \\
               NGC3031 &  18.58$\pm$0.36 &  12.6$\pm$0.12 &   0.83 &  141.88 &  40.87 &  1.02$\pm$0.17 &     0.64 &  2.76 &   0.00 &  \checkmark \\
     NGC4696/Centaurus &  18.33$\pm$0.35 &  14.6$\pm$0.14 &   8.44 &  302.22 &  21.65 &  0.80$\pm$0.18 &     0.47 &  4.50 &   6.60 &  \checkmark \\
               NGC1399 &  18.30$\pm$0.37 &  13.8$\pm$0.13 &   4.11 &  236.62 & -53.88 &  0.89$\pm$0.17 &     0.45 &  3.87 &   0.72 &  \checkmark \\
                IC0356 &  18.26$\pm$0.36 &  13.5$\pm$0.13 &   3.14 &  138.06 &  12.70 &  0.92$\pm$0.17 &     0.43 &  3.51 &   0.02 &             \\
               NGC4594 &  18.26$\pm$0.35 &  13.3$\pm$0.13 &   2.56 &  299.01 &  51.30 &  0.94$\pm$0.17 &     0.43 &  3.36 &   0.00 &  \checkmark \\
               IC1613 &  18.17$\pm$0.37 &  10.6$\pm$0.10 &   0.17 &  129.74 & -60.58 &  1.18$\pm$0.17 &     0.48 &  1.67 &   1.72 &             \\
  Norma &  18.16$\pm$0.33 &  15.1$\pm$0.15 &  17.07 &  325.29 &  -7.21 &  0.74$\pm$0.18 &     0.39 &  5.17 &   0.00 &  \checkmark \\
               NGC4736 &  18.12$\pm$0.36 &  12.2$\pm$0.12 &   1.00 &  124.83 &  75.76 &  1.05$\pm$0.17 &     0.38 &  2.58 &   0.00 &             \\
      NGC1275/Perseus &  18.12$\pm$0.33 &  15.0$\pm$0.15 &  17.62 &  150.58 & -13.26 &  0.75$\pm$0.18 &     0.37 &  5.16 &   0.93 &  \checkmark \\
               NGC3627 &  18.11$\pm$0.35 &  13.0$\pm$0.13 &   2.20 &  241.46 &  64.36 &  0.98$\pm$0.17 &     0.35 &  3.23 &  27.24 &             \\
        NGC1316/Fornax &  18.01$\pm$0.36 &  13.5$\pm$0.13 &   4.17 &  239.98 & -56.68 &  0.92$\pm$0.17 &     0.32 &  3.49 &   2.33 &             \\
               NGC5236 &  18.01$\pm$0.36 &  12.2$\pm$0.12 &   1.09 &  314.58 &  31.98 &  1.05$\pm$0.17 &     0.33 &  2.56 &  22.08 &             \\
                IC0342 &  18.00$\pm$0.37 &  11.8$\pm$0.11 &   0.73 &  138.52 &  10.69 &  1.09$\pm$0.17 &     0.34 &  2.33 &   1.92 &             \\
               NGC4565 &  17.97$\pm$0.35 &  13.1$\pm$0.13 &   2.98 &  229.92 &  86.07 &  0.96$\pm$0.17 &     0.30 &  3.28 &  41.15 &             \\
 Coma &  17.96$\pm$0.33 &  15.2$\pm$0.15 &  24.45 &   57.20 &  87.89 &  0.73$\pm$0.18 &     0.31 &  5.21 &   2.35 &  \checkmark \\
        NGC1553/Dorado &  17.94$\pm$0.36 &  13.4$\pm$0.13 &   4.02 &  265.56 & -43.51 &  0.94$\pm$0.17 &     0.30 &  3.41 &   0.08 &  \checkmark \\
         NGC3311/Hydra &  17.94$\pm$0.34 &  14.4$\pm$0.14 &  10.87 &  269.55 &  26.41 &  0.82$\pm$0.17 &     0.30 &  4.32 &   0.04 &  \checkmark \\
               NGC3379 &  17.93$\pm$0.37 &  12.9$\pm$0.12 &   2.42 &  233.64 &  57.77 &  0.99$\pm$0.17 &     0.29 &  3.11 &   0.00 &  \checkmark \\
               NGC5194 &  17.93$\pm$0.37 &  12.6$\pm$0.12 &   1.84 &  104.86 &  68.53 &  1.01$\pm$0.17 &     0.30 &  2.81 &   4.94 &  \checkmark \\
            ESO097-013 &  17.92$\pm$0.37 &  11.6$\pm$0.11 &   0.60 &  311.33 &  -3.81 &  1.11$\pm$0.17 &     0.32 &  2.12 &  13.45 &             \\
               NGC4258 &  17.92$\pm$0.36 &  12.5$\pm$0.12 &   1.64 &  139.02 &  68.89 &  1.03$\pm$0.17 &     0.30 &  2.71 &   0.50 &  \checkmark \\
               NGC1068 &  17.91$\pm$0.35 &  13.3$\pm$0.13 &   3.60 &  171.99 & -51.86 &  0.95$\pm$0.17 &     0.29 &  3.33 &   6.95 &  \checkmark \\
               NGC4261 &  17.90$\pm$0.37 &  13.9$\pm$0.13 &   7.16 &  281.87 &  67.45 &  0.88$\pm$0.17 &     0.28 &  4.03 &  12.56 &             \\
               NGC4826 &  17.87$\pm$0.36 &  12.1$\pm$0.12 &   1.16 &  315.69 &  84.42 &  1.06$\pm$0.17 &     0.29 &  2.53 &   3.31 &  \checkmark \\

        \bottomrule
\Xhline{3\arrayrulewidth}
\end{tabular}
\caption{The top thirty halos included from the T15~\cite{Tully:2015opa} and T17~\cite{2017ApJ...843...16K} catalogs, as ranked by inferred $J$-factor, which includes the boost factor.  For each group, we show the brightest central galaxy and the common name, if one exists, as well as the virial mass, cosmological redshift, Galactic coordinates, inferred concentration using Ref.~\cite{Correa:2015dva}, angular extension, boost factor using the fiducial model from Ref.~\cite{Bartels:2015uba}, and the maximum test statistic (TS$_\text{max}$) over all $m_\chi$ between the model with and without DM annihilating to $b \bar b$. A checkmark indicates that the halo satisfies the selection criteria and is included in the stacking analysis.  A complete listing of all the halos used in this study is provided as Supplementary Data.
}
\label{tab:tully_extended}
\end{table*}

\afterpage{
\begin{figure*}[p]
  \centering
  \includegraphics[width=0.9\textwidth]{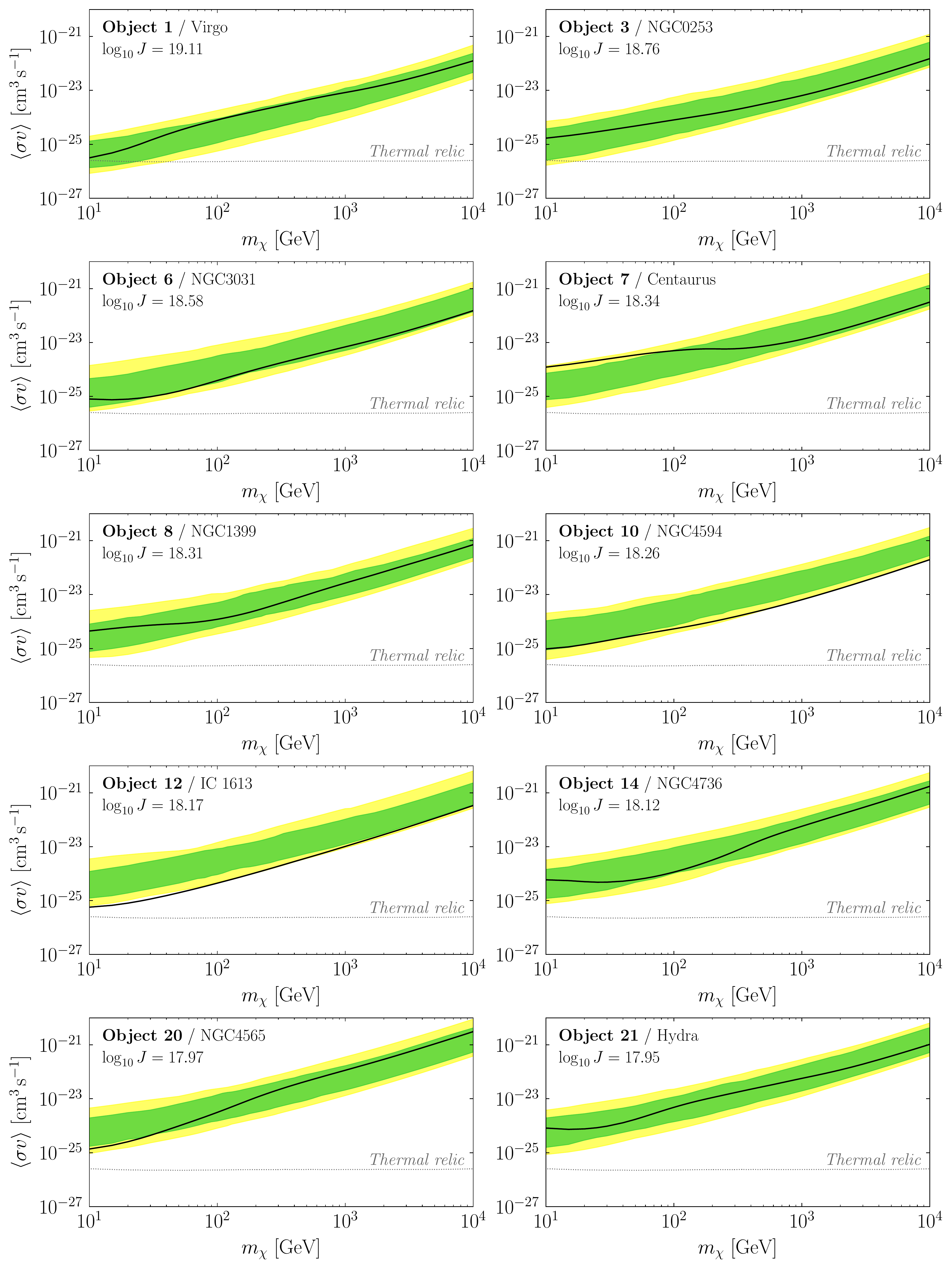}
  \caption{The 95\% confidence limit on the DM annihilation cross section to the $b \bar b$ final state for each of the top ten halos listed in Tab.~\ref{tab:tully_extended} that pass the selection cuts. For each halo, we show the 68\% and 95\% containment regions (green and yellow, respectively), which are obtained by placing the halo at 200 random sky locations.  The inferred ${J}$-factors, assuming the fiducial boost factor model~\cite{Bartels:2015uba}, are provided for each object.}
  \label{fig:individual_lims}
\end{figure*}
\clearpage}

\afterpage{
\begin{figure*}[htbp]
 \centering
  \includegraphics[width=0.9\textwidth]{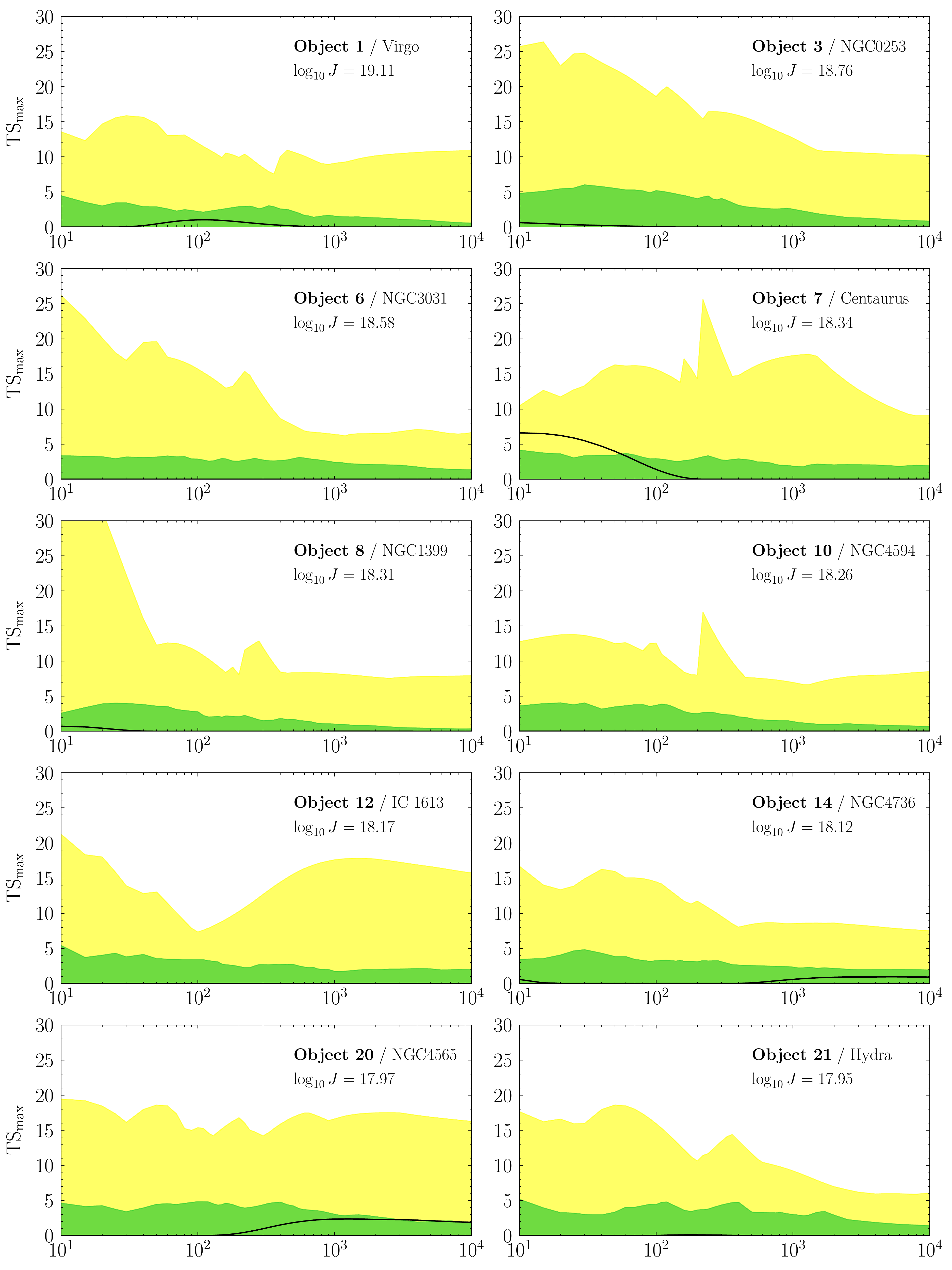}
 \caption{Same as Fig.~\ref{fig:individual_lims}, except showing the maximum test statistic (TS$_\text{max}$) for each individual halo, as a function of DM mass. These results correspond to the $b \bar b$ annihilation channel.}
  \label{fig:individual_maxts}
\end{figure*}
\clearpage}

\afterpage{
\begin{figure*}[htbp]
 \centering
  \includegraphics[width=0.9\textwidth]{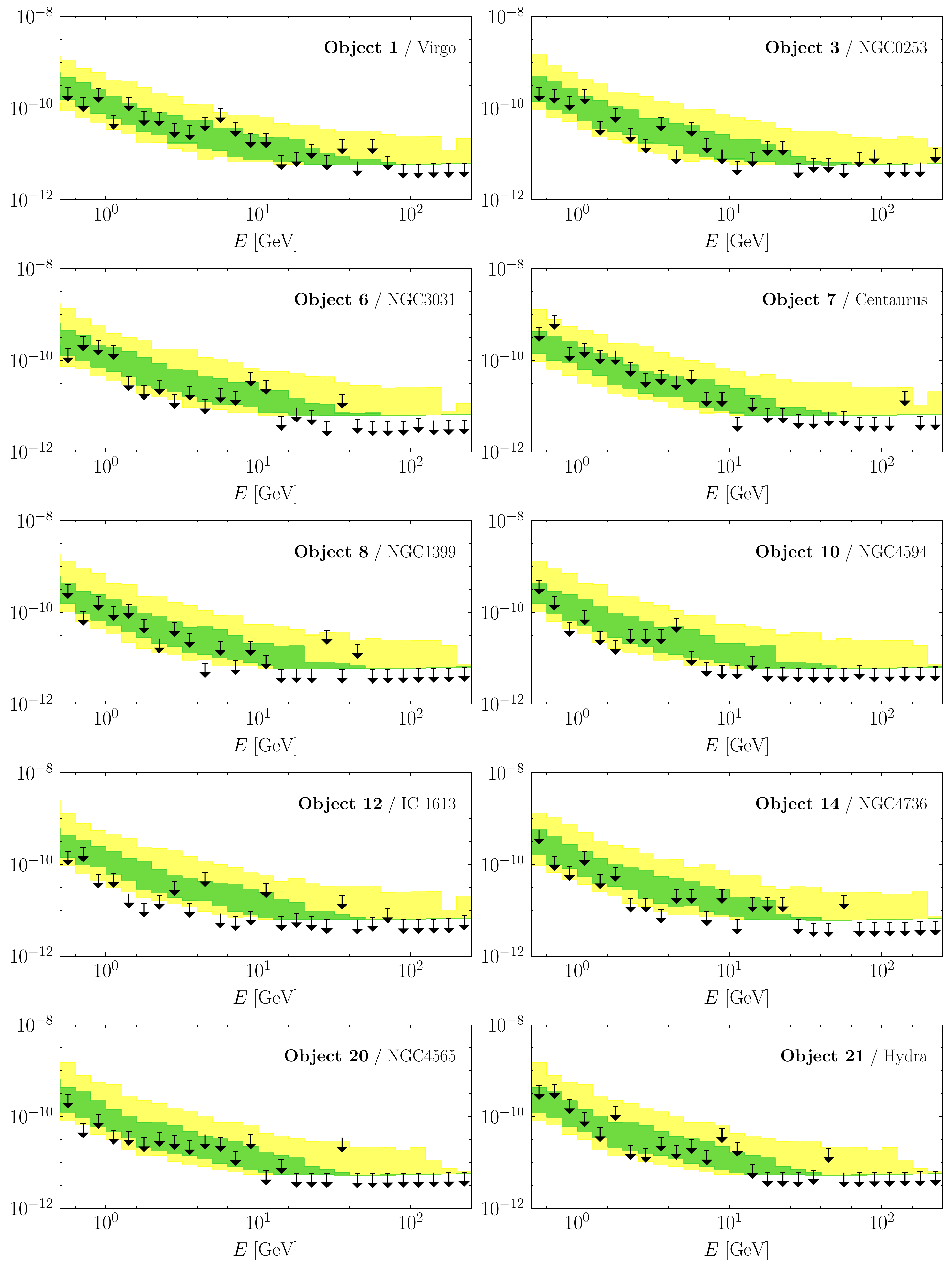}
 \caption{Same as Fig.~\ref{fig:individual_lims}, except showing the 95\% upper limit on the gamma-ray flux correlated with the DM annihilation profile in each halo.  We use 26 logarithmically spaced energy bins between 502~MeV and 251~GeV. 
 }
  \label{fig:individual_flux}
\end{figure*}
\clearpage}

\afterpage{
\begin{figure*}[htbp]
  \centering
  \includegraphics[width=0.95\textwidth]{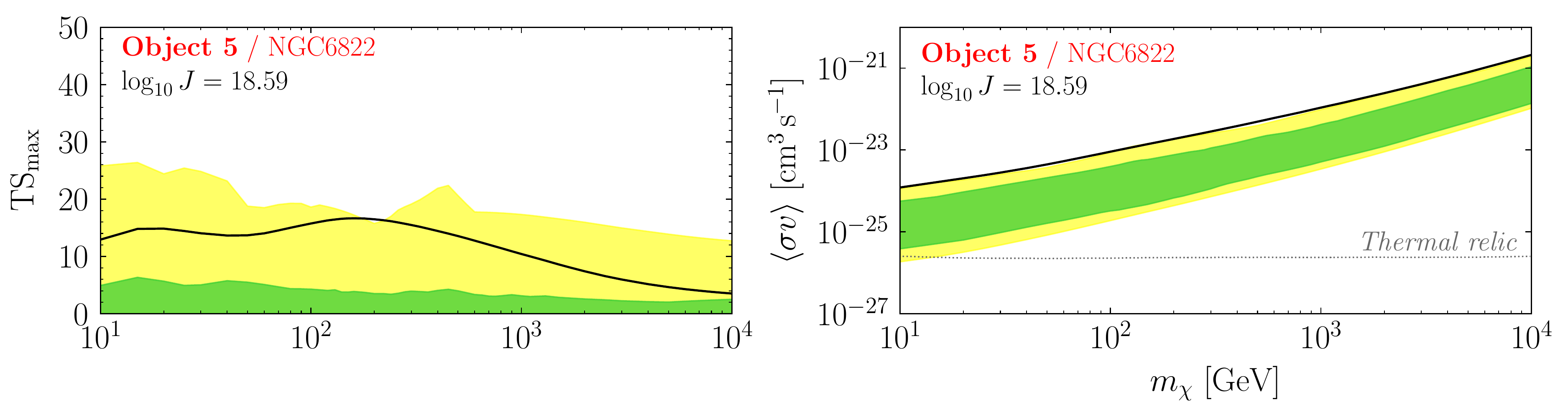}
  \caption{NGC6822 has one of the largest $J$-factors of the objects in the catalog, but it fails the selection requirements because of its proximity to the Galactic plane.  We show the analog of Fig.~\ref{fig:individual_maxts} (left) and Fig.~\ref{fig:individual_lims} (right). We see that this object  has a broad TS$_\text{max}$ excess over many masses and a weaker limit than expected from random sky locations.}
  \label{fig:maxTSoneobject}
\end{figure*}

\begin{figure*}[htbp]
 \centering
  \includegraphics[width=0.9\textwidth]{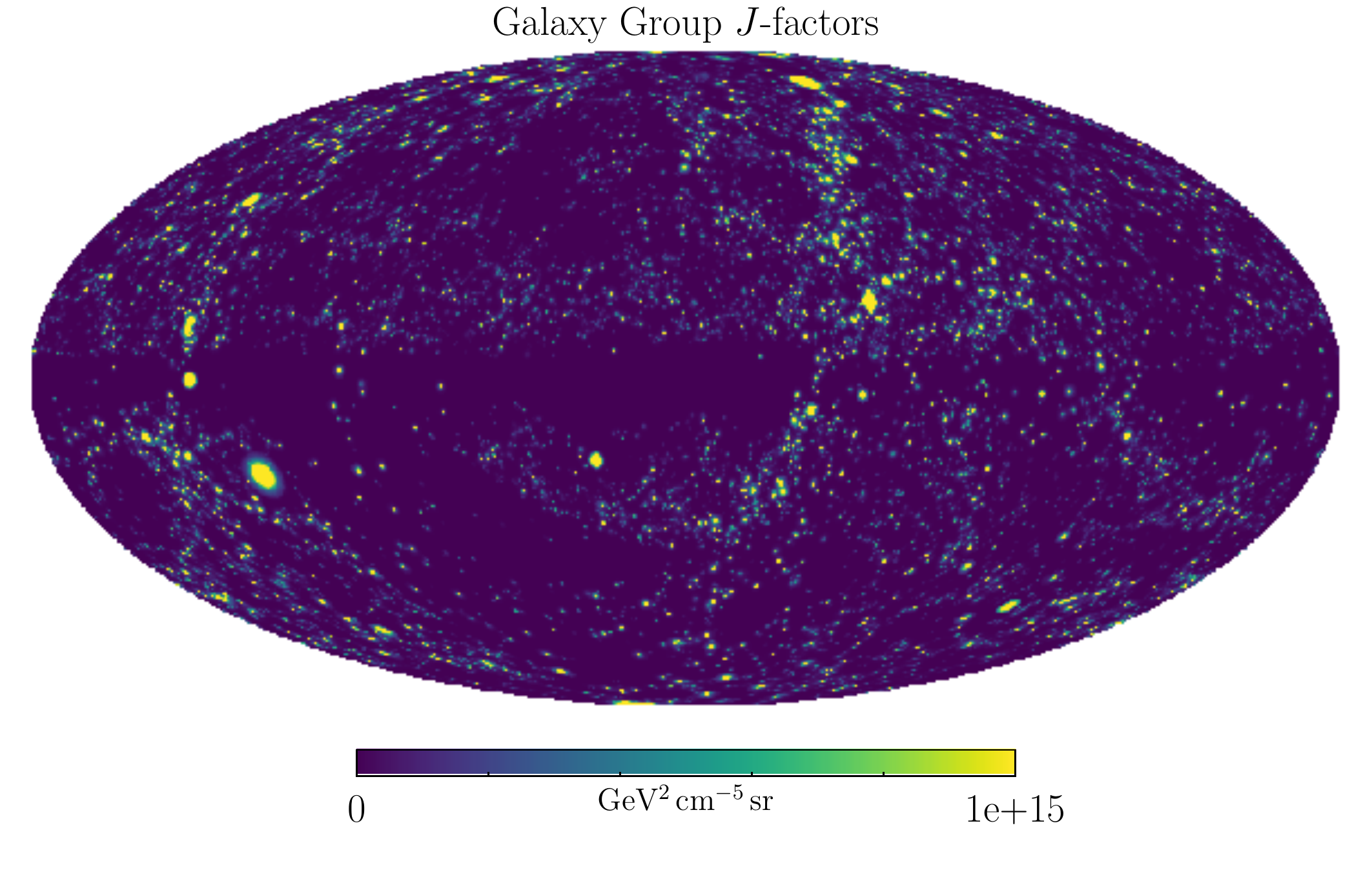}
 \caption{Mollweide projection of all the $J$-factors inferred using the T15 and T17 catalogs, smoothed at $2^\circ$ with a Gaussian kernel. If we could see beyond conventional astrophysics to an extragalactic DM signal, this is how it would appear on the sky.}
  \label{fig:jfactor_maps}
\end{figure*}
\clearpage}

\afterpage{
\begin{figure*}[htbp]
  \centering
  \includegraphics[width=0.8\textwidth]{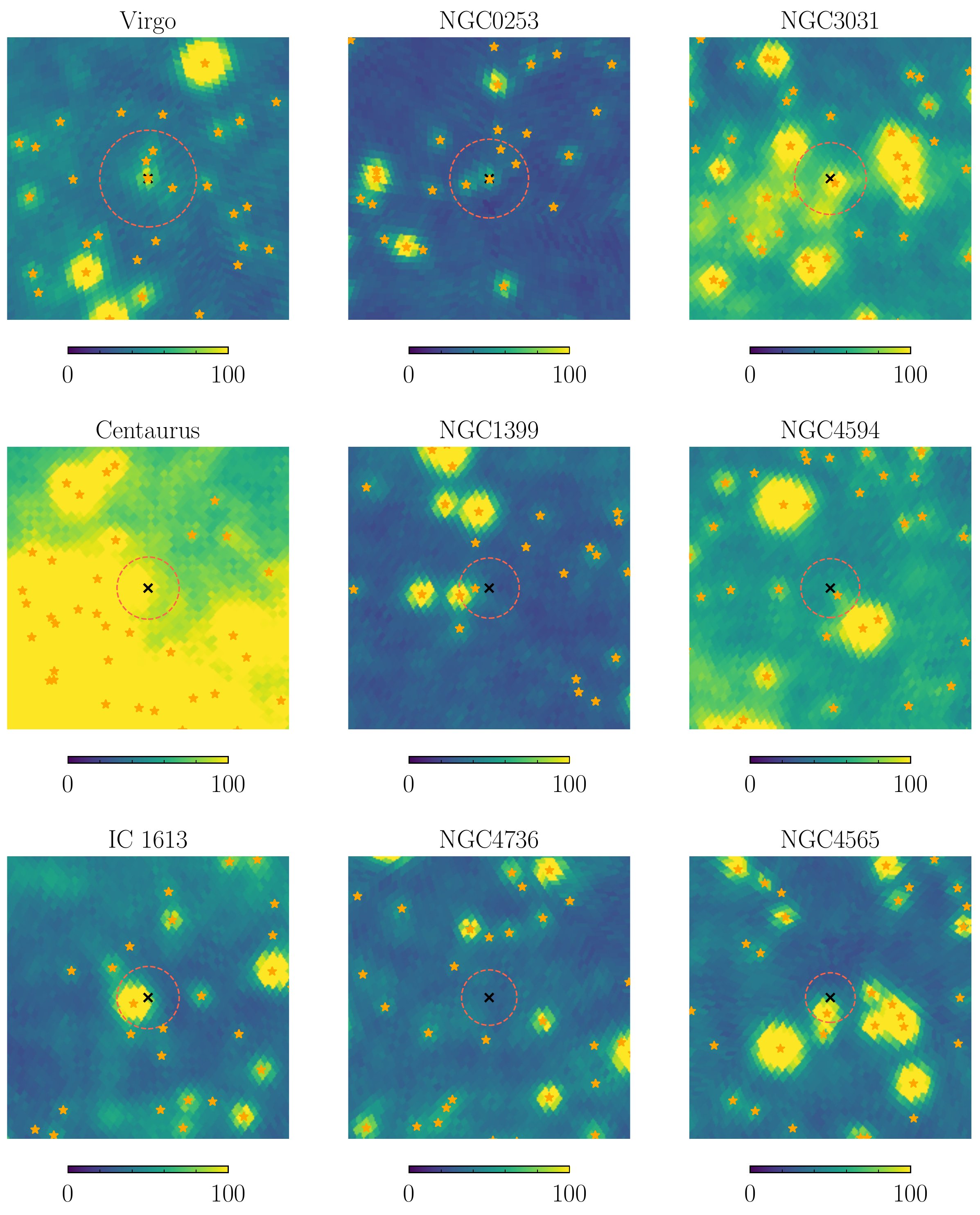}
  \caption{The \textit{Fermi}-LAT data centered on the top nine  halos that are included in the stacked sample. We show the photon counts (for the energies analyzed) within a $20^\circ$$\times$$20^\circ$ square centered on the region of interest. The dotted circle shows the scale radius $\theta_\mathrm{s}$, which is a proxy for the scale of DM annihilation, and the orange stars indicate the \emph{Fermi} 3FGL point sources.}
  \label{fig:individual_skyrois}
\end{figure*}
\clearpage}

\newpage

\section{Variations on the Analysis}
\label{sec:systematics}

We have performed a variety of systematic tests to understand the robustness of the results presented in the main body of the Letter.  Several of these uncertainties are discussed in detail in our companion paper~\cite{companion}; here, we focus specifically on how they affect the results of the data analysis.  \vspace{0.1in}

\noindent  {\bf Halo Selection Criteria.}  
Here, we demonstrate how variations on the halo selection conditions listed above affect the baseline results of Fig.~\ref{fig:bounds}.  In the left panel of Fig.~\ref{fig:cutsandhalos}, the red line shows the limit that is obtained when starting with 10,000 halos instead of 1000, but requiring the same selection conditions.  Despite the modest improvement in the limit, we choose to use 1000 halos in the baseline study because systematically testing the robustness of the analysis procedure, as done in Ref.~\cite{companion}, becomes computationally prohibitive otherwise. In order to calibrate the analysis for higher halo numbers, it would be useful to use semi-analytic methods to project the sensitivity, such as those discussed in Ref.~\cite{Cowan:2010js,Edwards:2017mnf}, although we leave the details to future work.

\begin{figure*}[b]
  \centering
	\includegraphics[width=.45\textwidth]{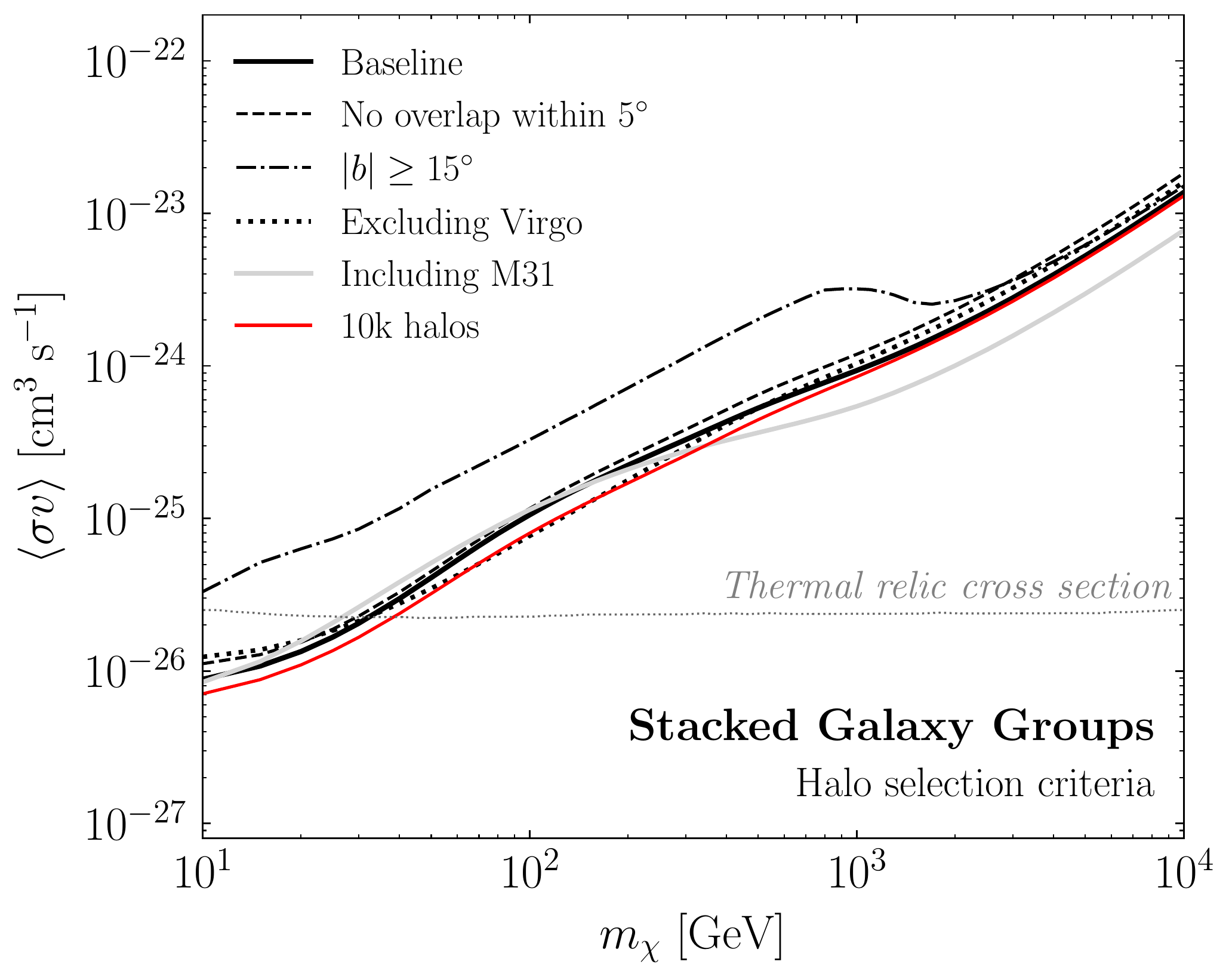} 
	\includegraphics[width=.45\textwidth]{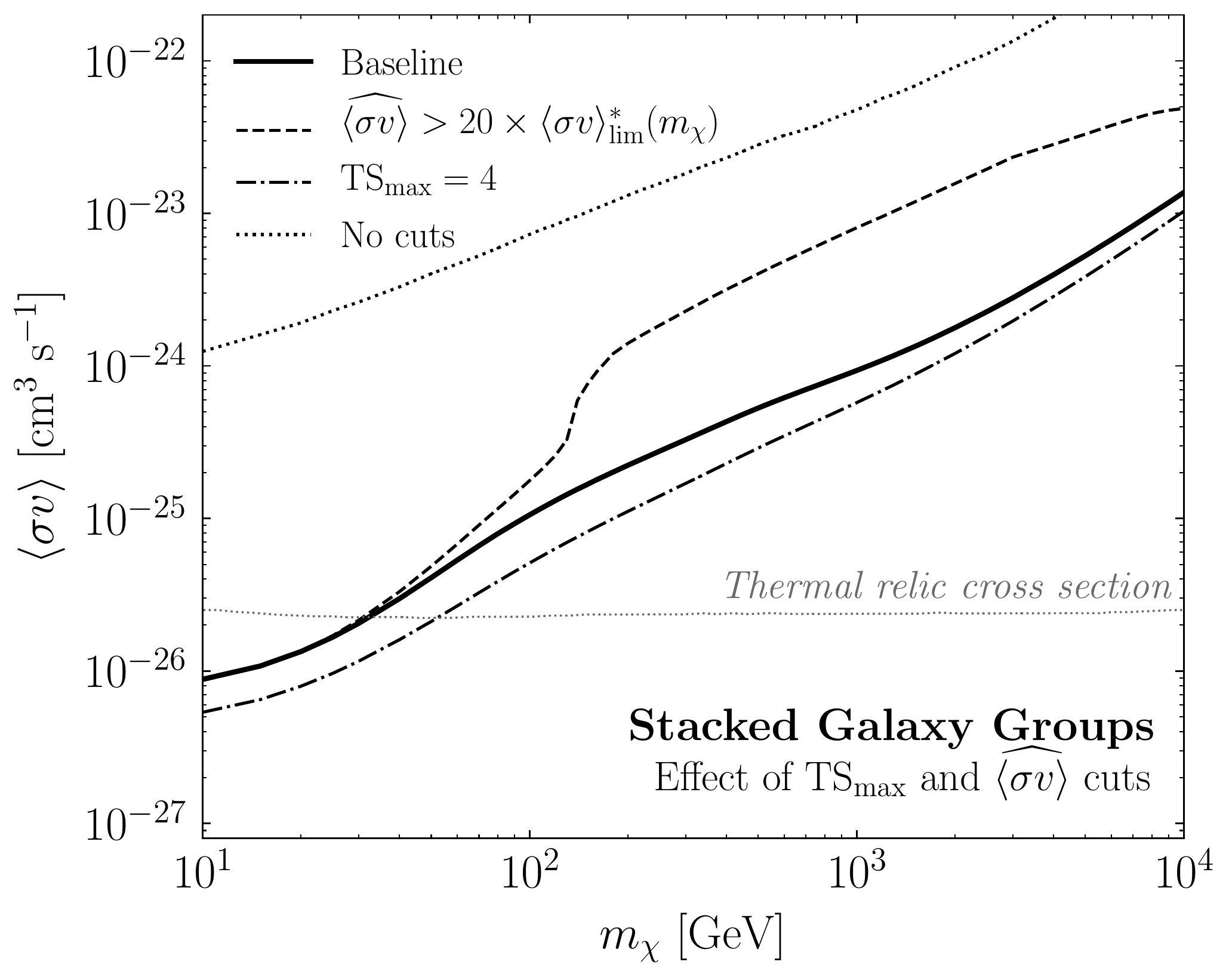} 
  \caption{The same as the baseline analysis shown in the left panel of Fig.~\ref{fig:bounds} of the main Letter, except varying several assumptions made in the analysis.  (Left) We show the effect of relaxing the overlapping halo criterion to $5^\circ$ (dashed), reducing the latitude cut to $|b|\geq 15^\circ$ (dot-dashed), excluding Virgo (dotted), and including Andromeda (gray).  The limit obtained when starting from an initial 10,000 halos is shown as the red line.  (Right) We show the effect of strengthening the cross section (dashed) or weakening the TS$_\text{max}$ (dot-dashed) selection criteria, as well as completely removing the TS$_\text{max}$ and cross section cuts (dotted). }
  \label{fig:cutsandhalos}
\end{figure*} 

Virgo is the object with the highest $J$-factor in the stacked sample. As made clear in the dedicated study of this object by the \emph{Fermi} Collaboration~\cite{Ackermann:2015fdi}, there are challenges associated with modeling the diffuse emission in Virgo's vicinity.  However, we emphasize that the baseline limit is not highly sensitive to any one halo, including the brightest in the sample.  For example, the dotted line in the left panel of Fig.~\ref{fig:cutsandhalos} shows the impact on the limit after removing Virgo from the stacking. Critically, we see that the limit is almost unchanged, highlighting that the stacked result is not solely driven by the object with the largest $J$-factor.

The effect of including Andromeda (M31) is shown as the gray solid line. We exclude Andromeda from the baseline analysis because of its large angular size, as discussed in detail above. Our analysis relies on the assumption that the DM halos are approximately point-like on the sky, which fails for Andromeda, and we therefore deem it to fall outside the scope of the systematic studies performed here.

The dashed line shows the effect of tightening the condition on overlapping halos from $2^\circ$ to $5^\circ$. Predictably, the limit is slightly weakened due to the smaller pool of available targets.  We also show the effect of decreasing the latitude cut to $b\geq 15^\circ$ (dot-dashed line). In this case, the number of halos included in the stacked analysis increases, but the limit is weaker---considerably so below $m_\chi \sim 10^3$~GeV.  The weakened limits are likely due to enhanced diffuse emission along the plane as well as contributions from unresolved point sources, both of which are difficult to accurately model. In cases with such mismodeling, the addition of a DM template can generically improve the quality of the fit, which leads to excesses at low energies, in particular.  The baseline latitude cut ameliorates  precisely these concerns.

The right  panel of Fig.~\ref{fig:cutsandhalos} illustrates the effects of changing, or removing completely, the cross section and TS$_\text{max}$ cuts on the halos.  Specifically, the dashed black line shows what happens when we require that a halo's excess be even more inconsistent with the limits set by other galaxy groups; specifically, requiring that $(\sigma v)_\text{best} > 20 \times (\sigma v)^*_\text{lim}$. The dot-dashed line shows the limit when we decrease the statistical significance requirement to $\text{TS}_\text{max} > 4$.  
Note that the two changes have opposite effects on the limits.  This is expected because more halos with excesses are included in the stacking procedure with the more stringent cross section requirement, which weakens the limit, whereas fewer are included if we reduce the TS$_\text{max}$ cut, strengthening the limit.  

The dotted line in the right panel of Fig.~\ref{fig:cutsandhalos} shows  what happens when no requirement at all is placed on the TS$_\text{max}$ and cross section; in this case, the limit is dramatically weakened by several orders of magnitude.   We show the same result in Fig.~\ref{fig:systematics_nots_cuts} (dotted line), but with a comparison to the null hypothesis corresponding to no TS$_\text{max}$ and cross section cuts, which is shown as the 68\%~(95\%) red~(blue) bands.\footnote{We thank A.~Drlica-Wagner for suggesting this test.}    In the baseline case, the limit is consistent with the random sky locations---\emph{i.e.}, the solid black line falls within the green/yellow bands.  However, with no TS$_\text{max}$ and cross section cuts, this is no longer true---\emph{i.e.}, the dotted black line falls outside the red/blue bands.  Clear excesses are observed above the background expectation in this case, but they are inconsistent with a DM interpretation as they are strongly excluded by other halos in the stack.  When deciding on the TS$_\text{max}$ and cross section requirements that we used for the baseline analysis in Fig.~\ref{fig:bounds}, our goal was to maximize the sensitivity reach while simultaneously ensuring that an actual DM signal would not be excluded.  We verified the selection criteria thoroughly by performing injected signal tests on the data (discussed above) as well as on mock data (discussed in Ref.~\cite{companion}).  Ideally, galaxy groups would be excluded from the stacking based on the specific properties of the astrophysical excesses that they exhibit, as opposed to the TS$_\text{max}$ and cross section requirements used here.  For example, one can imagine excluding groups that are known to host  AGN or galaxies with high amounts of star-formation activity.  We plan to study such possibilities in future work.        \vspace{0.1in}

\begin{figure*}[tb]
  \centering
  \includegraphics[width=.5\textwidth]{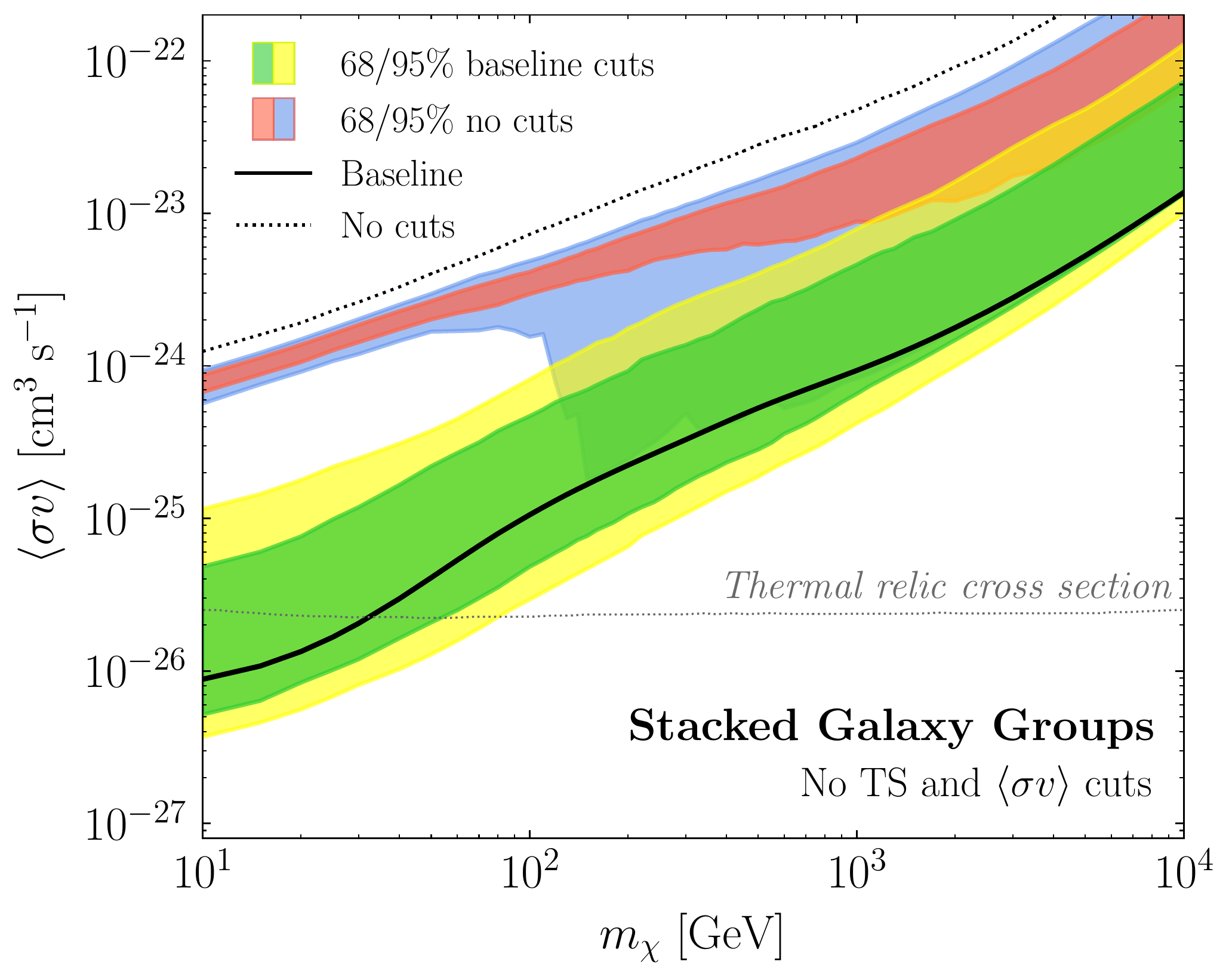}
  \caption{The results of the baseline analysis with the default cuts, as shown in the left of Fig.~\ref{fig:bounds}, compared to the corresponding result when no cuts are placed on the TS$_\text{max}$ or cross section of the halos in the catalog.  The significant offset between the limit obtained with no cuts (dotted line) and the corresponding expectation from random sky locations (red/blue band)  demonstrates that many of the objects that are removed by the TS$_\text{max}$ and cross section cuts are  legitimately associated with astrophysical emission. See text for details.}
  \label{fig:systematics_nots_cuts}
\end{figure*}

\begin{figure*}[t]
  \centering
  \includegraphics[width=.45\textwidth]{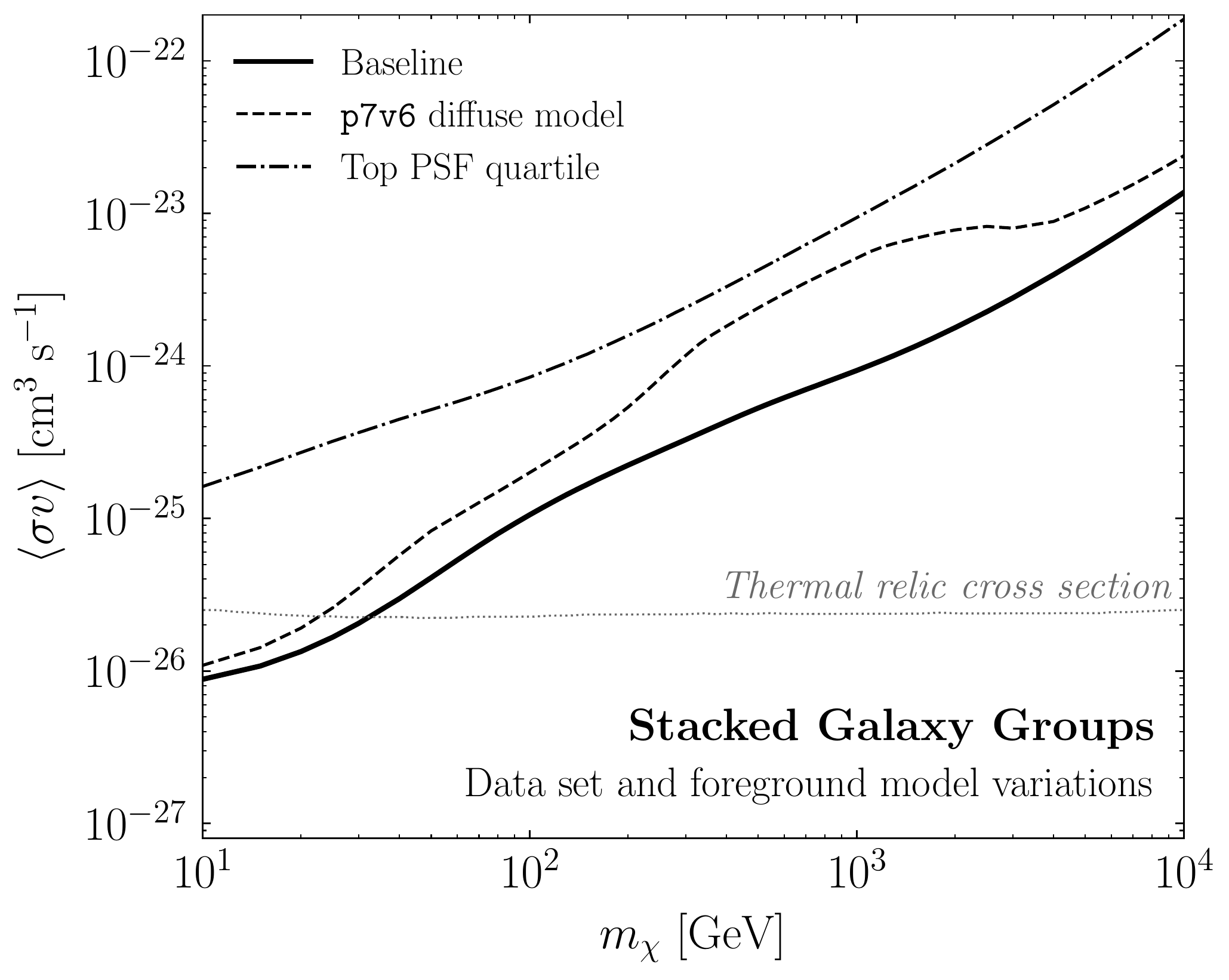}
  \includegraphics[width=.45\textwidth]{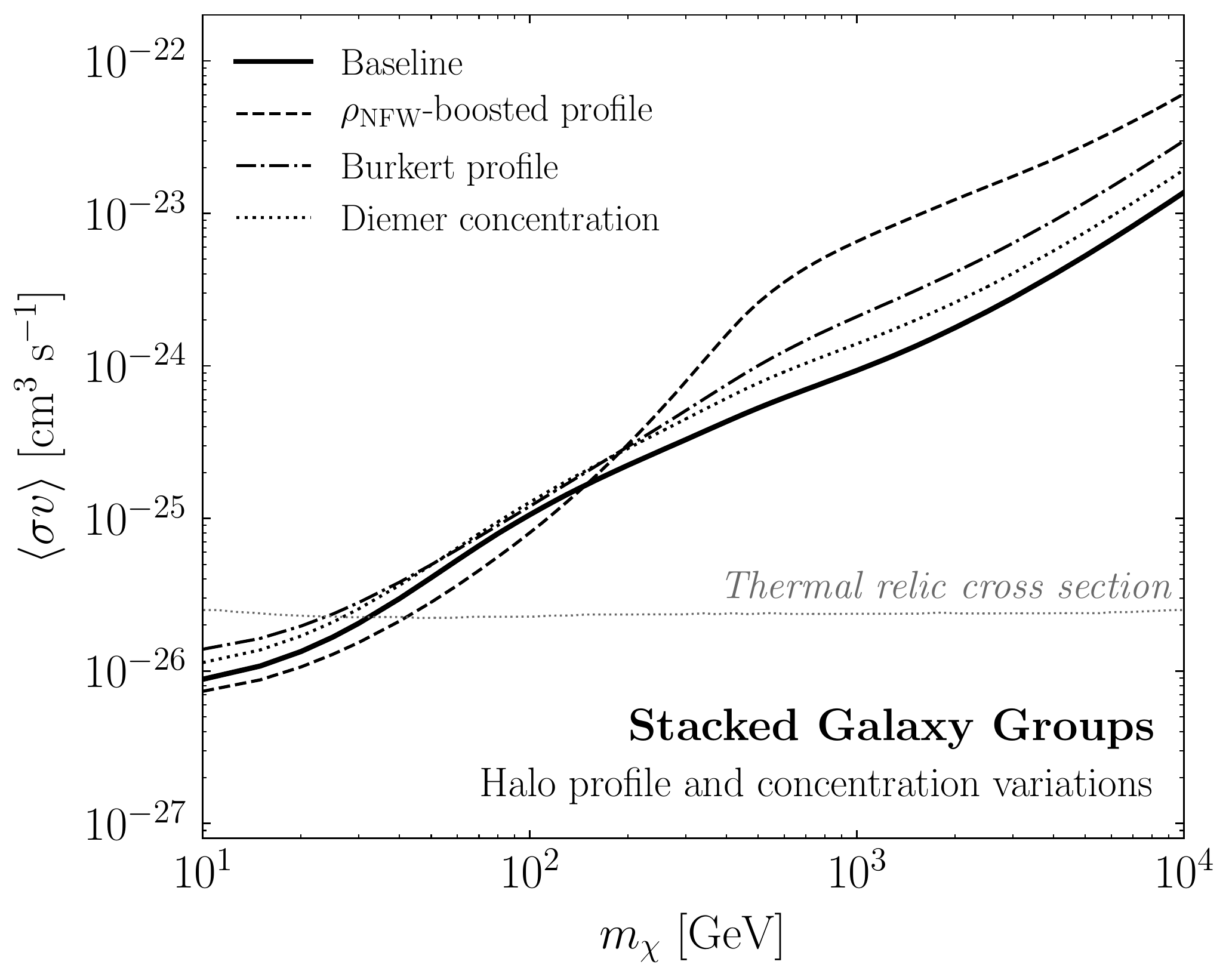}
  \caption{The same as the baseline analysis shown in the left panel of Fig.~\ref{fig:bounds} of the main Letter, except varying several assumptions made in the analysis.  (Left) We show the effect of using the top PSF quartile of the UltracleanVeto data set (dot-dashed) and the \texttt{p7v6} diffuse model (dashed).  (Right) We show the effect of using the cored Burkert profile~\cite{Burkert:1995yz} (dot-dashed) and the Diemer~and~Kravtsov concentration model~\cite{Diemer:2014gba} (dotted).  The ``$\rho_\text{NFW}$-boosted profile'' (dashed) shows what happens when the annihilation flux from the subhalo boost is assumed to follow the NFW profile (as opposed to a squared-NFW profile). }
  \label{fig:systematics_data_profile}
\end{figure*}

\noindent  {\bf Data Set and Foreground Models.}  
In the results presented thus far, we have used all quartiles of the UltracleanVeto event class of the {\it Fermi} data.  Alternatively, we can restrict ourselves to the top quartile of events, as ranked by PSF.  Using this subset of data has the advantage of improved angular resolution, but the disadvantage of a $\sim$75\% reduction in statistics.  The  left panel of Fig.~\ref{fig:systematics_data_profile} shows the limit (dot-dashed line) obtained by repeating the analysis with the top quartile of UltracleanVeto data; the bounds are weaker than in the all-quartile case, as would be expected.  However, the amount by which the limit weakens is not completely consistent with the decrease in statistics.  Rather, it appears that when we lower the photon statistics, more halos that were previously excluded by the cross section and TS$_\text{max}$ criteria in the baseline analysis are allowed into the stacking and collectively weaken the limit.

Another choice that we made for the baseline analysis was to use the \texttt{p8r2} foreground model for gamma-ray emission from cosmic-ray processes in the Milky Way.   In this model, the bremsstrahlung and boosted pion emission are traced with gas column-density maps and the IC emission is modeled using \texttt{Galprop}~\cite{Strong:2007nh}.  After fitting the data with these three components, any `extended emission excesses' are identified and added back into the foreground model~\cite{Acero:2016qlg}.  To study the dependence of the results on the choice of foreground model, we repeat the analysis using the Pass~7 \emph{gal\_2yearp7v6\_v0.fits} (\texttt{p7v6}) model, which includes large-scale structures like Loop~1 and the \emph{Fermi} bubbles---in addition to the bremsstrahlung, pion, and IC emission---but does not account for any data-driven excesses as is done in \texttt{p8r2}.  The results of the stacked analysis using the \texttt{p7v6} model are shown in the left panel of Fig.~\ref{fig:systematics_data_profile} (dashed line).  The limit is somewhat weaker to that obtained using \texttt{p8r2}, though it is broadly similar to the latter.  This is to be expected for stacked analyses, where the dependence on mismodeling of the foreground emission is reduced because the fits are done on small, independent regions of the sky, so that offsets in the point-to-point normalizations of the diffuse model can have less impact. For more discussion of this point, see Ref.~\cite{Daylan:2014rsa,Linden:2016rcf,Narayanan:2016nzy,Cohen:2016uyg}.\vspace{0.1in}

\noindent  {\bf Halo Density Profile and Concentration.} Our baseline analysis makes two assumptions about the profiles of gamma-ray emission from the extragalactic halos.  The first assumption is that the DM profile of the smooth halo is described by an NFW profile:
\begin{equation}
\rho_{\rm NFW}(r) = \frac{\rho_s}{r/r_s\,(1+r/r_s)^2}\,,
\end{equation}
where $\rho_s$ is the normalization and $r_s$ the scale radius~\cite{Navarro:1996gj}.  The NFW profile successfully describes the shape of cluster-size DM halos in $N$-body simulations with and without baryons (see, {\it e.g.}, Ref.~\cite{Springel:2008cc,Schaller:2014uwa}).  However, some evidence exists pointing to cored density profiles on smaller scales ({\it e.g.}, dwarf galaxies), and the density profiles in these systems may be better described by the phenomenological Burkert profile~\cite{Burkert:1995yz}:
\begin{equation}
\rho_{\rm Burkert}(r) = \frac{\rho_B}{(1+r/r_B)(1+(r/r_B)^2)}\,,
\end{equation}
where $\rho_B$ and $r_B$ are the Burkert corollaries to the NFW $\rho_s$ and $r_s$, but have numerically different values. While it appears unlikely that the Burkert profile is a good description of the DM profiles of the cluster-scale halos considered here, using this profile provides a useful systematic variation because it predicts less annihilation flux than the NFW profile does.  The right panel of Fig.~\ref{fig:systematics_data_profile} shows the effect of using the Burkert profile to describe the halos in the T15 and T17 catalogs (dot-dashed line); the limit is slightly weaker, as expected.

The second assumption we made is that the shape of the gamma-ray emission from DM annihilation follows the projected integral of the DM-distribution squared.  This is likely incorrect because the contribution from the boost factor, which can be substantial, should have the spatial morphology of the distribution of DM subhalos.  Neglecting tidal effects, we expect the subhalos to follow the DM distribution (instead of the squared distribution).  Including tidal effects is complicated, as subhalos closer to the halo center are more likely to be tidally stripped, which both increases their concentration and decreases their number density.  We do not attempt to model the change in the spatial morphology of the subhalo distribution from tidal stripping and instead consider the limit where the annihilation flux from the subhalo boost follows the NFW distribution.  This gives a much wider angular profile for the annihilation flux for large clusters,  compared to the case where the boost is simply a multiplicative factor.  The dashed line in the right panel of Fig.~\ref{fig:systematics_data_profile} shows the effect on the limit of modeling the gamma-ray emission in this way (labeled ``$\rho_\text{NFW}$-boosted profile").  The extended spatial profile leads to a minimal change in the limit over most of the mass range, which is to be expected given that most of the galaxy groups can be well-approximated as point sources.

A halo's virial concentration is an indicator of its overall density and is defined as $c_\text{vir} \equiv r_\text{vir}/r_s$, where $r_\text{vir}$ is the virial radius and $r_s$ the NFW scale radius of the halo.  A variety of models exist in the literature that map from halo mass to concentration.  Our fiducial case is the Correa \emph{et al.} model from Ref.~\cite{Correa:2015dva}.  Here we show how the limit (dotted line) changes when we use the model of Diemer and Kravtsov~\cite{Diemer:2014gba}, updated with the Planck 2015 cosmology~\cite{Ade:2015xua}.  The change to the limit is minimal, which is perhaps a reflection of the fact that the change in the mean concentrations between the concentration-mass models is small compared to the statistical spread predicted in these models, which is incorporated into the $J$-factor uncertainties.  We have also verified that increasing the dispersion on the concentration for the Correa~\emph{et al.} model to 0.24~\cite{Bullock:1999he}, which is above the 0.14--0.19 range used in the baseline study, worsens the limit by a $\mathcal{O}(1)$ factor.\vspace{0.1in}

\begin{figure}[t]
  \centering
  \includegraphics[width=0.46\textwidth]{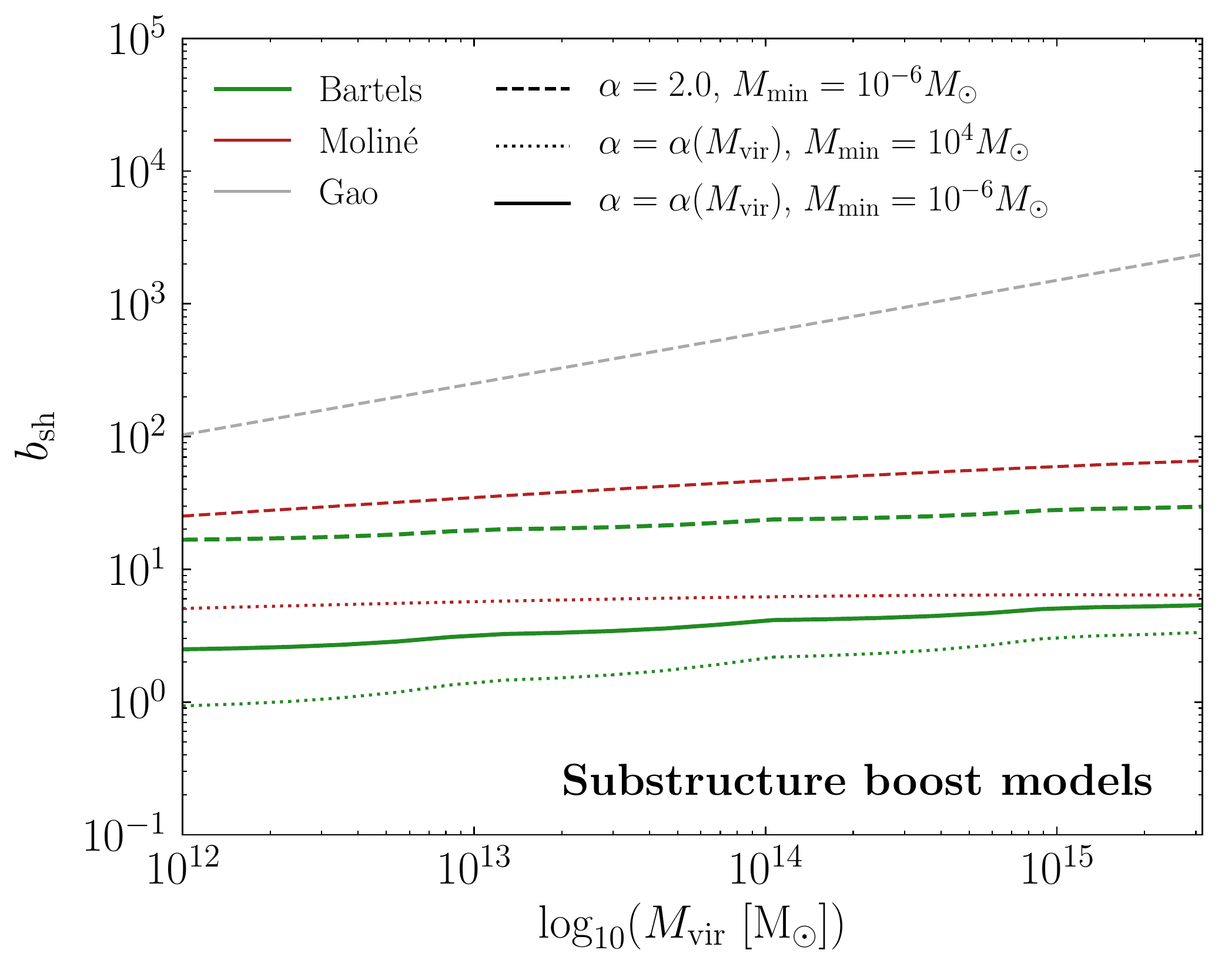} 
     \includegraphics[width=.45\textwidth]{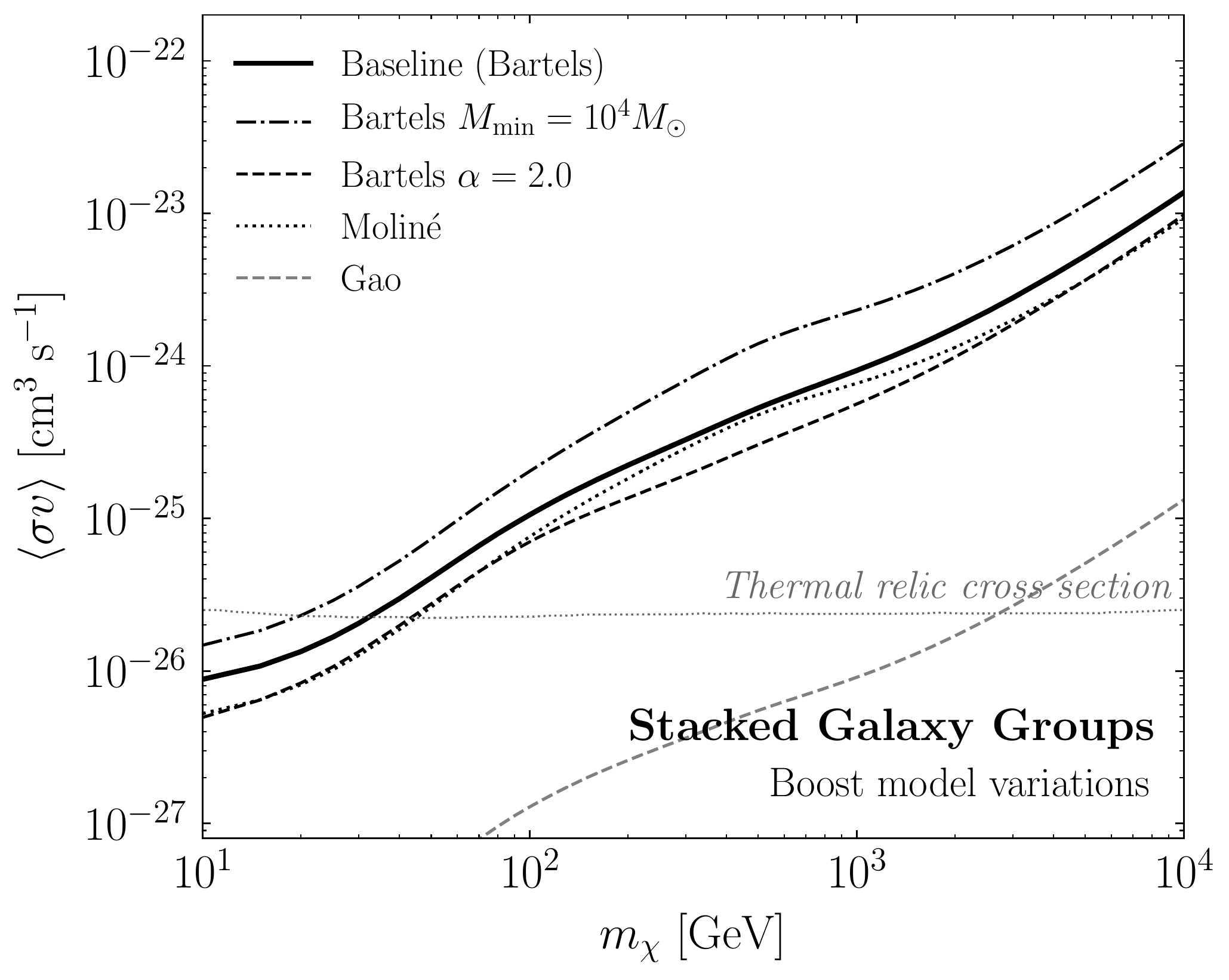}
  \caption{(Left) Examples of substructure boost models commonly used in the literature, reproduced from \cite{companion}. Our fiducial model, based on Ref.~\cite{Bartels:2015uba} using $M_\text{min} = 10^{-6}$~M$_\odot$ and self-consistently computing $\alpha$, is shown as the thick green solid line. Variations on $M_\text{min}$ and $\alpha$ are shown with the dotted and dashed lines, respectively. Also plotted are the boost models of Molin\'e~\cite{Moline:2016pbm} (red) and Gao~\cite{Gao:2011rf} (grey).  (Right) The same as the baseline analysis shown in the left panel of Fig.~\ref{fig:bounds} of the main Letter, except varying the boost model.
  }
  \label{fig:systematics_boost}
\end{figure}

\noindent  {\bf Substructure Boost.}  
Hierarchical structure formation implies that larger structures can host smaller substructures, the presence of which can significantly enhance signatures of DM annihilation in host halos. Although several models exist in the literature to characterize this effect, the precise enhancement sensitively depends on the methods used as well as the astrophysical and particle physics properties that are assumed.  Phenomenological extrapolation of subhalo properties (\emph{e.g.}, the concentration-mass relation) over many orders of magnitude down to very small masses $\mathcal O(10^{-6}$)~M$_{\odot}$ lead to large enhancements of $\mathcal O(10^{2})$ and $\mathcal O(10^{3})$ for galaxy- and cluster-sized halos, respectively~\cite{Gao:2011rf}. Recent numerical simulations and analytic studies~\cite{Anderhalden:2013wd,Correa:2015dva,Ludlow:2013vxa} suggest that the concentration-mass relation flattens at smaller masses, yielding boosts that are much more modest, about an order-of-magnitude below phenomenological extrapolations~\cite{Nezri:2012tu,Sanchez-Conde:2013yxa}.  In addition, the concentration-mass relation for field halos cannot simply be applied to subhalos, because the latter undergo tidal stripping as they fall into and orbit their host.  Such effects tend to make the subhalos more concentrated---and therefore more luminous---than their field-halo counterparts, though the number-density of such subhalos is also reduced~\cite{Bartels:2015uba}.    

When taken together, the details of the halo formation process shape the subhalo mass function $dn/dM_\text{sh}\propto M_\text{sh}^{-\alpha}$, where $\alpha  \in \left[1.9, 2.0\right]$.  The mass function does not follow a power-law to arbitrarily low masses, however, because the underlying particle physics model for the DM can place a minimum cutoff on the subhalo mass, $M_\text{min}$.  For example, DM models with longer free-streaming lengths wash out smaller-scale structures, resulting in higher cutoffs.

The left panel of Fig.~\ref{fig:systematics_boost} shows a variety of boost models commonly used in DM studies. The fiducial boost model used here~\cite{Bartels:2015uba} is shown as the thick green solid line and variations on $M_\text{min}$ and $\alpha$ are also plotted. The right panel of Fig.~\ref{fig:systematics_boost} shows that the expected limit when $M_\text{min} = 10^4$~M$_\odot$ instead of $M_\text{min} = 10^{-6}$~M$_\odot$ (dot-dashed) is weaker across all masses.  While a minimum subhalo mass of  $10^{4}$~M$_\odot$ is likely inconsistent with bounds on the kinetic decoupling temperature of thermal DM, this example illustrates the importance played by $M_\text{min}$ in the sensitivity reach.  Additionally, Fig.~\ref{fig:systematics_boost} demonstrates the case where $\alpha=2.0$ (dashed line).  Increasing the inner slope of the subhalo mass function leads to a correspondingly stronger limit, however observations tend to favor a slope closer to $\alpha = 1.9$ (which is what the most massive halos correspond to in our fiducial case).

Ref.~\cite{Sanchez-Conde:2013yxa} derived a boost factor model that accounts for the flattening of the concentration-mass relation at low masses, but does not include the effect of tidal stripping.  They assume a minimum sub-halo mass of $10^{-6}$~M$_\odot$ and a halo-mass function $dN/dM \sim M^{-2}$.  This was updated by Ref.~\cite{Moline:2016pbm} to account for the effect of tidal disruption. This updated boost factor model, which takes $\alpha = 1.9$, gives the constraint shown in Fig.~\ref{fig:systematics_boost} labeled ``Molin\'e" (dotted).  This model is to be contrasted with the boost factor model of Ref.~\cite{Gao:2011rf}, labeled ``Gao" in Fig.~\ref{fig:systematics_boost} (grey-dashed), which uses a phenomenological power-law extrapolation of the concentration-mass relation to low sub-halo masses.  Because the annihilation rate increases with increasing concentration parameter, the model in Ref.~\cite{Gao:2011rf} predicts substantially larger boosts than other scenarios that take into account a more realistic flattening of the concentration-mass relation at low subhalo masses.\vspace{0.1in}

\noindent  {\bf Galaxy Group Catalog.}  
We now explore the dependence of the results on the group catalog that is used to select the halos.  In this way, we can better understand how the DM bounds are affected by uncertainties on galaxy clustering algorithms and the inference of the virial mass of the halos.  The baseline limits are based on the T15 and T17 catalogs, but here we repeat the analysis using the Lu~\emph{et al.} catalog~\cite{Lu:2016vmu}, which solely relies on 2MRS observations.  The group-finding algorithm used by Ref.~\cite{Lu:2016vmu} is different to that of T15 and T17 in many ways, relying on a friends-of-friends algorithm as opposed to one based on matching group properties at different scales to $N$-body simulations. Lu~\emph{et al.} also use a different halo mass determination.  For these reasons, it provides a good counterpoint to T15 and T17 for estimating systematic uncertainties associated with the identification of galaxy groups. While T17 includes measured distances for nearby groups, the Lu catalog corrects for the effect of peculiar velocities following the prescription in Ref.~\cite{1996AJ....111..794K} and the effect of Virgo infall as in Ref.~\cite{2014ApJ...782....4K}. Figure~\ref{fig:lucatalog} is a repeat of Fig.~\ref{fig:bounds} in the main Letter, except using the Lu~\emph{et al.} catalog.  Despite important differences between the group catalogs used, the Lu~\emph{et al.} results are very similar to the baseline case. \\

\begin{figure*}[t]
  \centering
  \includegraphics[width=.45\textwidth]{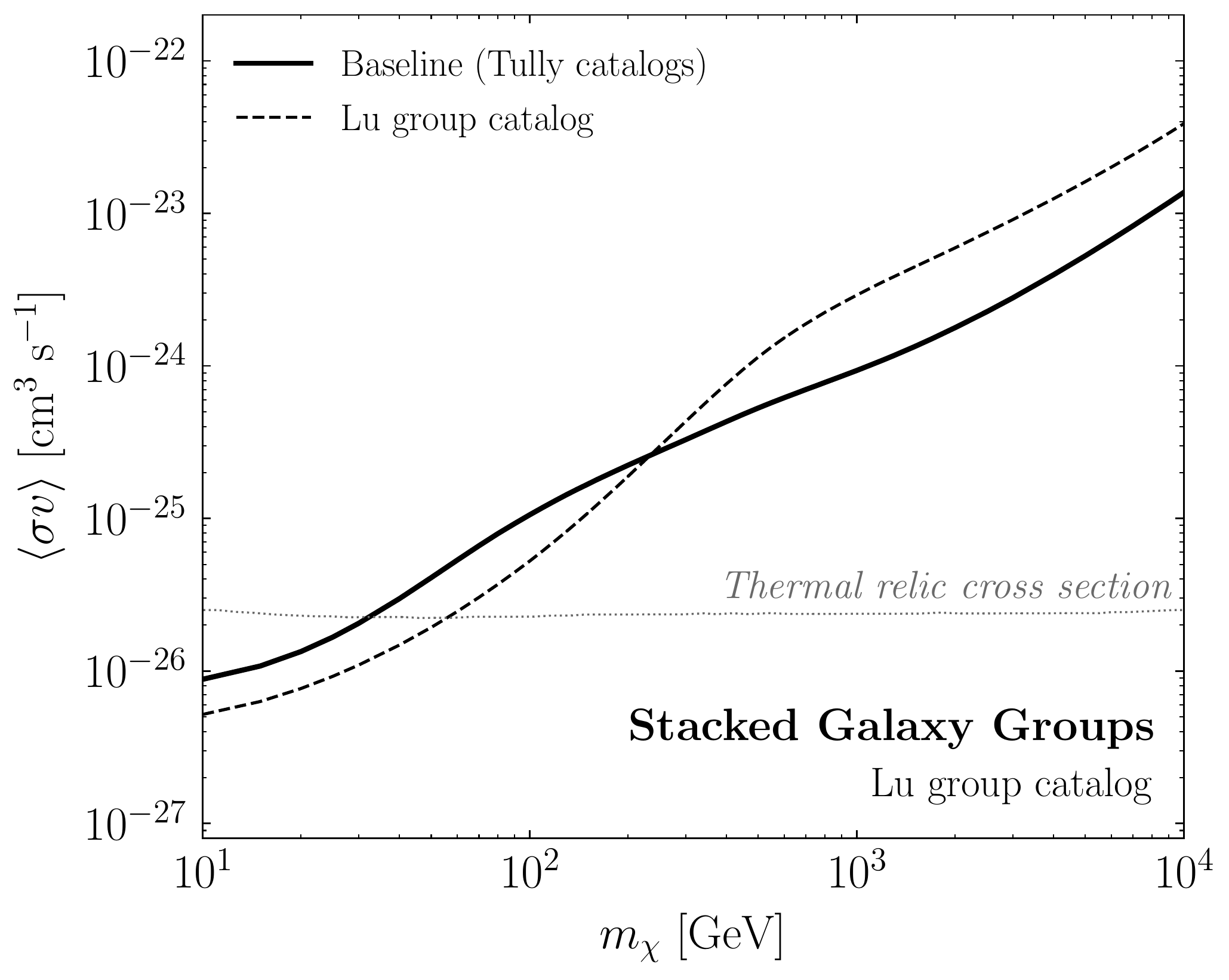}
   \includegraphics[width=.45\textwidth]{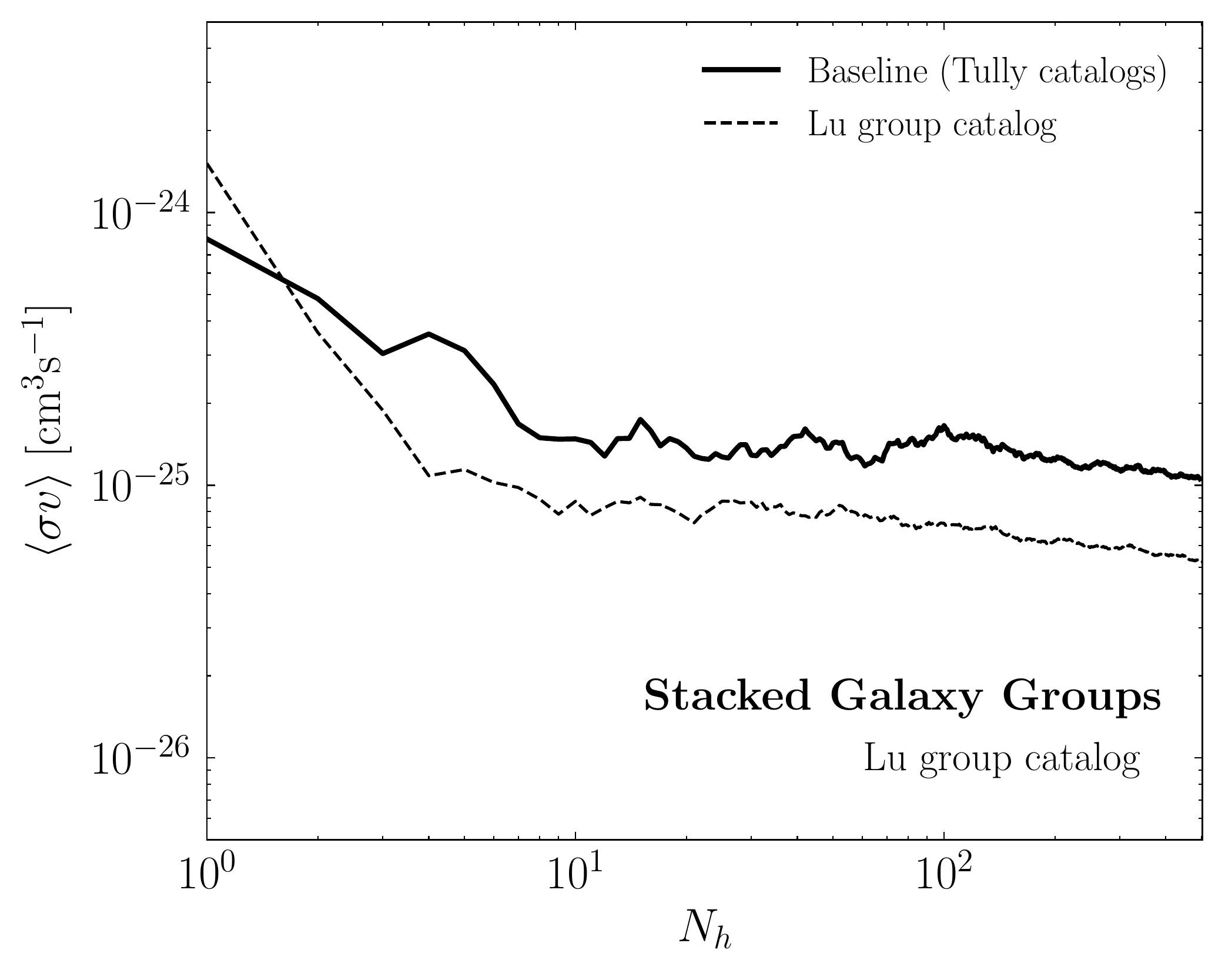} 
  \caption{The same as Fig.~\ref{fig:bounds} of the main Letter, except using the Lu~\emph{et al.} galaxy group catalog~\cite{Lu:2016vmu} (dashed) instead of the T15 and T17 catalogs in the baseline analysis. }
  \label{fig:lucatalog}
\end{figure*}

\noindent There are a variety of sources of systematic uncertainty beyond those described here that  deserve further study.  For example, a systematic bias in the $J$-factor determination due to offsets in either the mass inference or the concentration-mass relation can be a potential source of uncertainty. A better understanding of the galaxy-halo connection and the small-scale structure of halos is required to mitigate this. Furthermore, we assumed distance uncertainties to be subdominant in our analysis. While this is certainly a good assumption over the redshift range of interest---nearby groups have measured distances, while groups further away come with spectroscopic redshift measurements with small expected peculiar velocity contamination---uncertainties on these do exist. We have also assumed that our targets consist of virialized halos and have not accounted for possible out-of-equilibrium effects in modeling these~\cite{1993AJ....105.2035D}.

\end{document}